\pdfoutput=1

\documentclass[11pt,twoside,a4paper,cmspaper,final,collab]{cms-tdr}
\def\svnVersion{397531}\def\cmsCernNoTag{CERN-EP/2016-196}\def\cmsCernDate{\today}\def\cmsMessage{Published in the Journal of High Energy Physics as \href{http://dx.doi.org/10.1007/JHEP03(2017)156}{\doi{10.1007/JHEP03(2017)156.}}}

\begin{document}\cmsNoteHeader{SMP-14-001}

\hyphenation{had-ron-i-za-tion}
\hyphenation{cal-or-i-me-ter}
\hyphenation{de-vices}
\RCS$Revision: 385072 $
\RCS$HeadURL: svn+ssh://svn.cern.ch/reps/tdr2/papers/SMP-14-001/trunk/SMP-14-001.tex $
\RCS$Id: SMP-14-001.tex 385072 2017-02-04 21:12:42Z alverson $
\newlength\cmsFigWidth
\ifthenelse{\boolean{cms@external}}{\setlength\cmsFigWidth{0.85\columnwidth}}{\setlength\cmsFigWidth{0.4\textwidth}}
\ifthenelse{\boolean{cms@external}}{\providecommand{\cmsLeft}{top}}{\providecommand{\cmsLeft}{left}}
\ifthenelse{\boolean{cms@external}}{\providecommand{\cmsRight}{bottom}}{\providecommand{\cmsRight}{right}}
\providecommand{\HERAFITTER} {{\textsc{HERA}fitter}\xspace}
\providecommand{\POWHEGBOX} {{\textsc{POWHEG-BOX}}\xspace}
\providecommand{\HERWIGPP} {{\textsc{herwig++}}\xspace}
\providecommand{\NLOJETPP} {{\textsc{NLOJet++}}\xspace}
\providecommand{\fastNLO} {{\textsc{fastNLO}}\xspace}
\providecommand{\fastjet} {{\textsc{FastJet}}\xspace}
\providecommand{\PYTHIAS} {{\textsc{pythia6}}\xspace}
\providecommand{\PYTHIAE} {{\textsc{pythia8}}\xspace}
\providecommand{\RooUnfold} {{\textsc{RooUnfold}}\xspace}
\providecommand{\lumieff} {\ensuremath{\mathcal{L}_\mathrm{int,eff}}\xspace}
\providecommand{\alpsmz}{\ensuremath{\alpha_\mathrm{S}(M_\mathrm{Z})}\xspace}
\providecommand{\rbthm}{\rule[-2ex]{0ex}{5ex}}
\providecommand{\rbtrr}{\rule[-0.8ex]{0ex}{3.2ex}}
\newcommand{\x}{\ensuremath{\phantom{0}}}
\newcommand{\xx}{\ensuremath{\phantom{00}}}
\newcommand{\as}{\ensuremath{\alpha_\mathrm{S}}\xspace}
\newcommand{\asq}{\ensuremath{\alpha_\mathrm{S}(Q)}\xspace}
\cmsNoteHeader{SMP-14-001}
\title{Measurement and QCD analysis of double-differential inclusive
jet cross sections in pp collisions at
\texorpdfstring{$\sqrt{s}=8\TeV$ and cross section ratios to 2.76 and 7\TeV}{sqrt(s) = 8 TeV and
ratios to 2.76 and 7 TeV}}

\date{\today}
\abstract{
A measurement of the double-differential inclusive jet cross section
as a function of the jet transverse momentum $\pt$ and
the absolute jet rapidity $\abs{y}$ is presented. Data from LHC
proton-proton collisions at $\sqrt{s}=8\TeV$, corresponding
to an integrated luminosity of 19.7\fbinv, have been collected
with the CMS detector. Jets are reconstructed using the
anti-$\kt$ clustering algorithm with a size parameter
of 0.7 in a phase space region covering jet $\pt$ from
74\GeV up to 2.5\TeV and jet absolute rapidity up to $\abs{y}=3.0$. The
low-$\pt$ jet range between 21 and 74\GeV is also
studied up to $\abs{y}=4.7$, using a dedicated data sample
corresponding to an integrated luminosity of 5.6\pbinv. The
measured jet cross section is corrected for detector effects and
compared with the predictions from perturbative QCD at
next-to-leading order (NLO) using various sets of parton
distribution functions (PDF). Cross section ratios to the
corresponding measurements performed at 2.76 and 7\TeV are
presented. From the measured double-differential jet cross section,
the value of the strong coupling constant evaluated at the $\PZ$ mass is
$\alpha_\mathrm{S}(M_{\PZ}) = 0.1164^{+0.0060}_{-0.0043}$,
where the errors include the PDF, scale, nonperturbative effects and
experimental uncertainties, using the CT10 NLO PDFs. Improved
constraints on PDFs based on the inclusive jet cross section
measurement are presented.}
\hypersetup{%
pdfauthor={CMS Collaboration},%
pdftitle={Measurement and QCD analysis of double-differential
  inclusive jet cross sections in pp collisions at sqrt(s) = 8 TeV and
  cross section ratios to 2.76 and 7 TeV},%
pdfsubject={CMS},%
pdfkeywords={QCD, jet cross sections, strong coupling, parton distribution functions}}
\maketitle

\section{Introduction}

Measurement of the cross sections for inclusive jet production in
proton-proton collisions is an ultimate test of quantum chromodynamics
(QCD). The process p + p $\rightarrow$ jet + X probes the
parton-parton interaction as described in perturbative QCD (pQCD), and
is sensitive to the value of the strong coupling constant,
\as. Furthermore, it provides important constraints on the description
of the proton structure, expressed by the parton distribution
functions (PDFs).

In this analysis, the double-differential inclusive jet cross section
is measured at the centre-of-mass energy $\sqrt{s} = 8\TeV$ as a
function of jet transverse momentum \pt and absolute jet rapidity
$\abs{y}$. Similar measurements have been carried out at the CERN LHC by
the ATLAS and CMS Collaborations at
2.76~\cite{Aad:2013lpa,Khachatryan:2015luy} and
7\TeV~\cite{Aad:2010wv,CMSInclusive,Aad:2011fc,Chatrchyan:2012bja},
and by experiments at other hadron colliders
~\cite{Banner:1982kt,Arnison:1983gw,Abulencia:2007ez,Abazov:2008hu,Aaltonen:2008eq}.

The measured inclusive jet cross section at $\sqrt{s} = 7\TeV$ is well
described by pQCD calculations at next-to-leading order (NLO) at small
$\abs{y}$, but not at large $\abs{y}$. The larger data sample at $\sqrt{s} =
8\TeV$ allows QCD to be probed with higher precision extending the
investigations to yet unexplored kinematic regions.  In addition, the
ratios of differential cross sections at different centre-of-mass
energies can be determined. In Ref.~\cite{Mangano:2012mh} an increased
sensitivity of such ratios to PDFs was suggested.

The data were collected with the CMS detector at the LHC during 2012
and correspond to an integrated luminosity of 19.7\fbinv. The average
number of multiple collisions within the same bunch crossing (known as
pileup) is 21. A low-pileup data sample corresponding to an integrated
luminosity of 5.6\pbinv is collected with an average of four
interactions per bunch crossing; this is used for a low-\pt jet cross
section measurement. The measured cross sections are corrected for
detector effects and compared to the QCD prediction at NLO. The
high-\pt part of the differential cross section, where the sensitivity
to the value of \as is maximal, is measured more accurately than
before. Also, the kinematic region of small \pt and large $y$ is
probed. The measured cross section is used to extract the value of the
strong coupling constant at the \PZ~boson mass scale, \alpsmz, and to
study the scale dependence of \as in a wider kinematic range than
is accessible at $\sqrt{s} = 7\TeV$. Further, the impact of the present
measurements on PDFs is illustrated in a QCD analysis using the
present measurements and the cross sections of deep-inelastic
scattering (DIS) at HERA~\cite{Abramowicz:2015mha}.

\section{The CMS detector}

The central feature of the CMS apparatus is a superconducting solenoid
of 6\unit{m} internal dia\-meter, providing a magnetic field of
3.8\unit{T}. Within the solenoid volume are a silicon pixel and strip
tracker, a lead tungstate crystal electromagnetic calorimeter (ECAL),
and a brass and scintillator hadron calorimeter (HCAL), each composed
of a barrel and two endcap sections. Forward calorimeters extend the
pseudorapidity ($\eta$) coverage~\cite{Chatrchyan:2008aa} provided by
the barrel and endcap detectors. Muons are measured in gas-ionization
detectors embedded in the steel flux-return yoke outside the solenoid.

The silicon tracker measures charged particles within the
pseudorapidity range $\abs{\eta}< 2.5$. It consists of 1440 silicon
pixel and 15\,148 silicon strip detector modules.
For non-isolated particles of $1 <
\pt < 10\GeV$ and $\abs{\eta} < 1.4$, the track resolutions are
typically 1.5\% in \pt and 25--90 (45--150)\unit{\mum} in the
transverse (longitudinal) impact parameter \cite{TRK-11-001}.  The
ECAL consists of 75\,848 lead tungstate crystals, which provide
coverage in $\abs{ \eta } < 1.479 $ in a barrel region (EB) and $1.479
< \abs{ \eta } < 3.0$ in two endcap regions (EE). A preshower detector
consisting of two planes of silicon sensors interleaved with a total
of $3 X_0$ of lead is located in front of the EE. In the region $\abs{
\eta }< 1.74$, the HCAL cells have widths of 0.087 in $\eta$ and
0.087 radians in azimuth ($\phi$). In the $\eta$--$\phi$ plane, and
for $\abs{\eta}< 1.48$, the HCAL cells map on to $5 \times 5$ arrays
of ECAL crystals to form calorimeter towers projecting radially
outwards from close to the nominal interaction point. For $\abs{\eta}
> 1.74$, the coverage of the towers increases progressively to a
maximum of 0.174 in $\Delta\eta$ and $\Delta\phi$. The hadronic forward
(HF) calorimeters consist of iron absorbers with embedded
radiation-hard quartz fibres, located at 11.2\unit{m} from the interaction
point on both sides of the experiment covering the region of $2.9 <
|\eta| < 5.2$. Half of the HF fibres run over the full depth of the
absorber, while the other half start at a depth of 22\cm from the
front of the detector to allow for a separation between
electromagnetic and hadronic showers. The $\eta$--$\phi$ tower
segmentation of the HF calorimeters is $0.175 \times 0.175$, except
for $\eta$ above 4.7, where the segmentation is $0.175 \times 0.35$.

The first level of the CMS trigger system, composed of custom hardware
processors, uses information from the calorimeters and muon detectors
to select events in a fixed time interval of less than
4\unit{\mus}. The high-level trigger (HLT) processor farm further
decreases the event rate from 100\unit{kHz} to around 400\unit{Hz},
before data storage. A more detailed description of the CMS detector,
together with a definition of the coordinate system used and the
relevant kinematic variables, can be found in
Ref.~\cite{Chatrchyan:2008aa}.

\section{Jet reconstruction and event selection}

The high-\pt jet measurement is based on data sets collected with six
single-jet triggers in the HLT system that require at least one jet
in the event with jet $\pt >$ 40, 80, 140, 200, 260, and 320\GeV,
respectively. All triggers were prescaled during the 2012 data-taking
period except the highest threshold trigger. The efficiency of each
trigger is estimated using triggers with lower \pt thresholds, and
each is found to exceed 99\% above the nominal \pt threshold.
The \pt thresholds of each trigger and the corresponding effective
integrated luminosity are listed in Table~\ref{table:lumi}. The jet
\pt range, reconstructed in the offline analysis, where the trigger
with the lowest \pt threshold becomes fully efficient is also shown.
This analysis includes jets with $74 < \pt < 2500\GeV$.

\begin{table}[htb]
\centering
\topcaption{HLT trigger ranges and effective integrated
luminosities used in the jet cross section measurement. The
luminosity is known with a 2.6\% uncertainty.}
\resizebox{\textwidth}{!}{
\begin{tabular}{cccccccc}
\hline
\begin{tabular}{c}Trigger \pt \\ threshold (\GeVns) \end{tabular}&  40 & 80 & 140 & 200 &  260 & 320 \\
\hline
\begin{tabular}{c}Offline analysis \\ \pt range (\GeVns) \end{tabular}&  74--133 & 133--220 & 220--300 & 300--395 & 395--507 & 507--2500 \\
\hline
\begin{tabular}{c}Effective integrated \\ luminosity ($\mathrm{pb}^{-1}$)  \end{tabular} &   $7.9 \times 10^{-2}$ & 2.12                & $55.7$ & $2.61 \times 10^{2}$ & $1.06 \times 10^{3}$ & $1.97 \times 10^{4}$ \\
\hline
\end{tabular}
}
\label{table:lumi}
\end{table}

Events for the low-\pt jet analysis are collected with a trigger that
requires at least two charged tracks reconstructed in the pixel
detector in coincidence with the nominal bunch crossing
time. This selection is highly efficient for finding jets (${\simeq}
100\%$) and also rejects noncollision background. The \pt range
considered in the low-\pt jet analysis is 21--74\GeV.

The particle-flow (PF) event algorithm reconstructs and identifies
each individual particle with an optimized combination of information
from the various elements of the CMS
detector~\cite{CMS-PAS-PFT-09-001,CMS-PAS-PFT-10-001}. Selected events
are required to have at least one reconstructed interaction vertex,
and the primary interaction vertex (PV) is defined as the
reconstructed vertex with the largest sum of $\pt^2$ of its
constituent tracks. The PV is required to be reconstructed from at
least five tracks and to lie within 24\cm in the longitudinal
direction from the nominal interaction point~\cite{TRK-11-001}, and to
be consistent with the measured transverse position of the beam. The
energy of photons is obtained directly from the ECAL measurement and
is corrected for zero-suppression effects. The energy of electrons is
determined from a combination of the electron momentum at the PV as
determined by the tracker, the energy of the corresponding ECAL
cluster, and the energy sum of all bremsstrahlung photons spatially
compatible with originating from the electron track. The transverse
momentum of muons is obtained from the curvature of the corresponding
track. The energy of charged hadrons is determined from a combination
of their momentum measured in the tracker and the matching ECAL and
HCAL energy deposits, corrected for zero-suppression effects and for
the response function of the calorimeters to hadronic
showers. Finally, the energy of neutral hadrons is obtained from the
corresponding corrected ECAL and HCAL energies. In the forward region,
the energies are measured in the HF detector.

For each event, hadronic jets are clustered from the reconstructed
particles with the infrared and collinear safe anti-$\kt$
algorithm~\cite{Cacciari:2008gp}, as implemented in the \fastjet
package~\cite{Cacciari:2011ma}, with a size parameter $R$ of 0.7. Jet
momentum is determined as the vector sum of the momenta of all
particles in the jet, and is found from simulation to be within 5\% to
10\% of the true momentum over the whole \pt spectrum and detector
acceptance, before corrections are applied. In order to suppress the
contamination from pileup, only reconstructed charged particles
associated to the PV are used in jet clustering. Jet energy scale
(JES) corrections are derived from simulation, by using events
generated with \PYTHIAS and processed through the CMS detector
simulation that is based on the \GEANT4~\cite{GEANT4} package, and
from in situ measurements by exploiting the energy balance in dijet,
photon+jet, and $\PZ$+jet events~\cite{JES,Khachatryan:2016kdb}.  The
\PYTHIAS version 4.22~\cite{PYTHIA} is used, with the Z2$^*$ tune. The
Z2$^*$ tune is derived from the Z1 tune~\cite{Field:2010bc} but uses
the CTEQ6L~\cite{Pumplin:2002vw} parton distribtion set whereas the Z1
tune uses the CTEQ5L set.  The Z2$^*$ tune is the result of retuning
the \PYTHIAS parameters PARP(82) and PARP(90) by means of the
automated PROFESSOR tool~\cite{Buckley:2009bj}, yielding
PARP(82)=1.921 and PARP(90)=0.227. The JES corrections account for
residual nonuniformities and nonlinearities in the detector
response. An offset correction is required to account for the extra
energy clustered into jets due to pileup. The JES correction, applied
as a multiplicative factor to the jet four momentum vector, depends on
the values of jet $\eta$ and \pt. For a jet with a \pt of 100\GeV the
typical correction is about 10\%, and decreases with increasing
\pt. The jet energy resolution (JER) is approximately 15\% at 10\GeV,
8\% at 100\GeV, and 4\% at 1\TeV.

The missing transverse momentum vector, \ptvecmiss, is defined as the
projection on the plane perpendicular to the beams of the negative
vector sum of the momenta of all reconstructed particles in an
event. Its magnitude is referred to as \ETmiss. A requirement is made
that the ratio of \ETmiss and the sum of the transverse energy of the
PF particles is smaller than 0.3, which removes background events and
leaves a negligible residual contamination. Additional selection
criteria are applied to each event to remove spurious jet-like
signatures originating from isolated noise patterns in certain HCAL
regions.  To suppress the noise patterns, tight identification
criteria are applied: each jet should contain at least two PF
particles, one of which is a charged hadron, and the jet energy
fraction carried by neutral hadrons and photons should be less than
90\%. These criteria have an efficiency greater than 99\% for genuine
jets. Events are selected that contain at least one jet with a \pt
higher than the \pt threshold of the lowest-threshold trigger that
recorded the event.

\section{Measurement of the jet differential cross section}

The double-differential inclusive jet cross section is defined as
\begin{equation}
\label{master_inclusive}
\frac{\rd^2\sigma}{\rd\pt\rd y} =
\frac{1}{\epsilon\lumieff}\,\frac{N_\mathrm{{jets}}}{\Delta\pt\,(2\Delta \abs{y})}\,~,
\end{equation}
where $\mathrm{N_{jets}}$ is the number of jets in a kinematic
interval (bin) of transverse momentum and rapidity, $\Delta \pt$ and
$\Delta \abs{y}$, respectively; \lumieff is the effective integrated
luminosity contributing to the bin; $\epsilon$ is the product of the
trigger and jet selection efficiencies, and is greater than
$99\%$. The widths of the \pt bins increase with \pt and are
proportional to the \pt resolution. The phase space in absolute
rapidity $\abs{y}$ is subdivided into six bins starting from
$y=0$ up to $\abs{y}=3.0$ with $\Delta \abs{y} = 0.5$. In the low-\pt jet
measurement an additional rapidity bin $3.2 < \abs{y} < 4.7$ is included. The
statistical uncertainty for each bin is computed according to the
number of events contributing to at least one entry per
event~\cite{Chatrchyan:2012bja}, corrected for possible multiple
entries per event. This correction is small, since at least 90\% of
the observed jets in each $\Delta \pt$ and $\Delta \abs{y}$ bin originate
from different events.

In order to compare the measured cross section with theoretical
predictions at particle level, the steeply falling jet \pt spectra
must be corrected for experimental \pt resolution.  An unfolding
procedure, based on the iterative D'Agostini
method~\cite{D'Agostini:1994zf}, implemented in the {\sc RooUnfold}
package~\cite{RooUnfold}, is used to correct the measured spectra for
detector effects. The response matrix is created by the convolution of
theoretically predicted spectra, discussed in
Section~\ref{sec:theory}, with the JER effects.  These effects are
evaluated as a function of \pt with the CMS detector simulation, after
correcting for the residual differences from data~\cite{JES}.  The
unfolding procedure induces statistical correlations among the
bins. The sizes of these correlations typically vary between 10\% and
20\%.

The dominant contribution to the experimental systematic uncertainty
in the measured cross section is from the JES corrections, determined
as in Ref.~\cite{JES,Khachatryan:2016kdb}. For the high-\pt
jet data set, this uncertainty is decomposed into 24 independent
sources, corresponding to the different components of the corrections:
pileup effects, relative calibration of JES versus $\eta$, absolute
JES including \pt dependence, and differences in quark- and
gluon-initiated jets. The set of components, used here, is discussed
in detail in Ref.~\cite{Khachatryan:2016kdb}, and represents an
evolution of the decomposition presented in
Ref.~\cite{Khachatryan:2014waa}. The low-pileup data set uses a
reduced number of components, since the pileup-related corrections are
negligible, and there is no JES time dependence.  Moreover, the
central values of the corrections, for the components common between
the two data sets, are not the same; the low-\pt jet analysis uses
corrections computed only on the initial part of the 2012 data
sample. The impact of the uncertainty induced by each correction
component on the measured cross section is evaluated separately. The
JES-induced uncertainty in the cross section depends on \pt and
$y$. For the high-\pt data, this ranges from 2\% to 4\% in the sub-TeV
region at central rapidity to about 20\% in the highest \pt bins for
rapidities $1.0 < \abs{y} < 2.0$. Due to the different set of corrections
used, the low-\pt jet cross section has a larger JES uncertainty than
the contiguous bins of the high-\pt part, and this effect becomes more
pronounced as the jet rapidity increases.

To account for the residual effects of small inefficiencies of less
than 1\% in the trigger performances and jet identification, an
uncertainty of 1\%, uncorrelated across all jet \pt and $y$ bins, is
assigned to each bin.

The unfolding procedure is affected by the uncertainties in the JER
parameterization, which are derived from the simulation. The JER parameters
are varied by one standard deviation up and down, and the corresponding response matrices
are used to unfold the measured spectra. The JER-induced uncertainty
amounts to 1--5\% in the high-\pt jet region, but can
exceed 30\% in the low-\pt jet region.

The uncertainties in the integrated luminosity, which propagate
directly to the cross section, are $2.6\%$~\cite{CMS-PAS-LUM-13-001}
and $4.4\%$~\cite{CMS-PAS-LUM-12-001} for normal and low-pileup data samples,
respectively. Other sources of uncertainty, such as the jet angular
resolution and the model dependence of the unfolding, arise from the
theoretical \pt spectrum used to calculate the response matrix and have
less than 1\% effect on the cross section. The total experimental
systematic uncertainty in the measured cross section is obtained as a
quadratic sum of contributions due to uncertainties in JES, JER, and integrated luminosity.

\section{Theoretical predictions}
\label{sec:theory}

Theoretical predictions for the jet cross section are known at NLO accuracy in
pQCD~\cite{Nagy:2001fj,Nagy:2003tz}, and the NLO electroweak corrections have been
computed in Ref.~\cite{EWK}. The pQCD NLO calculations are
performed by using the \NLOJETPP (version 4.1.3)
program~\cite{Nagy:2001fj,Nagy:2003tz} as implemented in the
\fastNLO(version 2.1) package~\cite{Britzger:2012bs}. The
renormalization ($\mu_\mathrm{R}$) and factorization
($\mu_\mathrm{F}$) scales are both set to the leading jet \pt. The calculations are
performed by using six PDF sets determined at NLO: CT10~\cite{CT10}, MSTW2008~\cite{Martin:2009ad},
NNPDF2.1~\cite{NNPDF}, NNPDF3.0~\cite{NNPDF30},
HERAPDF1.5~\cite{Aaron:2009aa}, and ABM11~\cite{ABKM11}. Each PDF set
is available for a range of \alpsmz values. The number of active
(massless) flavours chosen in \NLOJETPP is five in all of the PDF sets
except NNPDF2.1, where it is set to six.  All the PDF sets use a
variable flavour number scheme, except ABM11, which uses a fixed flavour
number scheme. The basic characteristics of each PDF set are
summarized in Table~\ref{tab:pdfsets}.

\begin{table}[htbp]
\topcaption{%
The PDF sets used in comparisons to the data together with the
corresponding number of active flavours $N_\mathrm{f}$, the assumed masses
$M_{\PQt}$ and $M_{\PZ}$ of the top quark and $\PZ$
boson, the default values of the strong coupling constant \alpsmz,
and the ranges in \alpsmz available for fits. For CT10 the updated
versions of 2012 are used.%
}
\label{tab:pdfsets}
\centering
\begin{tabular}{lllcclcc}
\hline
PDF set & Refs. & Order\ & $N_\mathrm{f}$ & $M_\PQt$ (\GeVns) &
$M_\PZ$ (\GeVns) &\alpsmz & \alpsmz range\rbthm\\
\hline
ABM11      & \cite{ABKM11} & NLO  &       5  & 180 & 91.174 & 0.1180 & 0.110--0.130\rbtrr\\
CT10       & \cite{CT10}     & NLO  & ${\leq} 5$ & 172 & 91.188 & 0.1180 & 0.112--0.127\rbtrr\\
HERAPDF1.5 & \cite{Aaron:2009aa}   & NLO  & ${\leq} 5$ & 180 & 91.187 & 0.1176 & 0.114--0.122\rbtrr\\
MSTW2008   & \cite{Martin:2009ad} & NLO  & ${\leq} 5$ & $10^{10}$ & 91.1876 & 0.1202 & 0.110--0.130\rbtrr\\
NNPDF2.1   & \cite{NNPDF}    & NLO  & ${\leq} 6$ & 175 & 91.2 & 0.1190 & 0.114--0.124\rbtrr\\
NNPDF3.0   & \cite{NNPDF30}    & NLO  & ${\leq} 5$ & 175 & 91.2 & 0.1180 & 0.115--0.121\rbtrr\\
\hline
\end{tabular}
\end{table}

\begin{figure}[ht]
\centering
\includegraphics[width=0.45\textwidth]{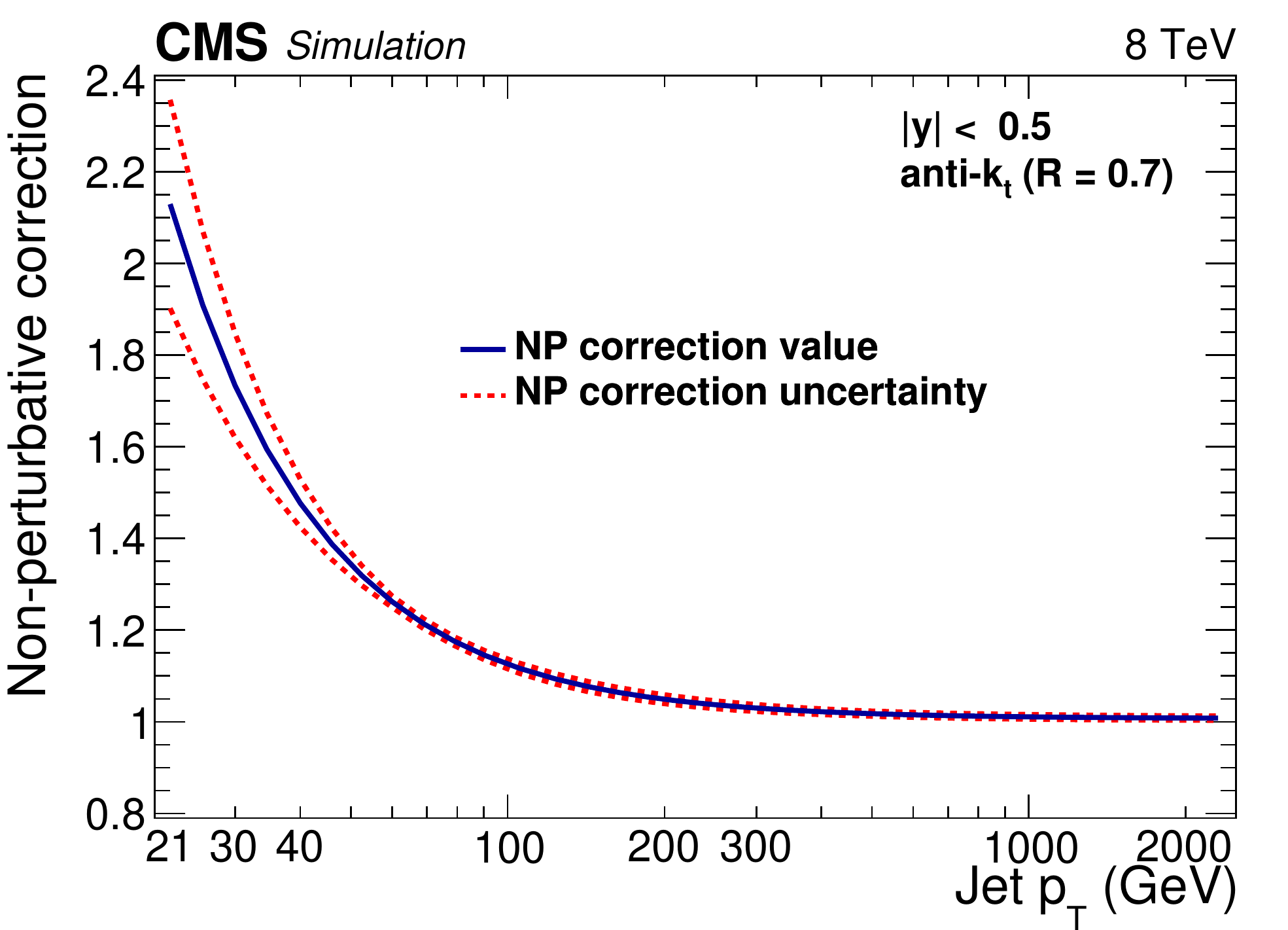}
\includegraphics[width=0.45\textwidth]{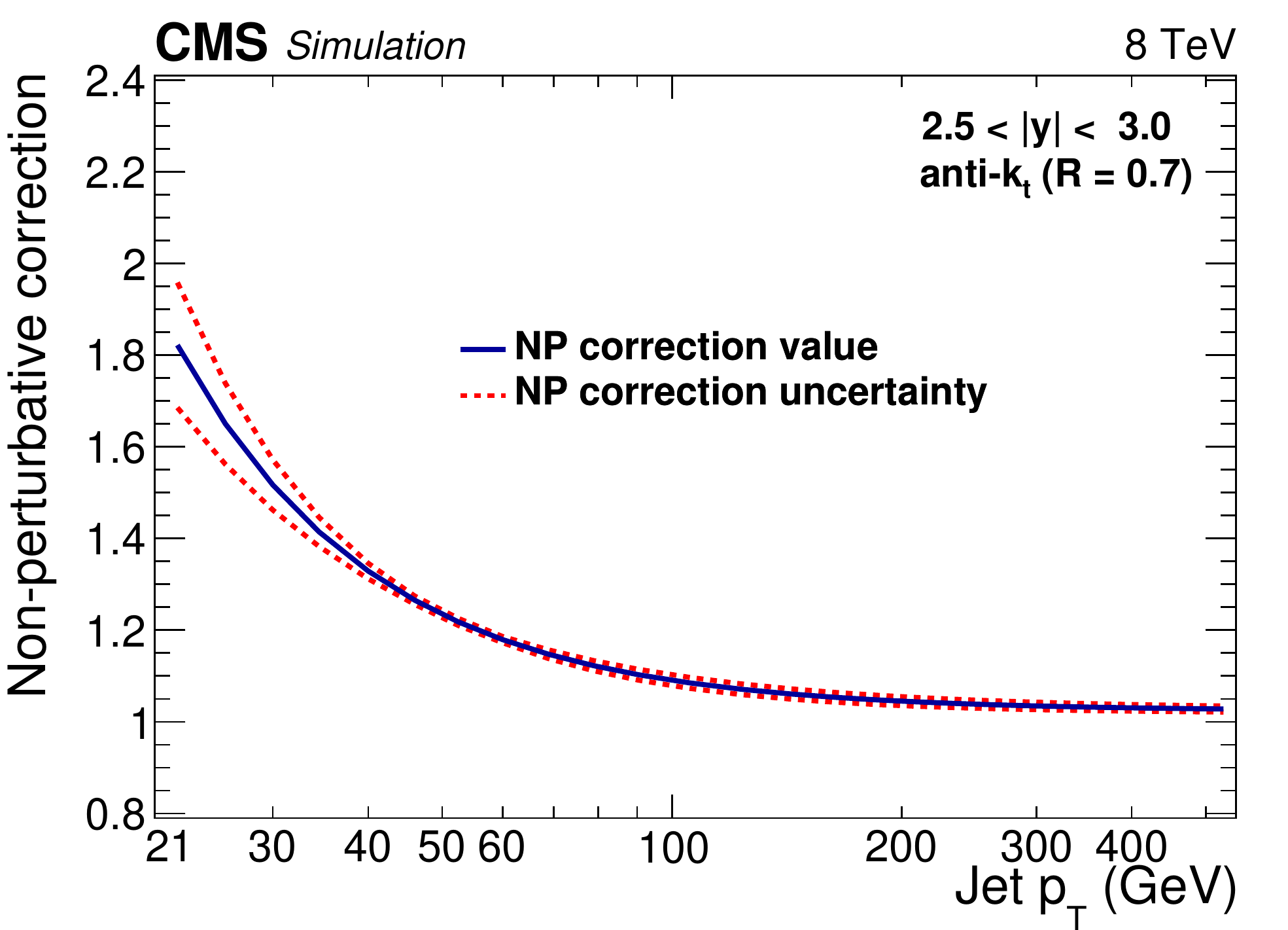}
\caption{The nonperturbative correction factor shown for the central (left) and
outermost (right) absolute rapidity bins as a function of jet \pt. The
correction is obtained by averaging LO- and NLO-based
predictions, and the envelope of these predictions is used as the
uncertainty band.}
\label{fig:NPCOR}
\end{figure}

The parton-level calculation at NLO has to be supplemented with
corrections due to non\-per\-tur\-ba\-tive (NP) effects,
i.e.\ hadronization and multiparton interactions (MPI). The nonperturbative effects
are estimated using both leading order (LO) and NLO event
generators. In the former case, the correction is evaluated by
averaging those provided by \PYTHIAS~\cite{PYTHIA} (version 4.26),
using tune Z2$^*$, and \HERWIGPP (version 2.4.2)~\cite{HERWIG}, using
tune UE~\cite{Seymour:2013qka}.  The size of these corrections ranges
from 20\% at low \pt to 1\% at the highest \pt of 2.5\TeV. The NLO nonperturbative
correction is derived using
\POWHEG~\cite{Nason:2004rx,Frixione:2007vw,Alioli:2010xd,Alioli:2010xa},
interfaced with \PYTHIAS for parton shower, MPI, and
hadronization. The nonperturbative correction factors are derived in this case by
averaging the results for two different tunes of \PYTHIAS, Z2$^*$ and
P11~\cite{Skands:2010ak}. Hadronization models have been tuned by
using LO calculations for the hard scattering, and applying these
tunes to NLO-based calculations is not expected to provide optimal
results.  On the other hand, the application of nonperturbative corrections based
on LO calculations to NLO predictions implicitly assumes that the
behaviour of nonperturbative effects is independent of the hard scattering
description. To take into account both facts, the final number used
for the nonperturbative correction, $C^{\mathrm {NP}}$, is an arithmetic average of
the LO- and NLO-based estimates.  Half the width of the envelope of
these predictions is used as the uncertainty due to the nonperturbative
correction. Figure \ref{fig:NPCOR} shows the nonperturbative correction factors
derived by combining both LO- and NLO-based calculations.

The uncertainty in the NLO pQCD calculation arising from missing
higher-order corrections is estimated by varying the renormalization
and factorization scales in the following six combinations of scale
factors: ($\mu_{\mathrm{R}}/\mu,\mu_{\mathrm{F}}/\mu) = (0.5,0.5)$,
$(2,2)$, $(1,0.5)$, $(1,2)$, $(0.5,1)$, $(2,1)$, where $\mu$ is the
default choice equal to the jet $\pt$, and considering the largest
variation in the prediction as the uncertainty.  The uncertainty
related to the choice of scale ranges from 5\% to 10\% for $\abs{y} < 1.5$
and increases to 40\% for the outer $\abs{y}$ bins and for high \pt. The
PDF uncertainties are estimated following the prescription from each
PDF group by using the provided eigenvectors (or replicas in case of
NNPDF). The corresponding uncertainty in the predicted cross section
varies from 5\% to 30\% in the entire \pt range for $\abs{y} \leq 1.5$.
Beyond $\abs{y}=1.5$, in the outer rapidity region, these uncertainties
become as large as 50\% at high \pt and even increase up to 100\% for
the CT10 and HERAPDF1.5 sets.  The nonperturbative correction induces an additional
uncertainty, which is estimated in the central rapidity bin to range
between $1.4\%$ at $\pt \sim 100\GeV$ to $0.06\%$ at ${\sim}2.5\TeV$.
Overall, the PDF uncertainty is dominant.

\begin{figure}[htb]
\centering
\includegraphics[width=0.45\textwidth]{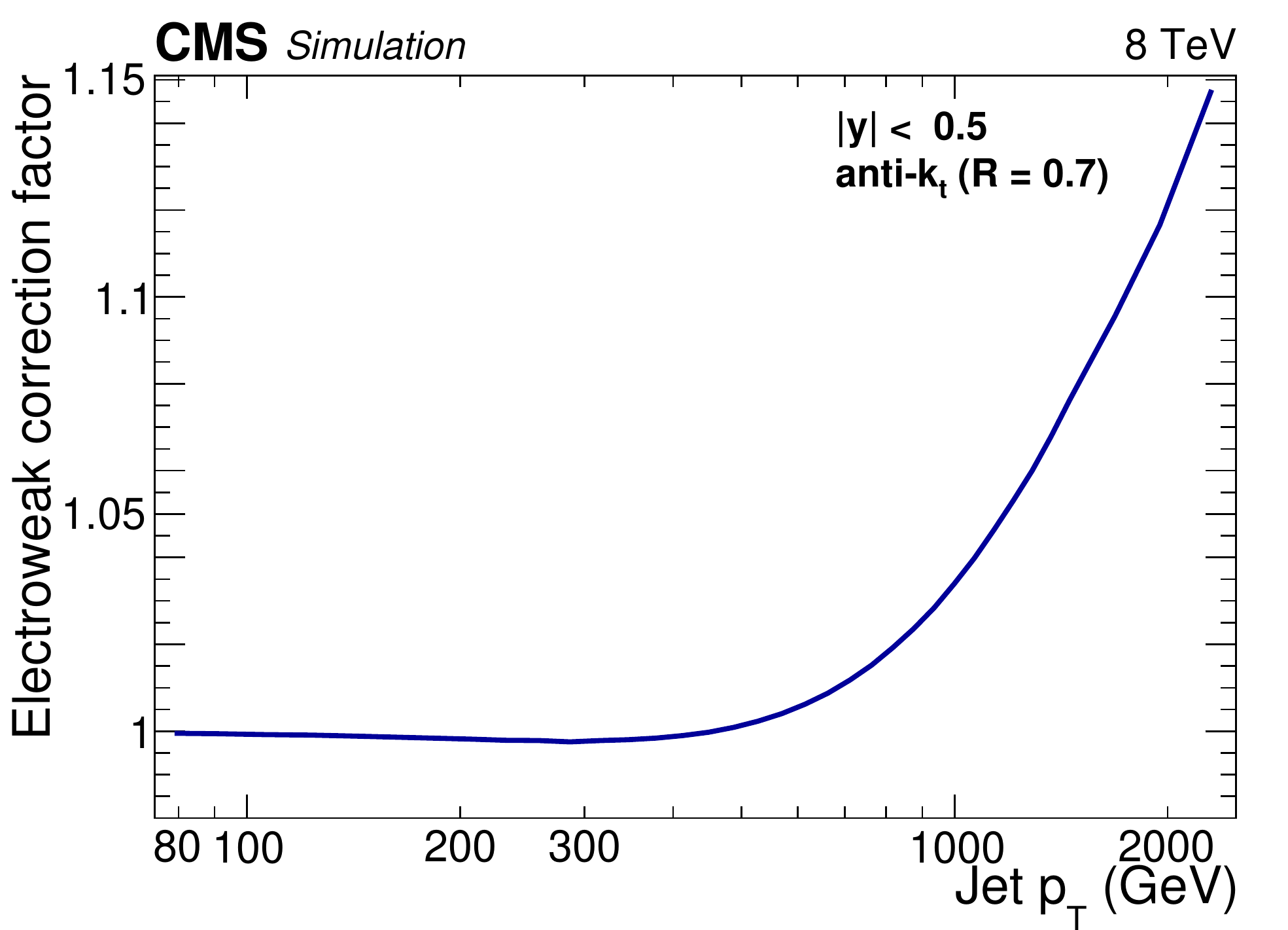}
\includegraphics[width=0.45\textwidth]{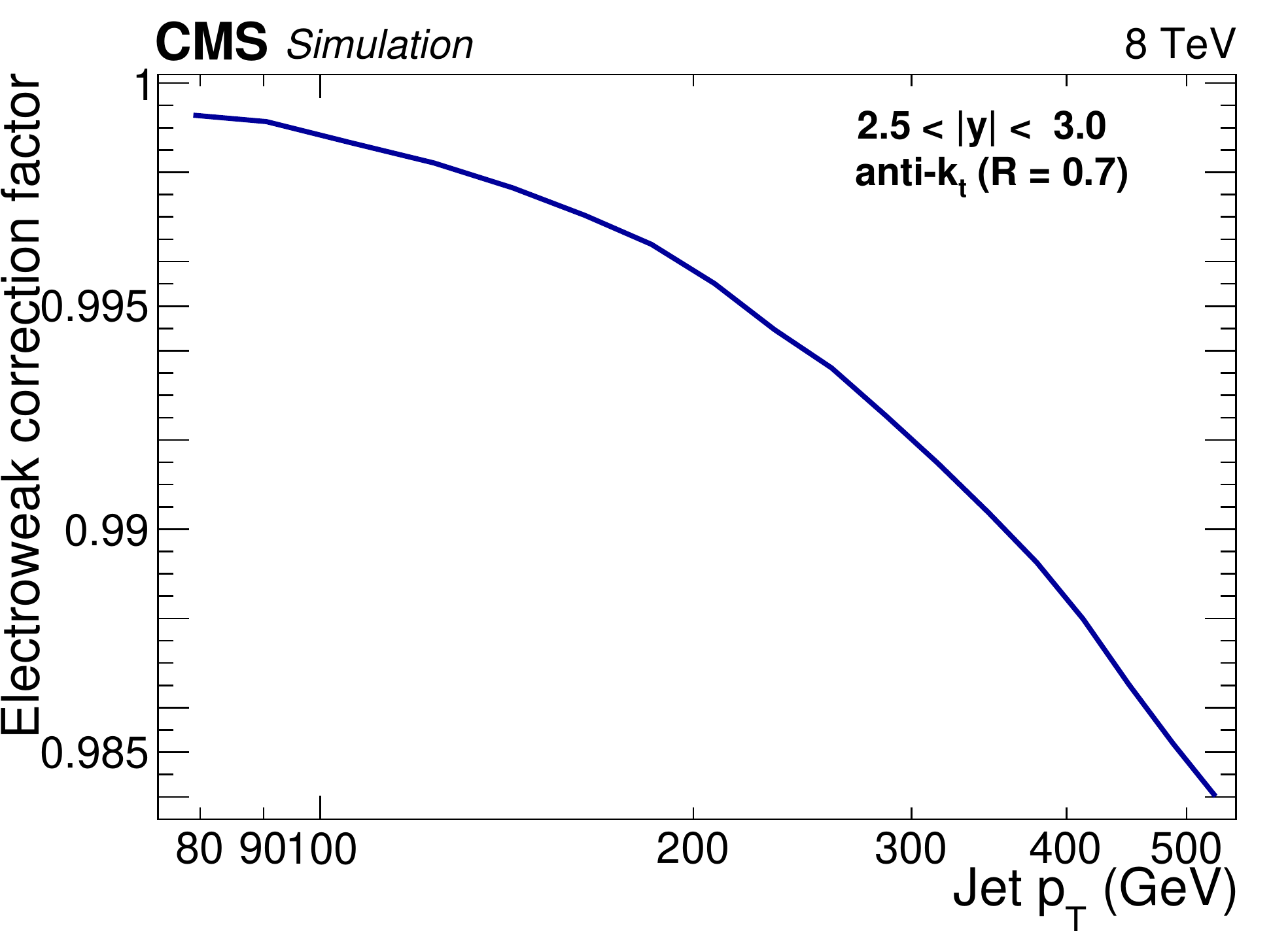}
\caption{Electroweak correction factor for the central (left) and
outermost (right) rapidity bins as a function of jet \pt.}
\label{fig:EWKCOR}
\end{figure}

Electroweak effects, which arise from the virtual exchange of the
massive $\PW$ and $\PZ$ gauge bosons, induce corrections with
magnitudes given by the Sudakov logarithmic factor \\
$\alpha_{\PW}\ln^2(Q^2/M_{\PW}^2)$, where $\alpha_{\PW}$ is the weak
coupling constant, $M_\PW$ is the mass of the $\PW$ boson, and $Q$ is
the energy scale of the interaction. For high-\pt jets, the values of
the logarithm, and therefore the correction, become large. The
derivation of the electroweak correction factor, applied to the NLO
pQCD spectrum corrected for nonperturbative effects, is provided in
Ref.~\cite{EWK}.  Figure \ref{fig:EWKCOR} shows the electroweak
correction for the two extreme rapidity regions as a function of jet
\pt.  In the most central rapidity bin for the high-\pt region, the
correction factor is as large as 14\%. Electroweak corrections are not
applied to the low-\pt results, where they are negligible.

\section{Comparison of theory and data}
\label{ref:compa}

\begin{figure}[hbt]
\centering
\includegraphics[width=0.75\textwidth]{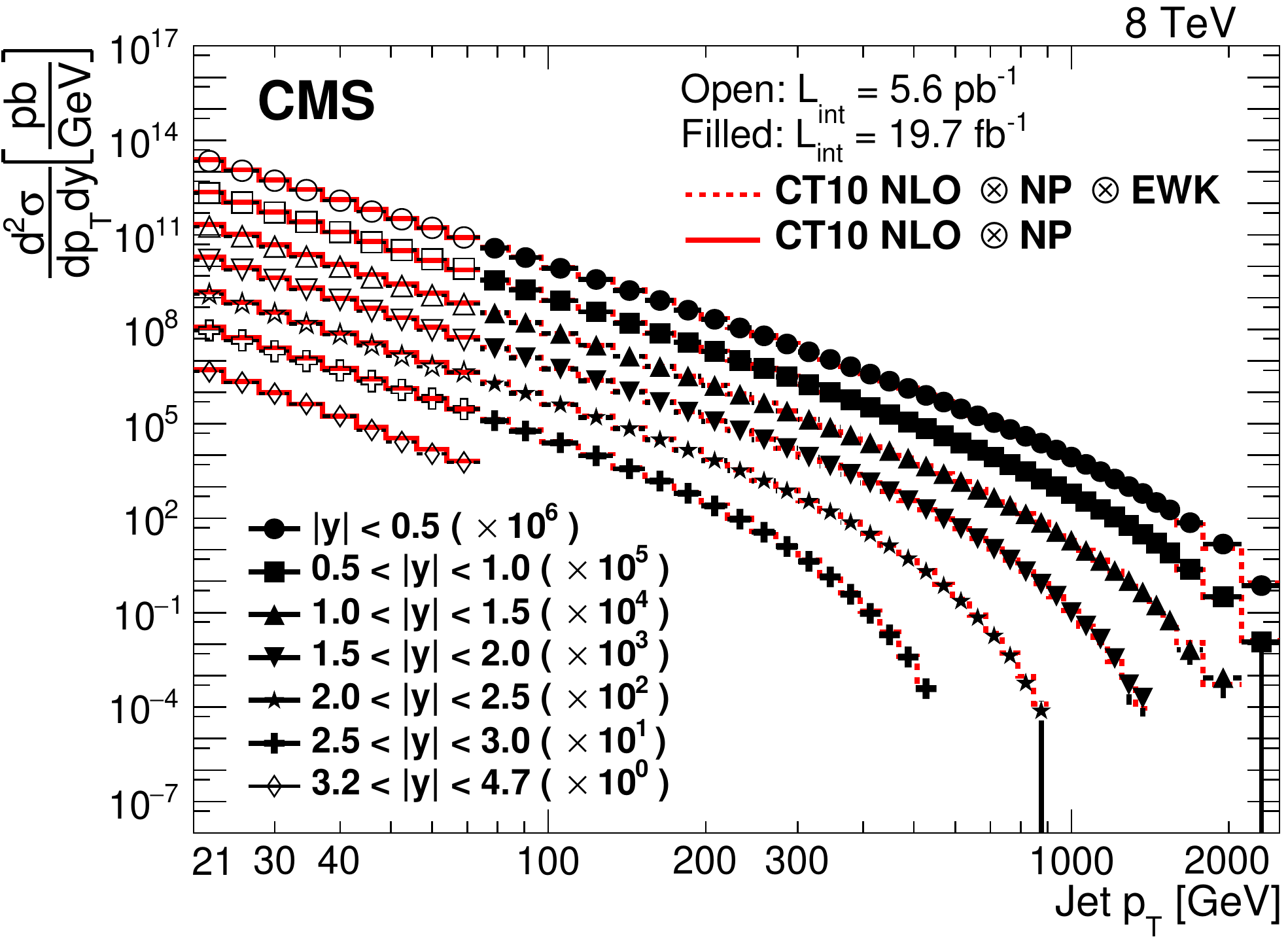}
\caption{Double-differential inclusive jet cross sections as function of jet
\pt. Data (open points for the low-\pt analysis, filled points for
the high-\pt one) and NLO predictions based on the CT10 PDF set
corrected for the nonperturbative factor for the low-\pt data (solid line) and the
nonperturbative and electroweak correction factors for the high-\pt data (dashed
line). The comparison is carried out for six different $\abs{y}$ bins
at an interval of $\Delta \abs{y} = 0.5$.}
\label{fig:ResultCT10}
\end{figure}

\begin{figure*}[hbpt]
\centering
\includegraphics[width=0.45\textwidth]{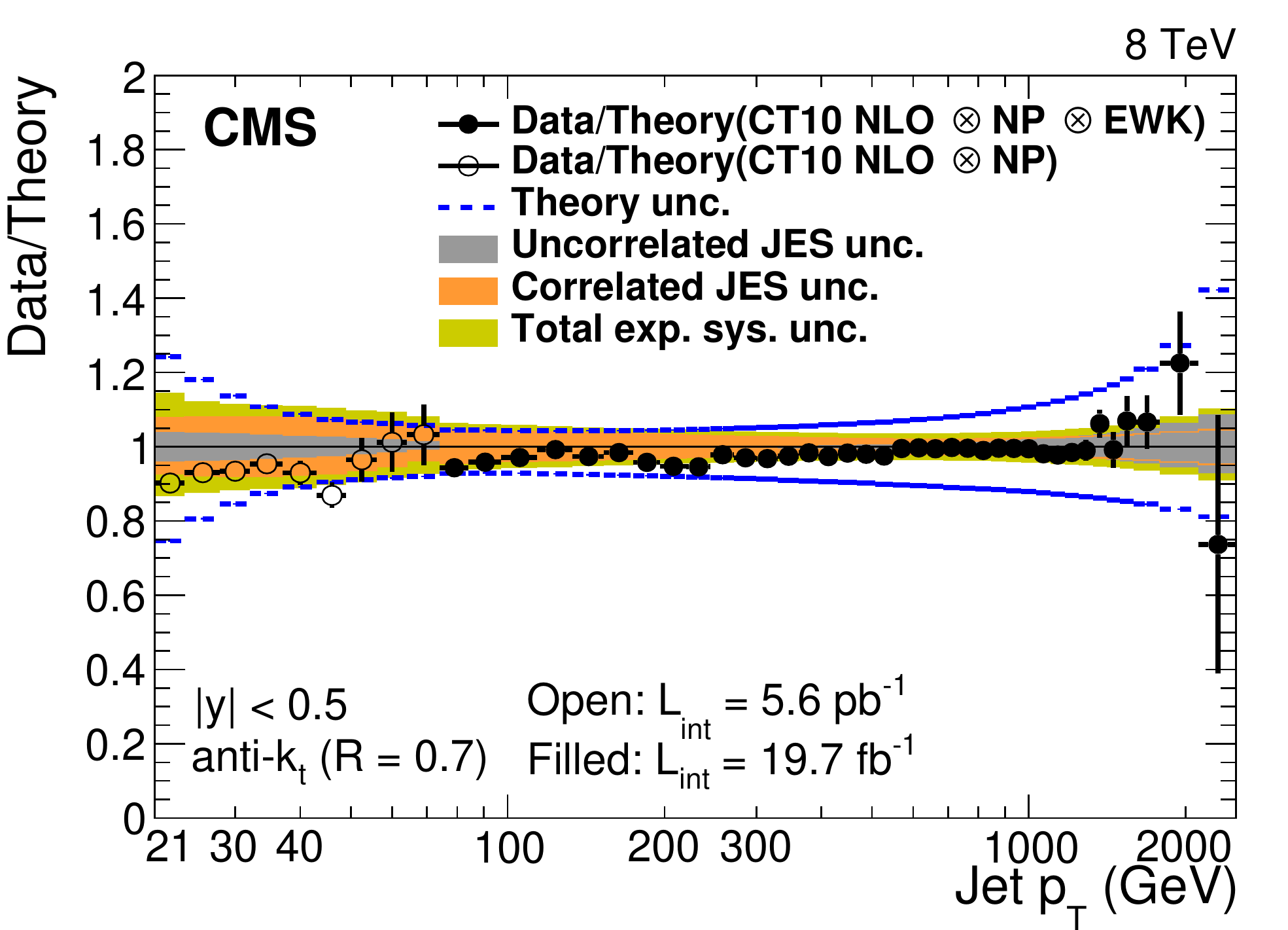}
\includegraphics[width=0.45\textwidth]{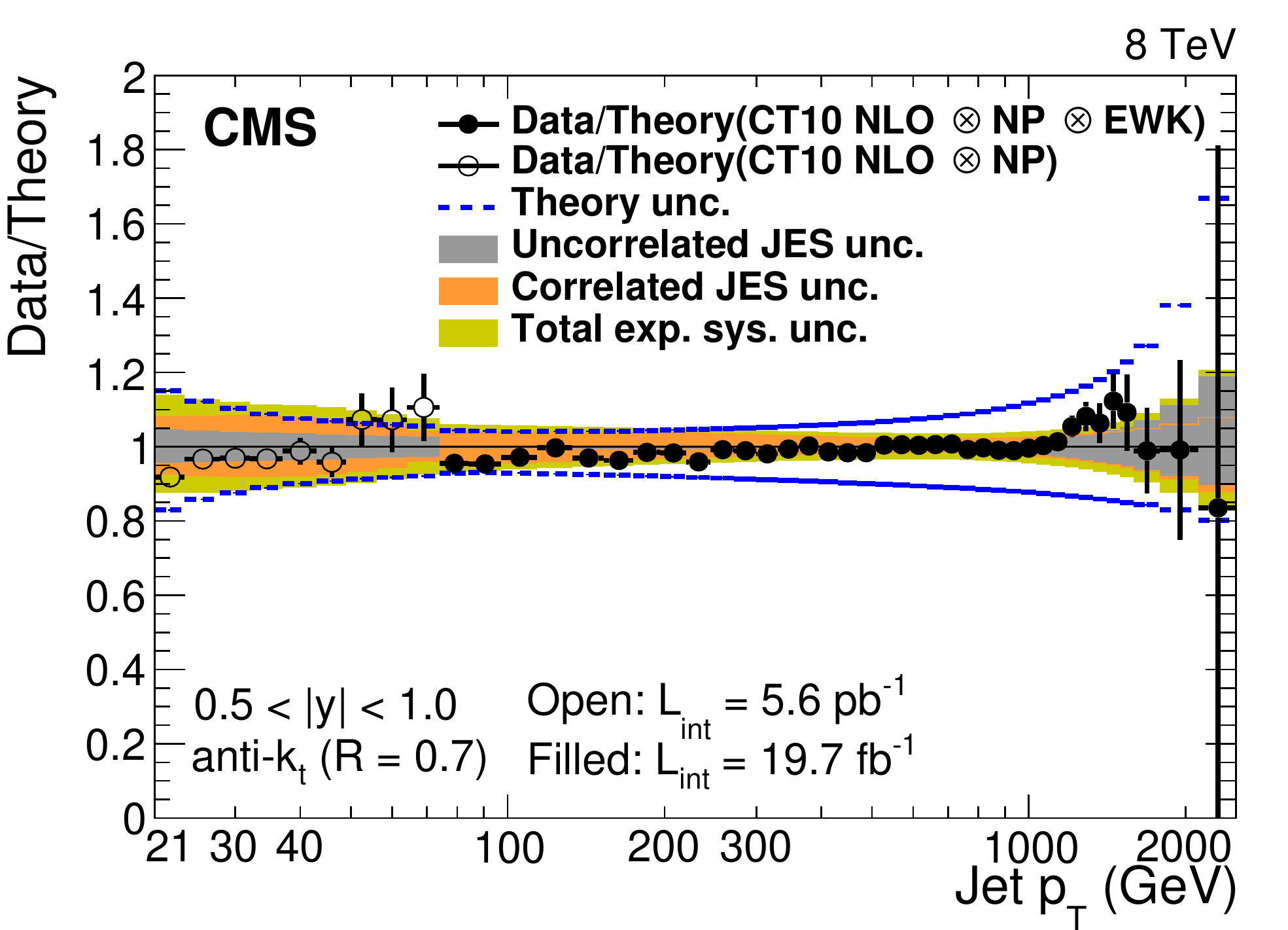}
\includegraphics[width=0.45\textwidth]{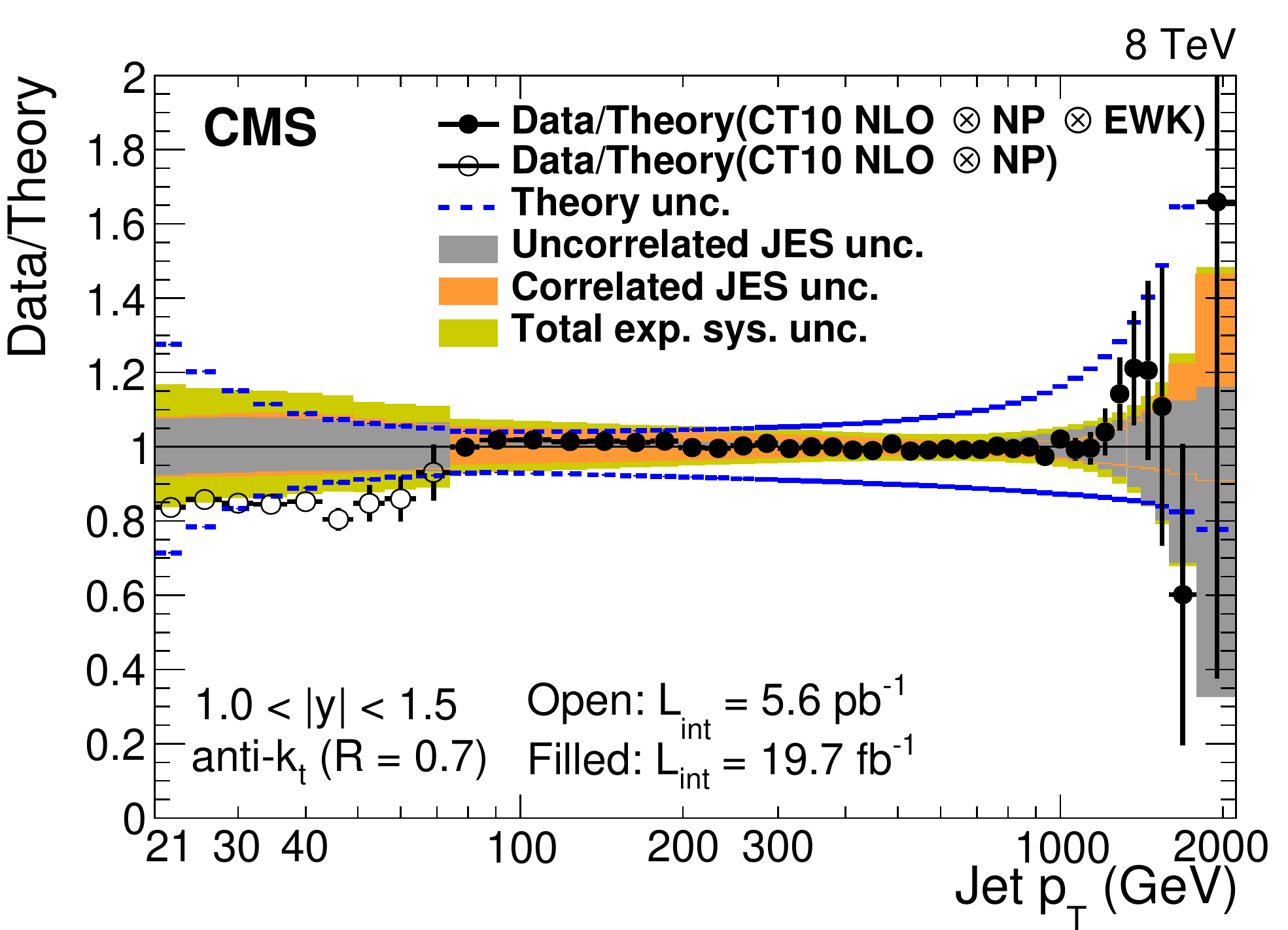}
\includegraphics[width=0.45\textwidth]{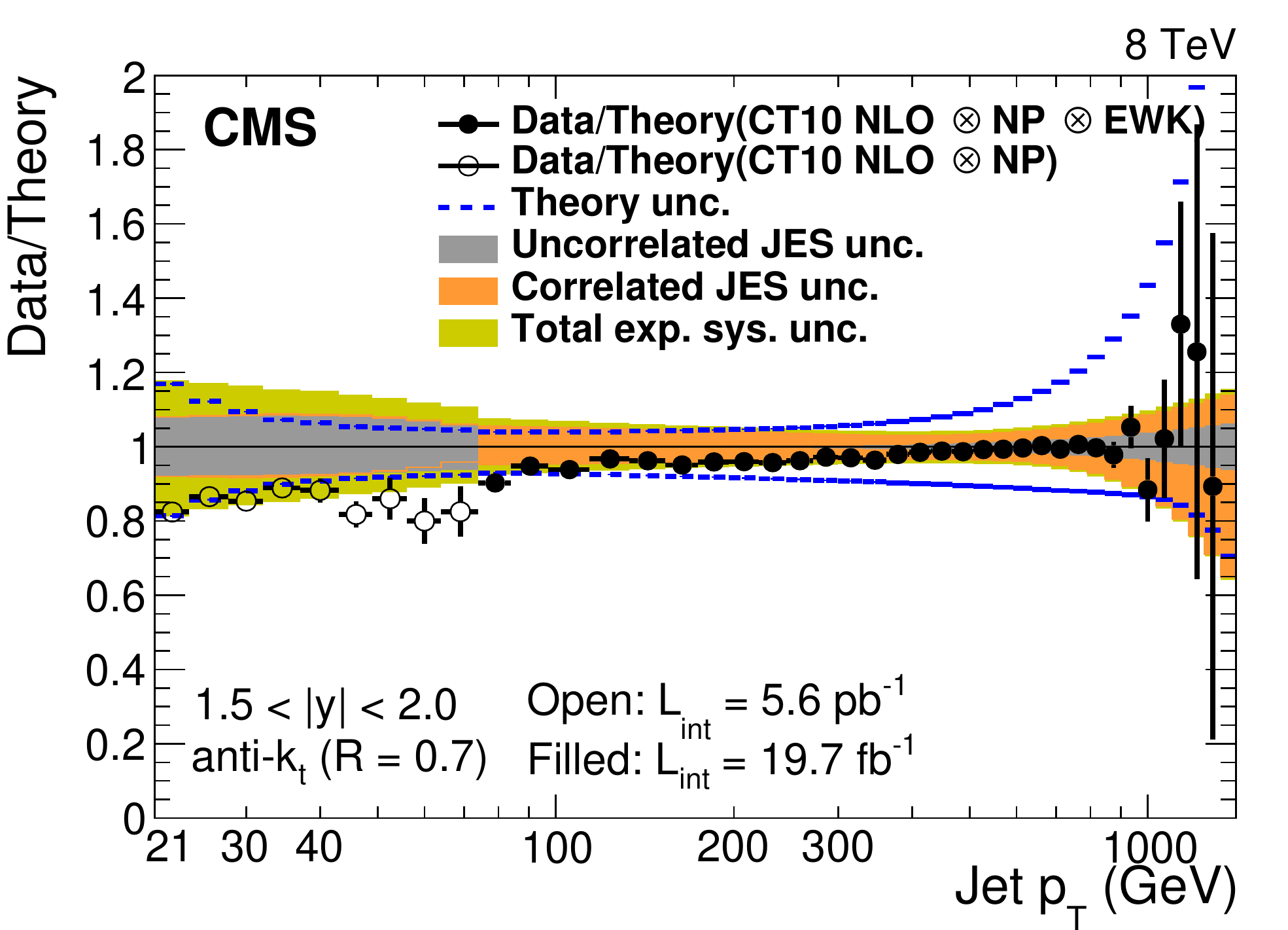}
\includegraphics[width=0.45\textwidth]{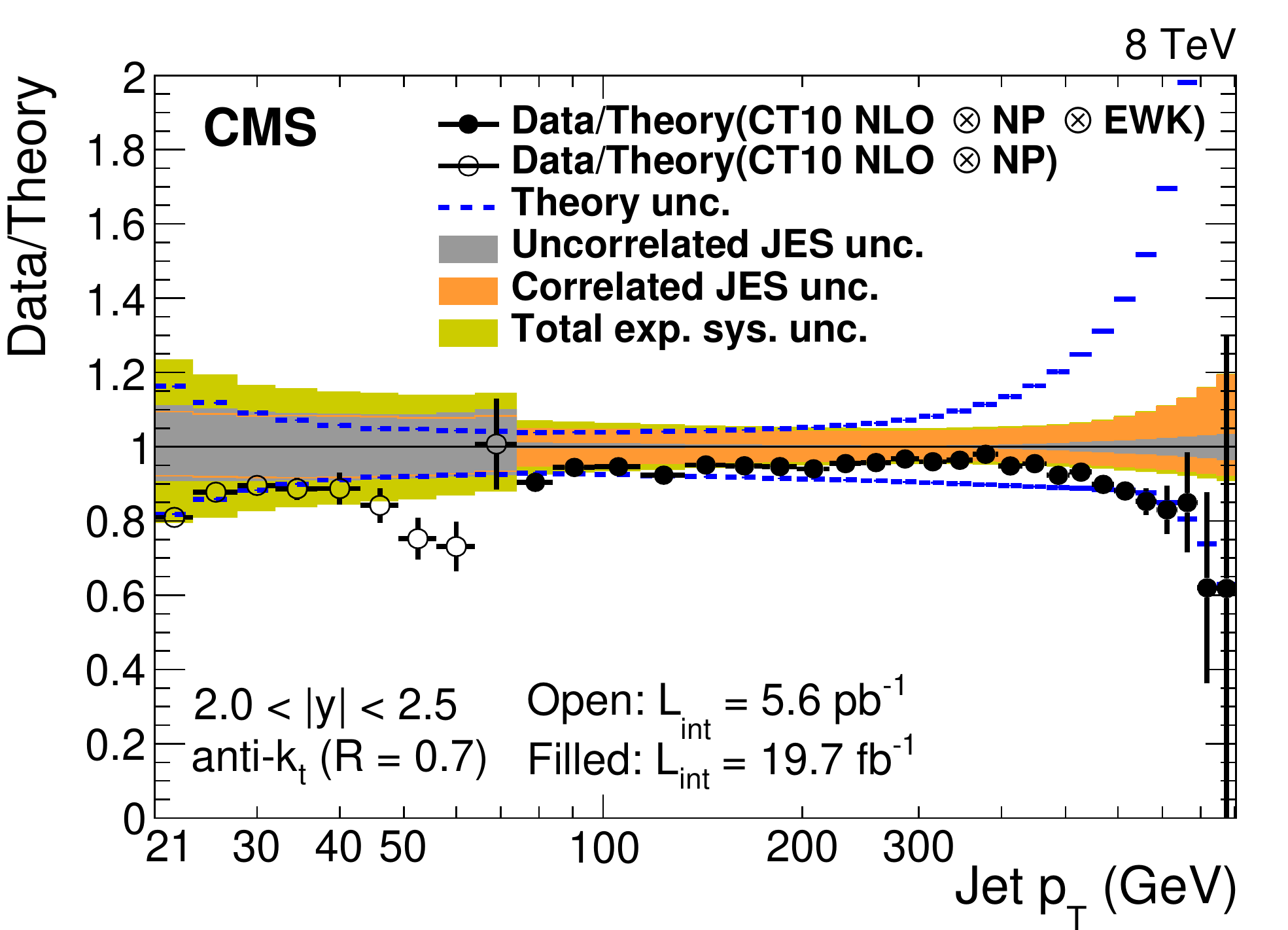}
\includegraphics[width=0.45\textwidth]{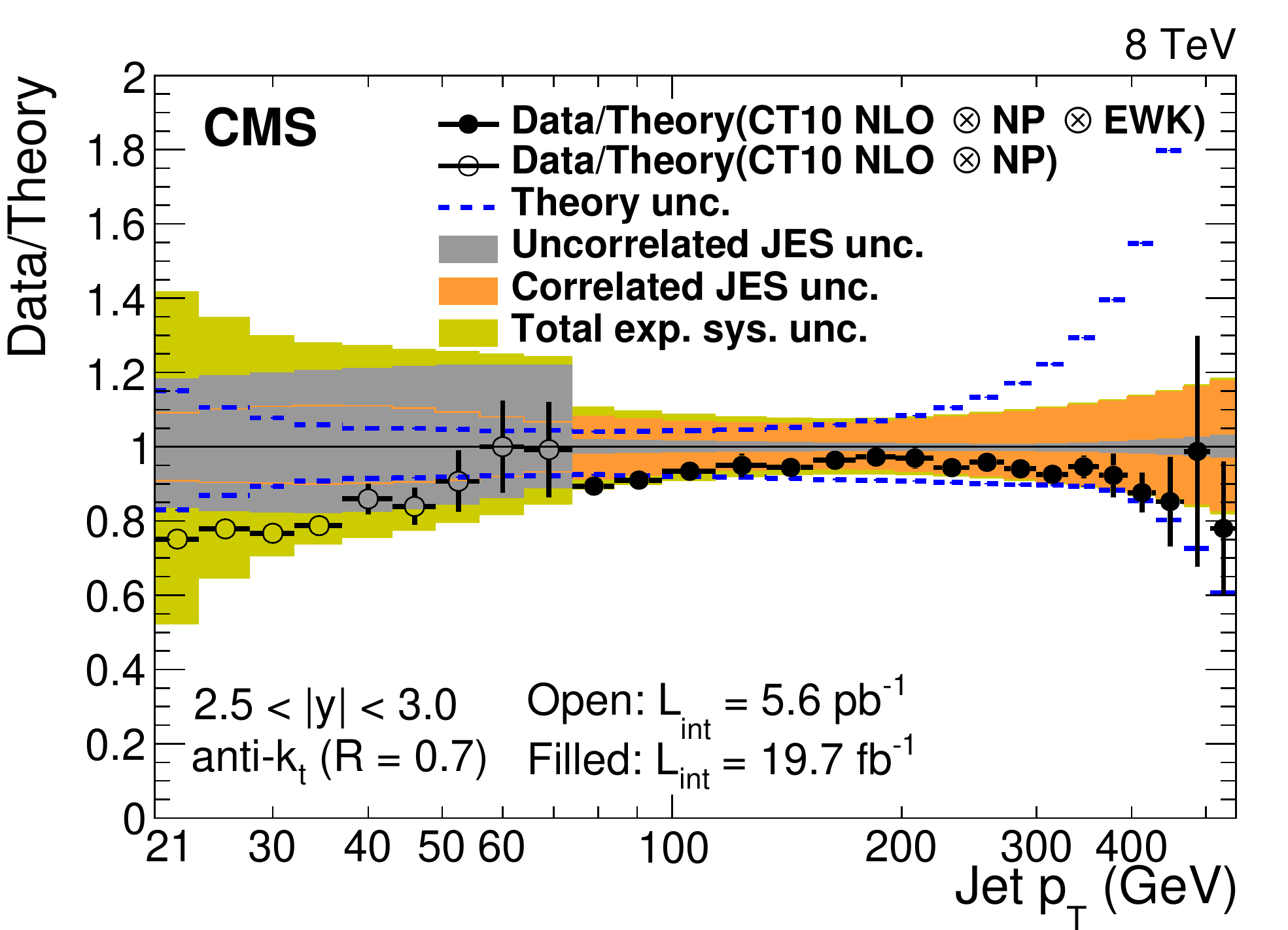}
\includegraphics[width=0.45\textwidth]{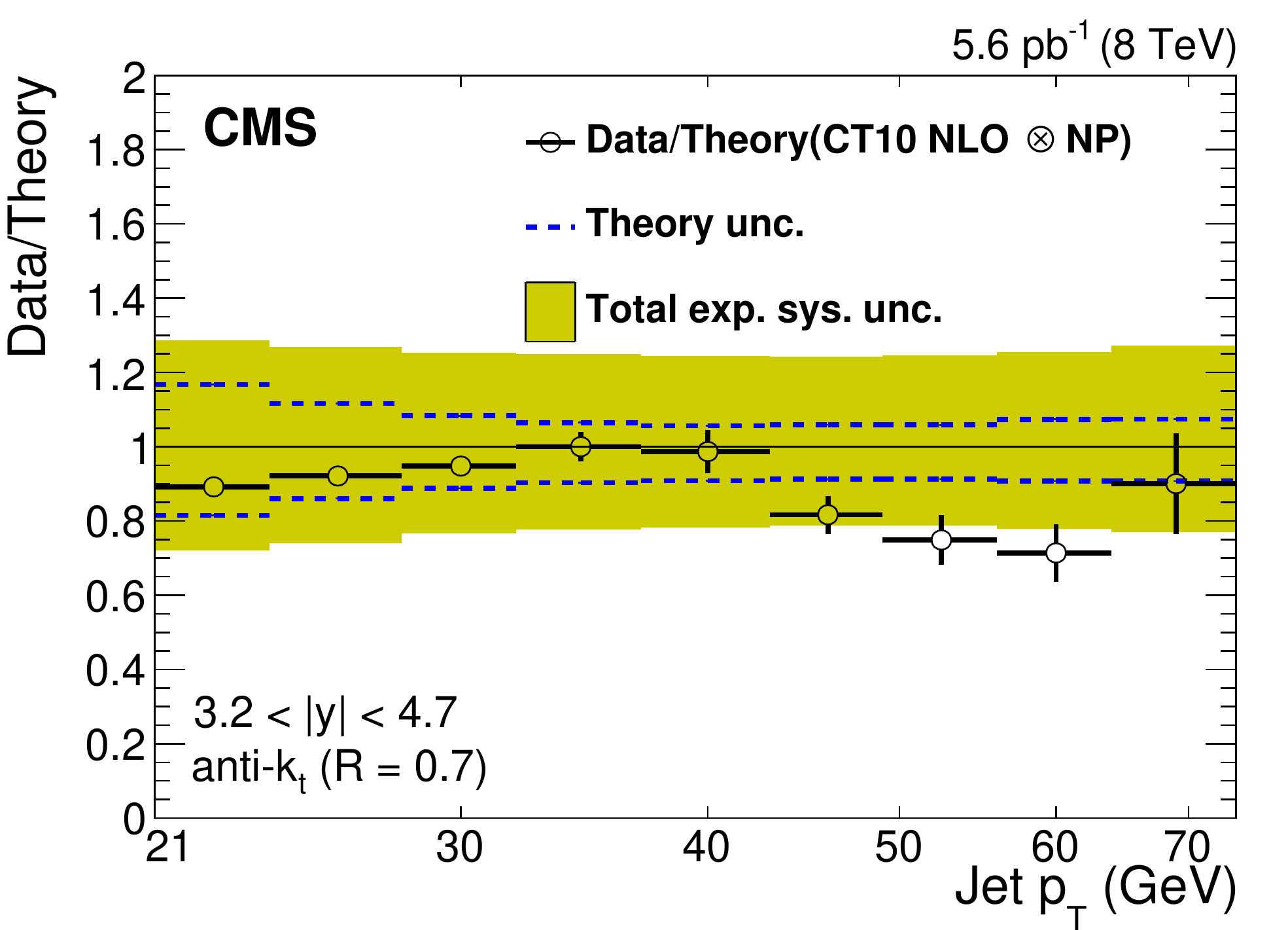}
\caption{Ratios of data to the theory prediction using the CT10 PDF
set. For comparison, the total theoretical (band enclosed by dashed
lines) and the total experimental systematic uncertainties (band
enclosed by full lines) are shown as well. The error bars
correspond to the statistical uncertainty in the data.}
\label{fig:DataTheoryRatioALLCT10}
\end{figure*}

The measured double-differential cross sections for inclusive jet
production are shown in Fig.~\ref{fig:ResultCT10} as a function of \pt
in the various $\abs{y}$ ranges after unfolding the detector effects. This
measurement is compared with the theoretical prediction discussed in
Section~\ref{sec:theory} using the CT10 PDF set. The ratios of the
data to the theoretical predictions in the various $\abs{y}$ ranges are
shown for the CT10 PDF set in
Fig.~\ref{fig:DataTheoryRatioALLCT10}. Good agreement is observed for
the entire kinematic range with some exceptions in the low-\pt region.

\begin{figure*}[hbpt]
\centering
\includegraphics[width=0.45\textwidth]{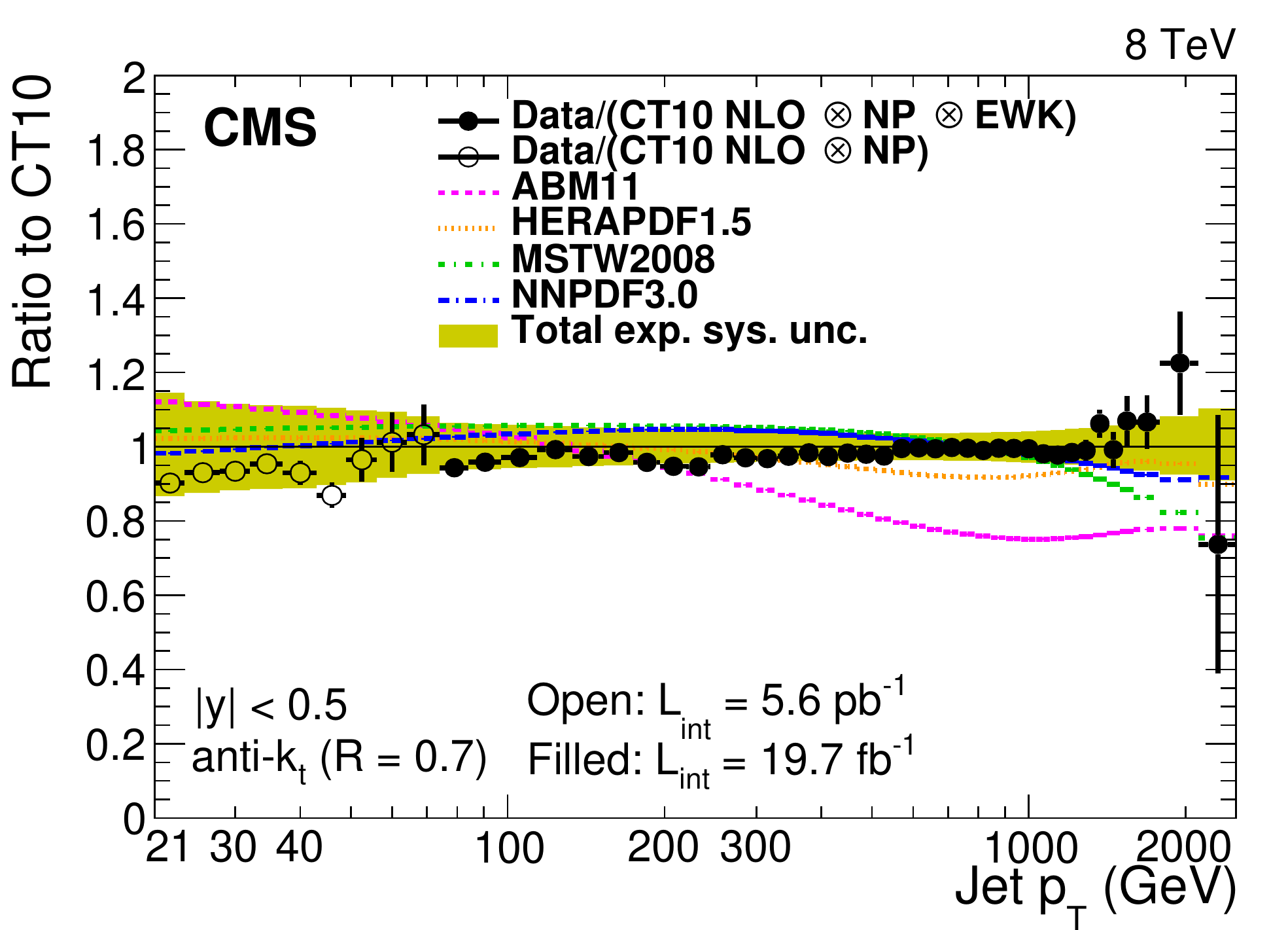}
\includegraphics[width=0.45\textwidth]{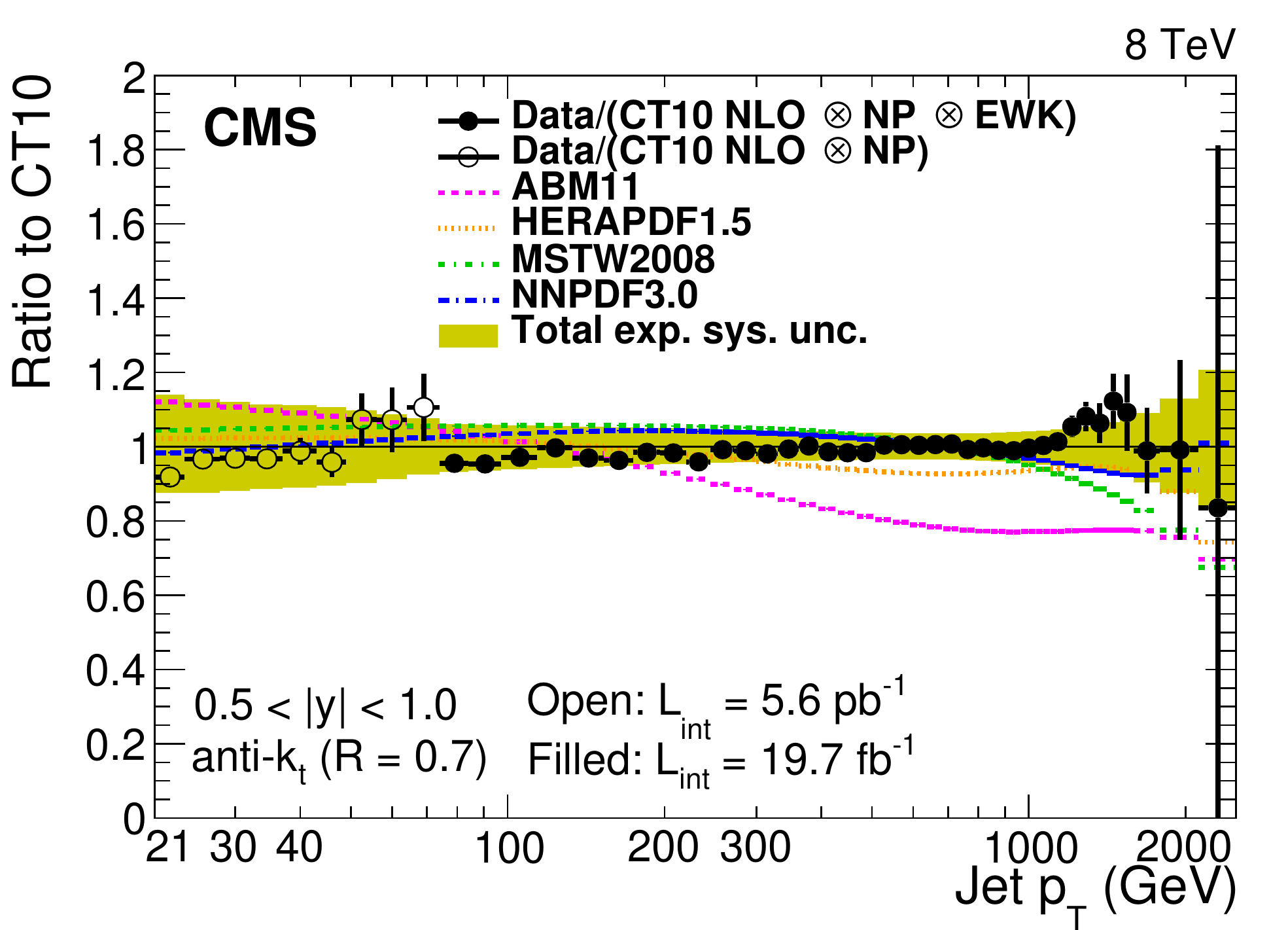}
\includegraphics[width=0.45\textwidth]{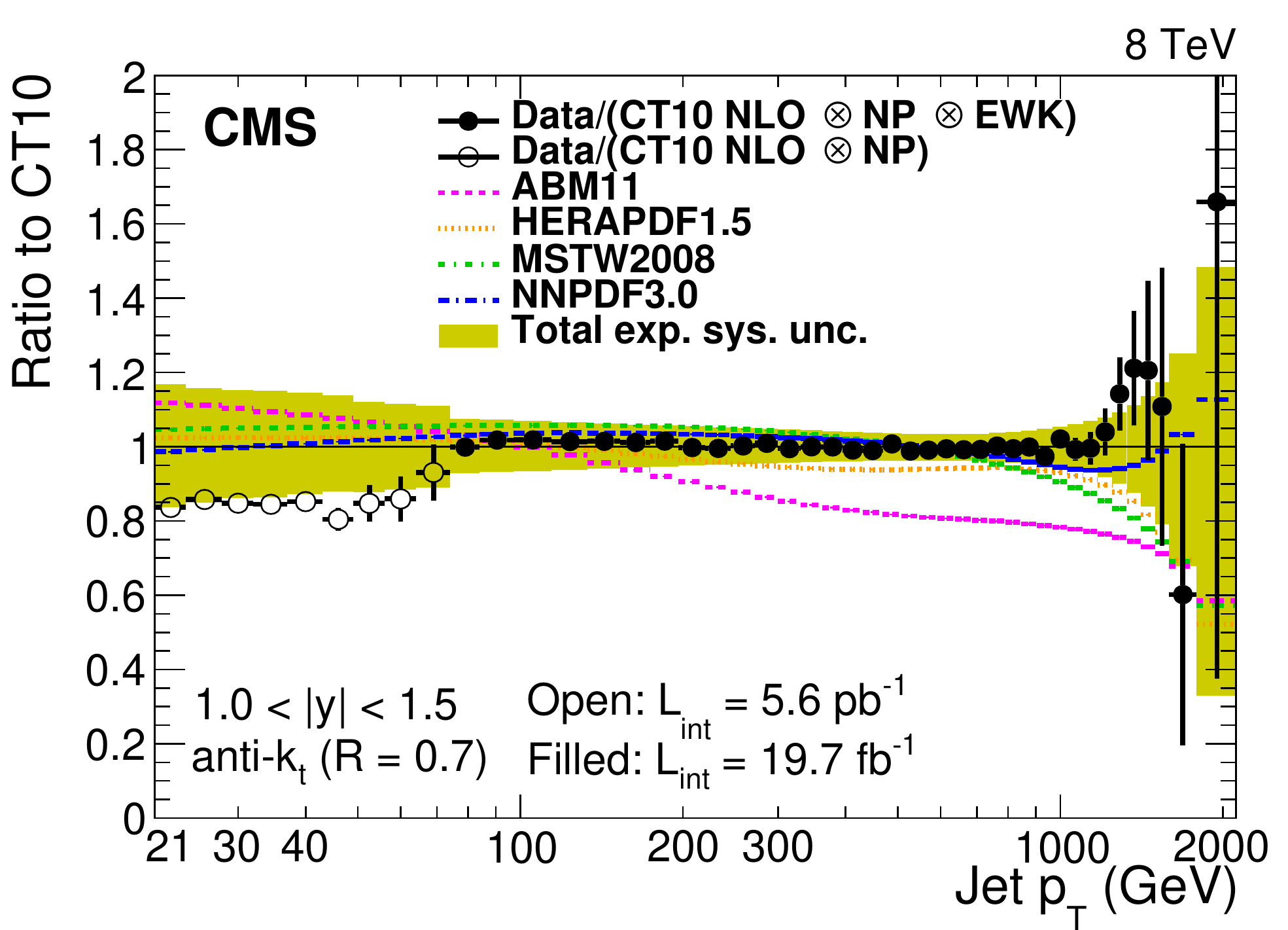}
\includegraphics[width=0.45\textwidth]{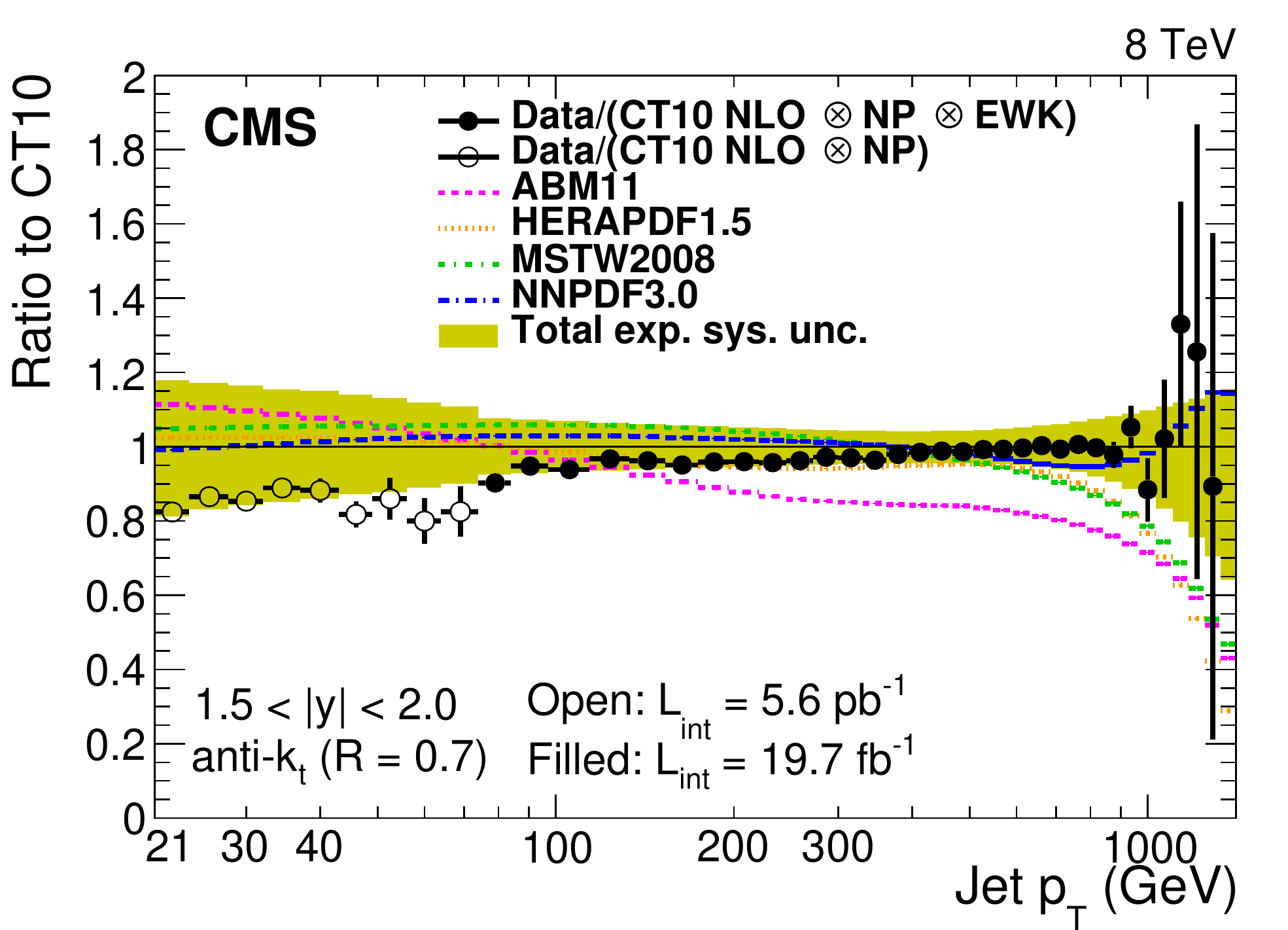}
\includegraphics[width=0.45\textwidth]{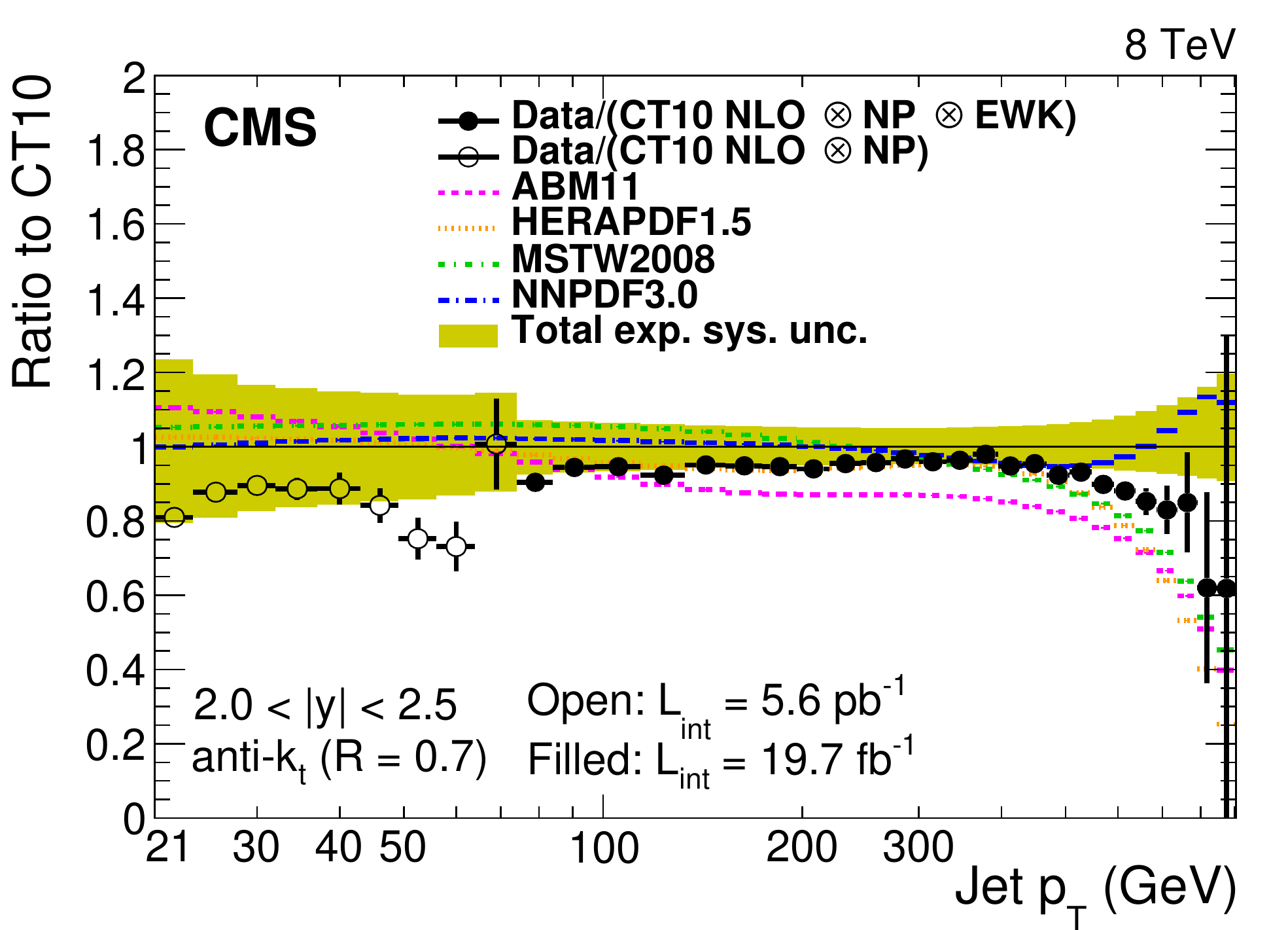}
\includegraphics[width=0.45\textwidth]{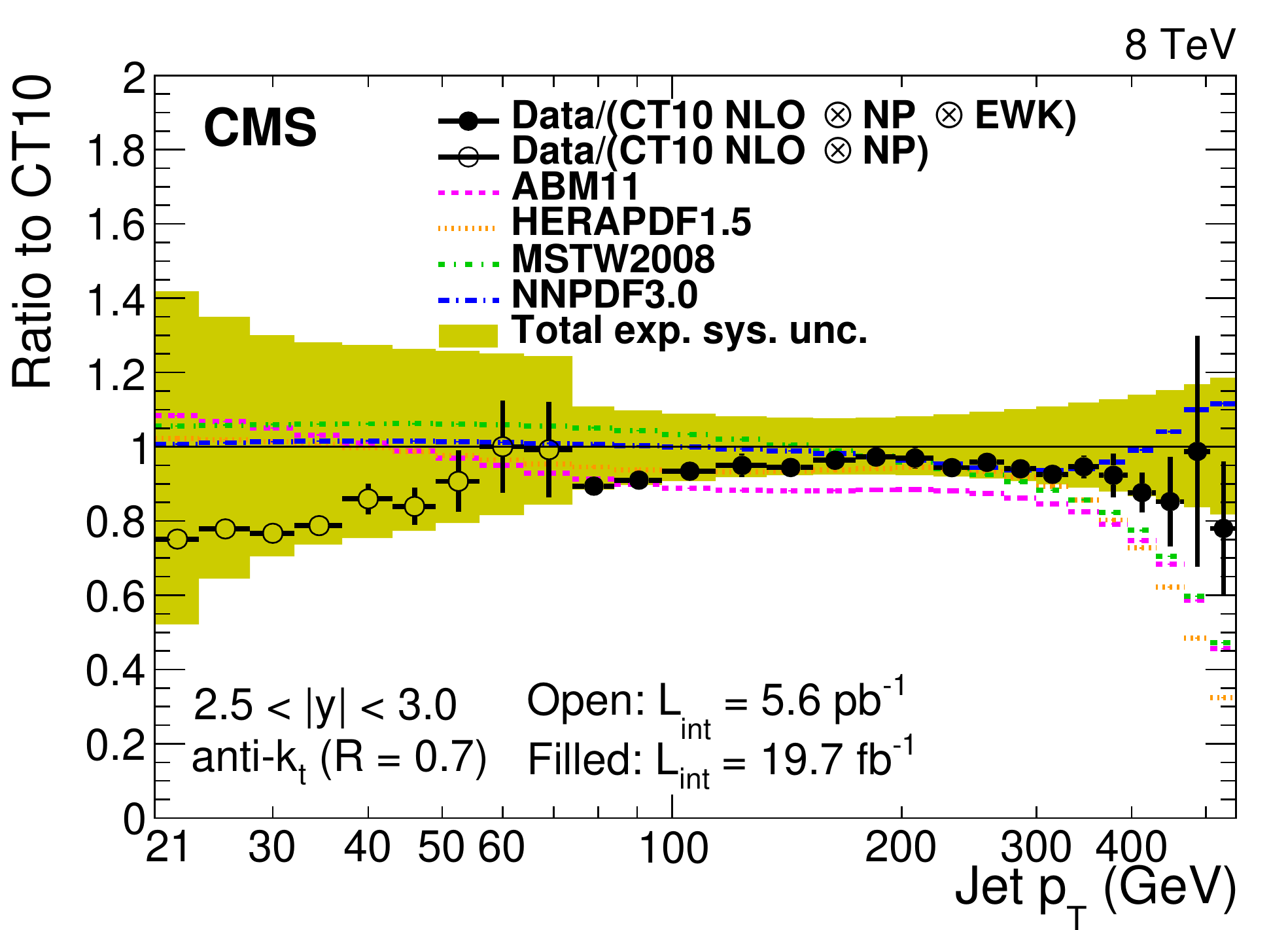}
\includegraphics[width=0.45\textwidth]{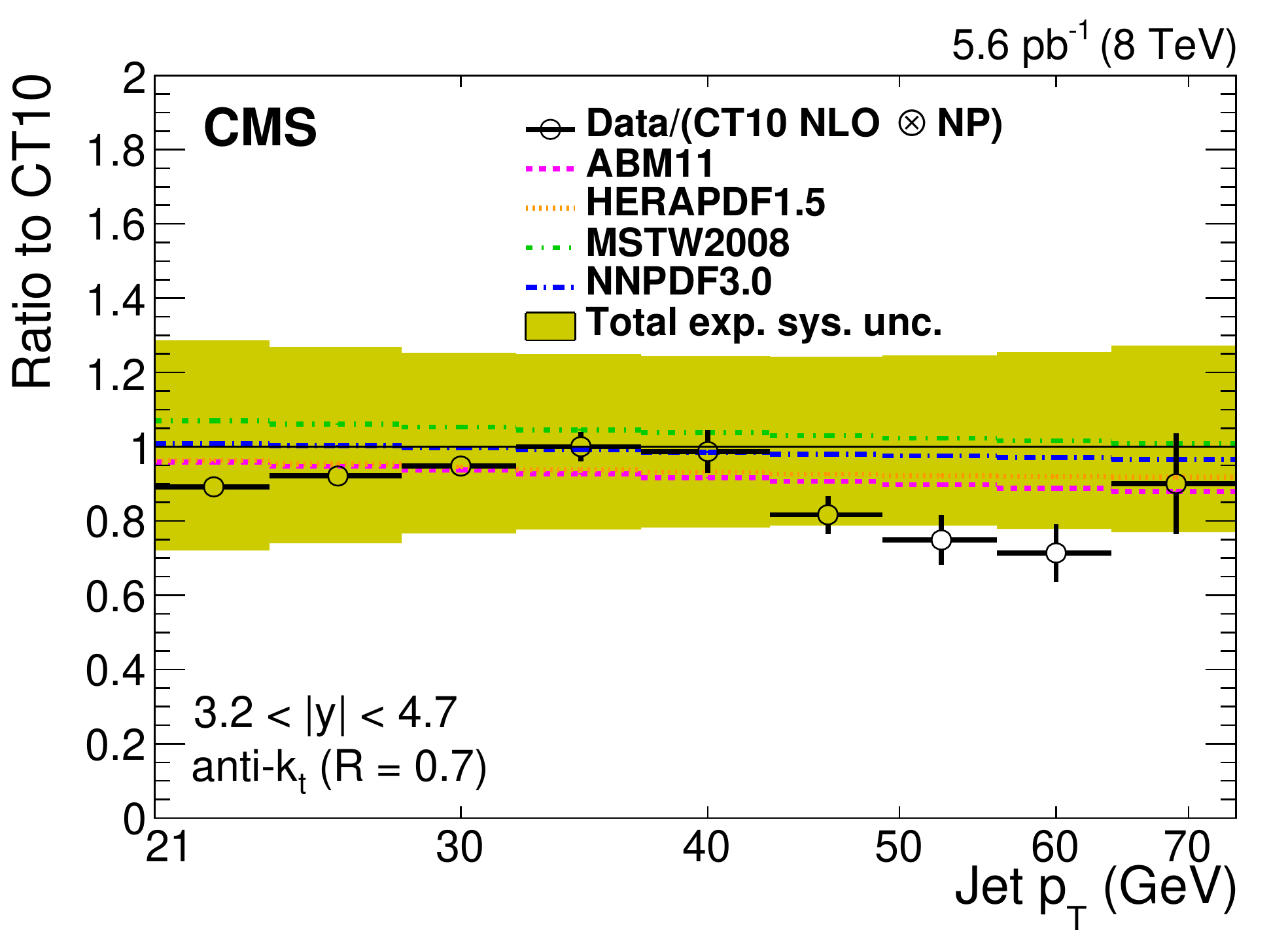}
\caption{Ratios of data and alternative predictions to the theory
prediction using the CT10 PDF set. For comparison, predictions
employing five other PDF sets are shown in addition to the total
experimental systematic uncertainties (band enclosed by full
lines). The error bars correspond to the statistical uncertainty
in the data.}
\label{fig:PDFRatioALLCT10}
\end{figure*}

Figure~\ref{fig:PDFRatioALLCT10} presents the ratios of the
measurements and a number of theoretical predictions based on
alternative PDF sets to the CT10 based prediction. A $\chi^2$ value is computed
based on the measurements, their covariance matrices, and the
theoretical predictions, as described in detail in
Section~\ref{sec:alphas}. The values for $\chi^2$ for the comparison between data and theory
based on different PDF sets for the high-\pt region are summarized in
Table~\ref{table:chi2pdf}.

\begin{table}[htb]
\centering
\topcaption{Summary of the $\chi^2$ values for the comparison of
data and theoretical predictions based on different PDF sets in
each $\abs{y}$ range, where cross sections are measured for a number
of \pt bins $N_{\mathrm{bins}}$.}
\begin{tabular}{cccccccc}
\hline
$\abs{y}$  & $N_{\mathrm{bins}}$ &  CT10   & HERAPDF1.5 & MSTW2008 & NNPDF2.1 &   ABM11  & NNPDF3.0 \\
\hline
0.0--0.5 & 37 & 49.2 & 66.3 &  68.0 &  58.3 & 136.6 &  62.5 \\
0.5--1.0 & 37 & 28.7 & 47.2 &  39.0 &  35.4 & 155.5 &  42.2 \\
1.0--1.5 & 36 & 19.3 & 28.6 &  27.4 &  20.2 & 111.8 &  25.9 \\
1.5--2.0 & 32 & 65.7 & 49.0 &  55.3 &  54.5 & 168.1 &  64.7 \\
2.0--2.5 & 25 & 38.7 & 32.0 &  53.1 &  34.6 &  \x80.2 &  36.0 \\
2.5--3.0 & 18 & 14.5 & 19.1 &  18.2 &  15.4 &  \x43.8 &  16.3 \\
\hline
\end{tabular}
\label{table:chi2pdf}
\end{table}

In most cases the theoretical predictions agree with the
measurements. The exception is the ABM11 PDF set, where significant
discrepancies are visible. Significant differences between the
theoretical predictions obtained by using different PDF sets are
observed in the high-\pt range.  The predictions based on CT10 PDF
show the best agreement with data, quantified by the lowest $\chi^2$
for most rapidity ranges, while predictions using MSTW, ABM11, and
HERAPDF1.5 exhibit differences compared to data and to the prediction
based on CT10, exceeding 100\% in the highest \pt range.

In the transition between the low- and high-\pt jet regions, some
discontinuity can be observed in the measured values, although
they are generally compatible within the total experimental
uncertainties. The highest \pt bins of the low-\pt jet range suffer
from a reduced sample size, and
therefore have a statistical uncertainty significantly larger than
the first bin of the high-\pt jet region. The JES corrections for
the low- and high-\pt regions are different, in particular in the
\pt-dependent components, and this also contributes to the observed
fluctuations in the matching region. The corresponding uncertainties
are treated as uncorrelated between the low- and high-\pt regions. The
overall estimated systematic uncertainties account for these residual
effects. The transition region between the low- and high-\pt jet measurements
has limited sensitivity to \as and no impact in constraining PDFs,
since it probes the $x$-range where the PDFs are well constrained
by more precise DIS data.

\section{Ratios of cross sections measured at different \texorpdfstring{$\sqrt{s}$}{sqrt(s)} values}

Ratios of cross sections measured at different energies may show a
better sensitivity to PDFs than cross sections at a single
energy, provided that the contributions to the theoretical and
experimental uncertainties from sources other than the PDFs themselves
are reduced. A calculation of the ratio of cross sections
measured at 7 and 8\TeV presented in Ref.~\cite{Mangano:2012mh}, for
instance, suggests a larger sensitivity to PDFs in the jet \pt range between
1 and 2\TeV. Therefore, it is interesting to study such cross section ratios.

Differential cross sections for the inclusive jet production have
been measured by the CMS Collaboration at $\sqrt{s} =
2.76$~\cite{Khachatryan:2015luy} and
7\TeV~\cite{Chatrchyan:2012bja}. Ratios are computed of the
double-differential cross section presented in this paper at 8\TeV to
the corresponding measurements at different energies.
For $\pt > 74\GeV$, the choice of jet \pt and rapidity
bins is identical for the various measurements, thus allowing an easy
computation of the ratio. Only the high-\pt jet data set at 8\TeV is
used, since no counterpart of the low-\pt jet analysis is available
for the other centre-of-mass energies.

\begin{figure*}[ht!]
\centering
\includegraphics[width=0.49\textwidth]{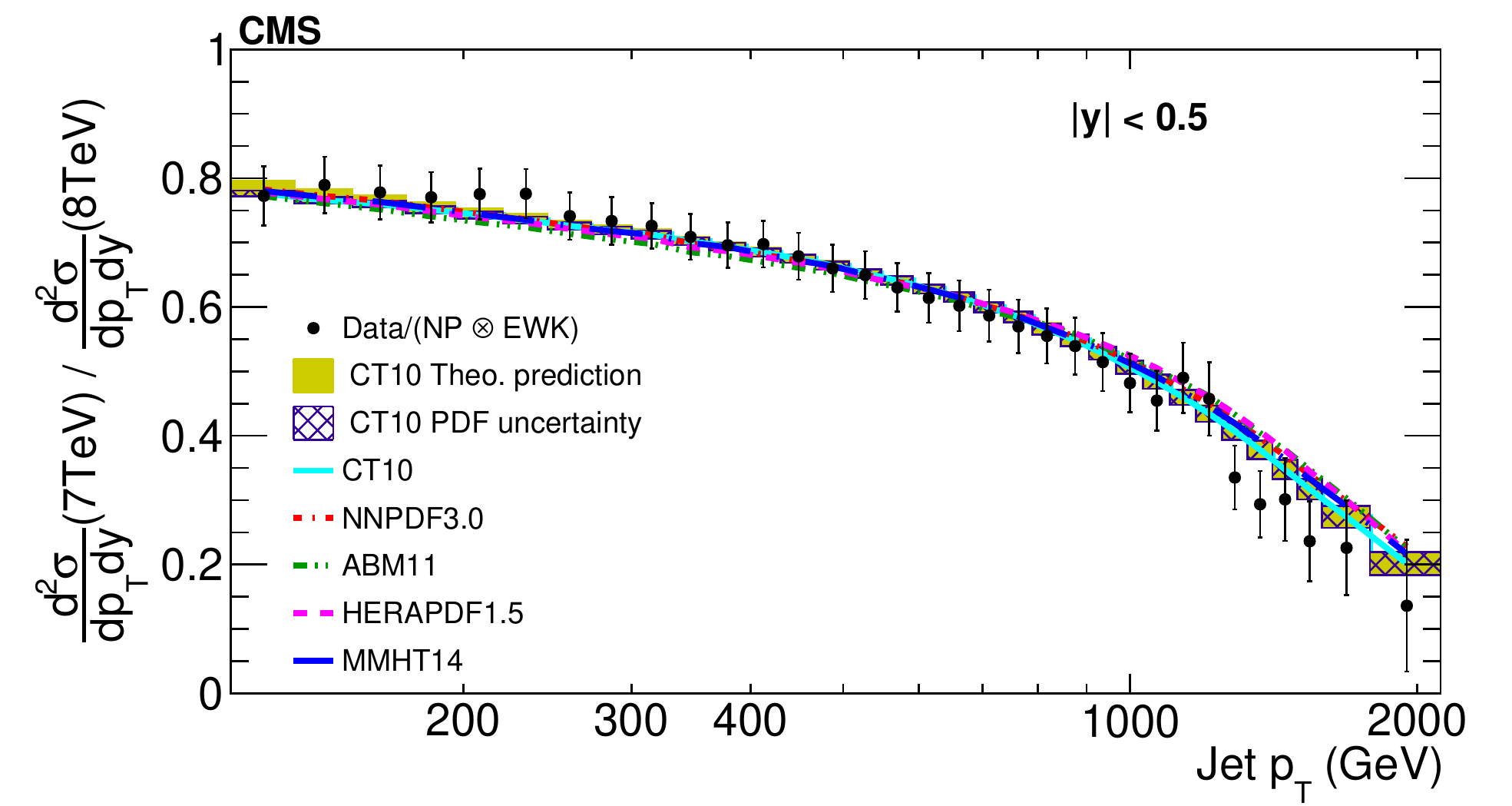}
\includegraphics[width=0.49\textwidth]{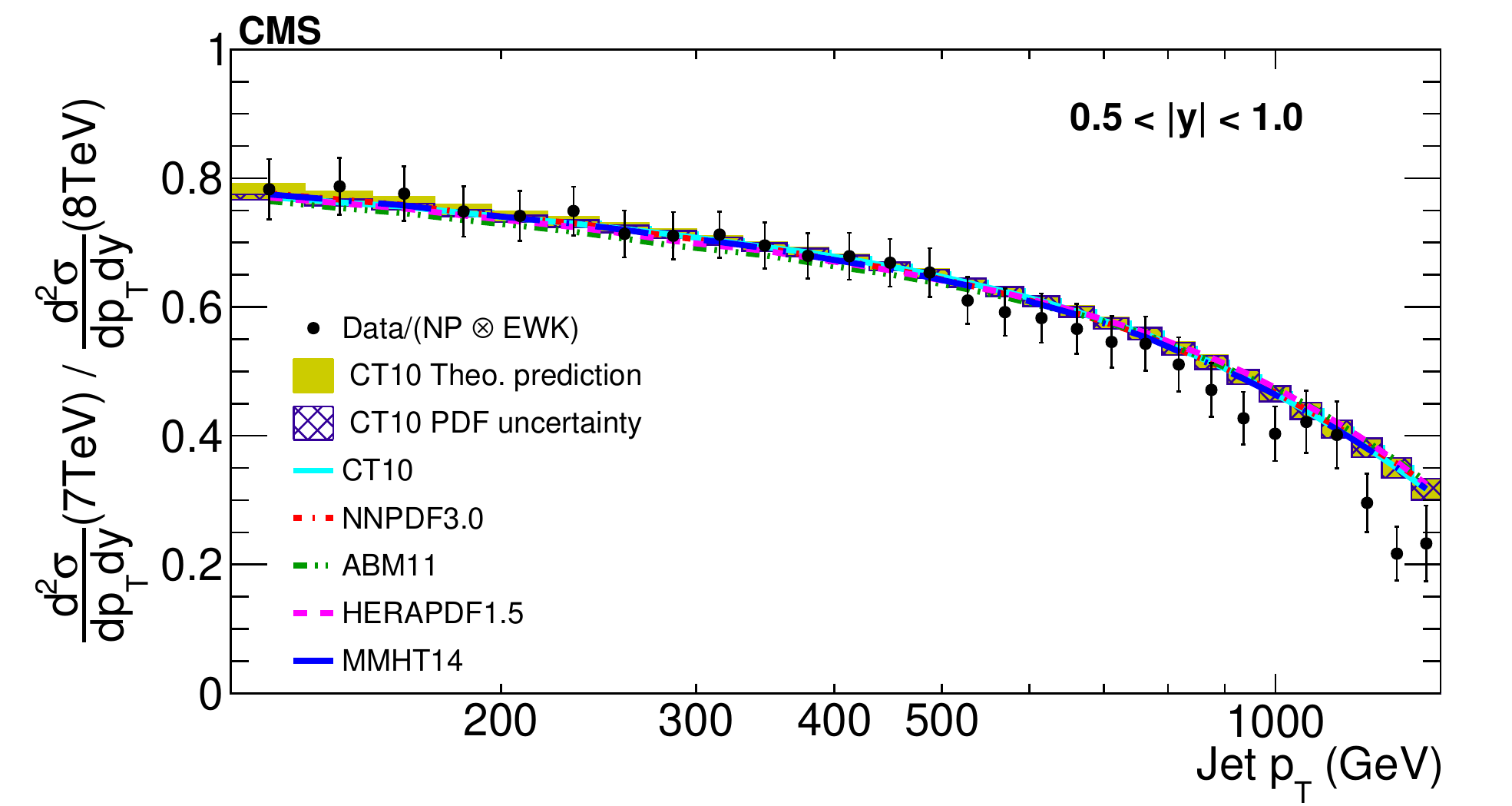}
\includegraphics[width=0.49\textwidth]{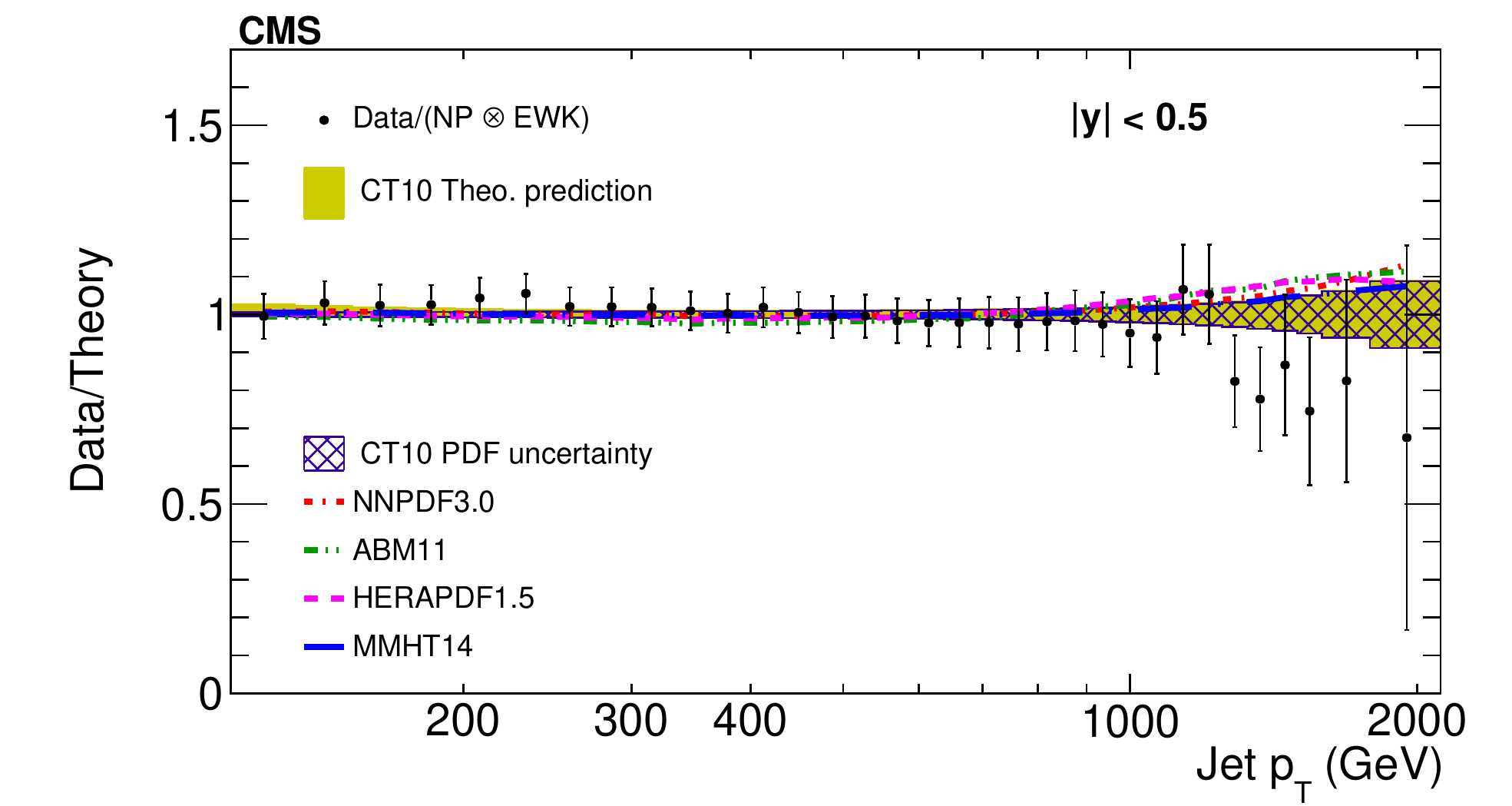}
\includegraphics[width=0.49\textwidth]{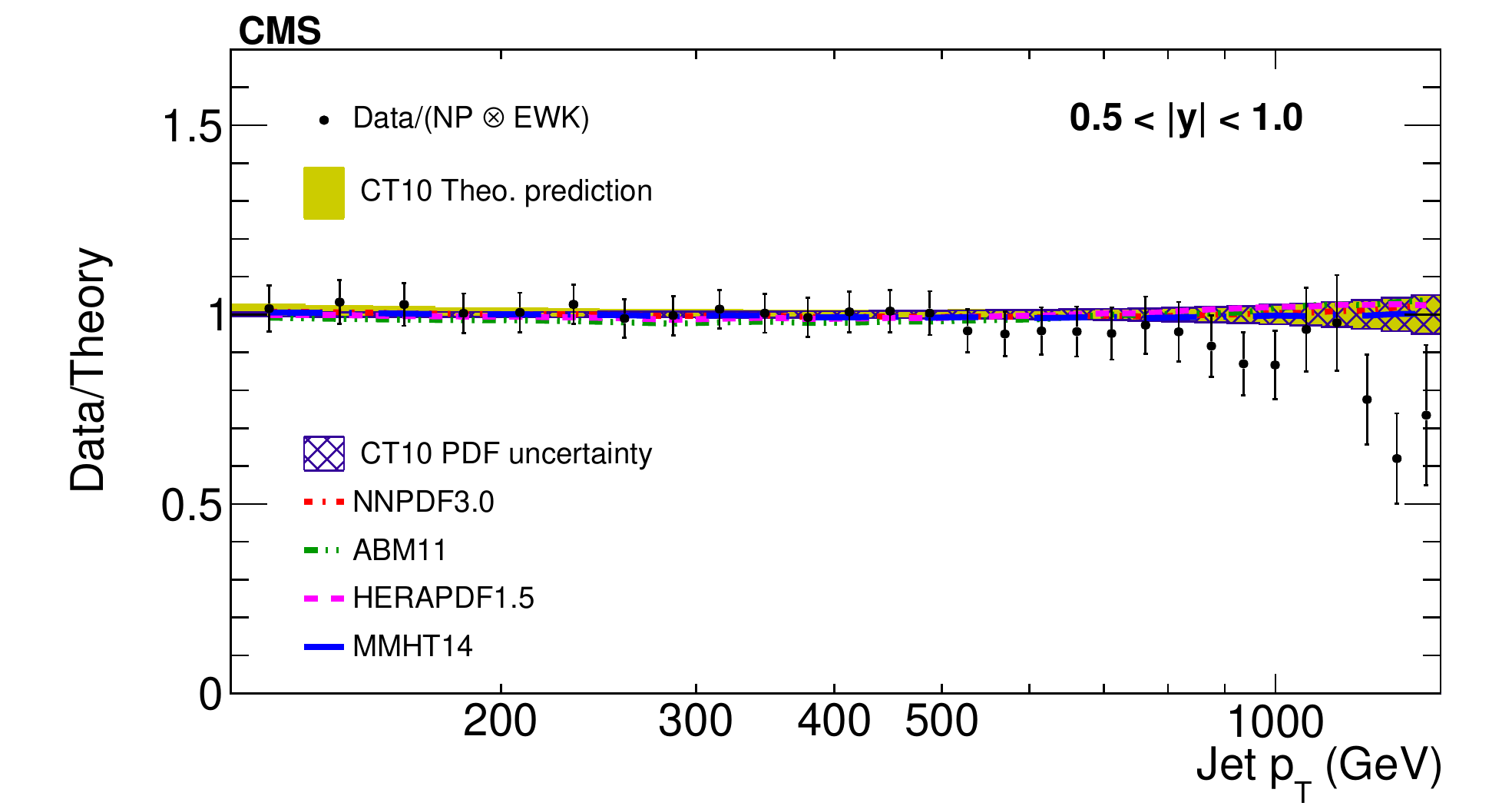}
\caption{The ratios (top panels) of the inclusive jet production cross
sections at $\sqrt{s}=7$ and 8\TeV, shown as a function of jet $p_T$ for
the absolute rapidity $\abs{y}< 0.5$ (left) and $0.5 < \abs{y} < 1.0$ (right). The data (closed
symbols) are shown with total uncertainties (vertical error bars). The NLO
pQCD prediction using the CT10 PDF is shown with its total uncertainty (shaded band) and
the contribution of the PDF uncertainty (hatched band). Predictions obtained using
alternative PDF sets are shown by lines of different styles without uncertainties.
The data to theory ratios (bottom panels) are shown by using the same notations for the
respective rapidities. The last bin for the $\abs{y} < 0.5$ region is wider
than the others in order to reduce the statistical uncertainty.}
\label{fig:8TeVOv7TeV1}
\end{figure*}

As a result of partial cancellation of the systematic uncertainties,
the relative precision of the ratios is improved compared with the cross
section. Experimental correlations between the measurements at
different centre-of-mass energies are taken into account in the
computation of the total experimental uncertainty. As a consequence of
the unfolding procedure, the results of the cross section measurements
at each energy are statistically correlated between different bins,
while the measurements at different energies are not statistically
correlated with each other. The statistical uncertainties in the ratio
measurement are calculated by using linear error propagation, taking
into account the bin-to-bin correlations in the unfolded
data. Correlations between the components of the jet energy
corrections at different energies are included, as well as
correlations in JER.  Uncertainties related to the determination of
luminosity are assumed to be uncorrelated.

The theoretical uncertainties are approached in a similar manner: the
uncertainties in nonperturbative corrections, PDFs, and those arising due to scale
variations are assumed to be fully correlated.

\begin{figure*}[ht!]
\centering
\includegraphics[width=0.49\textwidth]{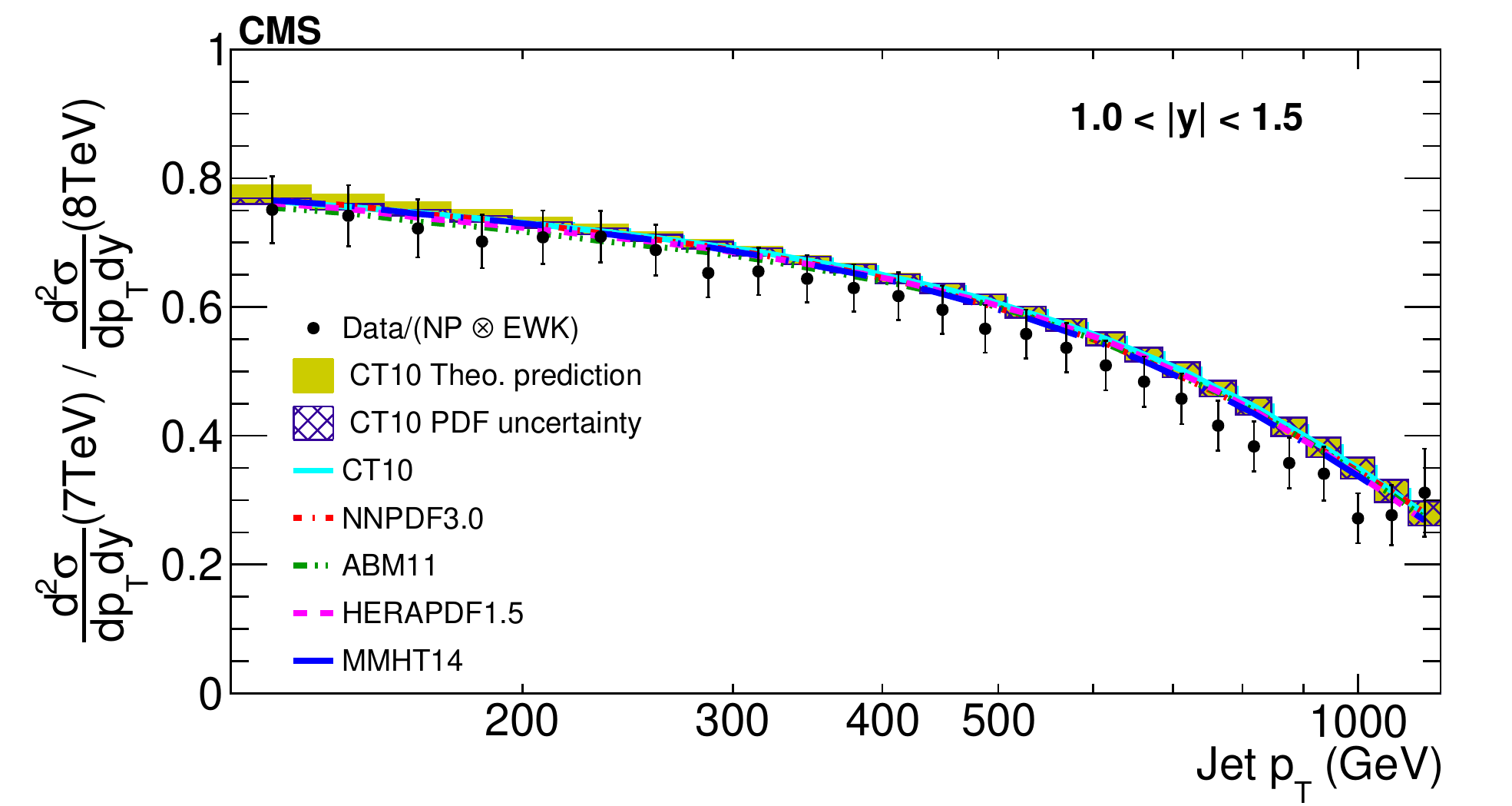}
\includegraphics[width=0.49\textwidth]{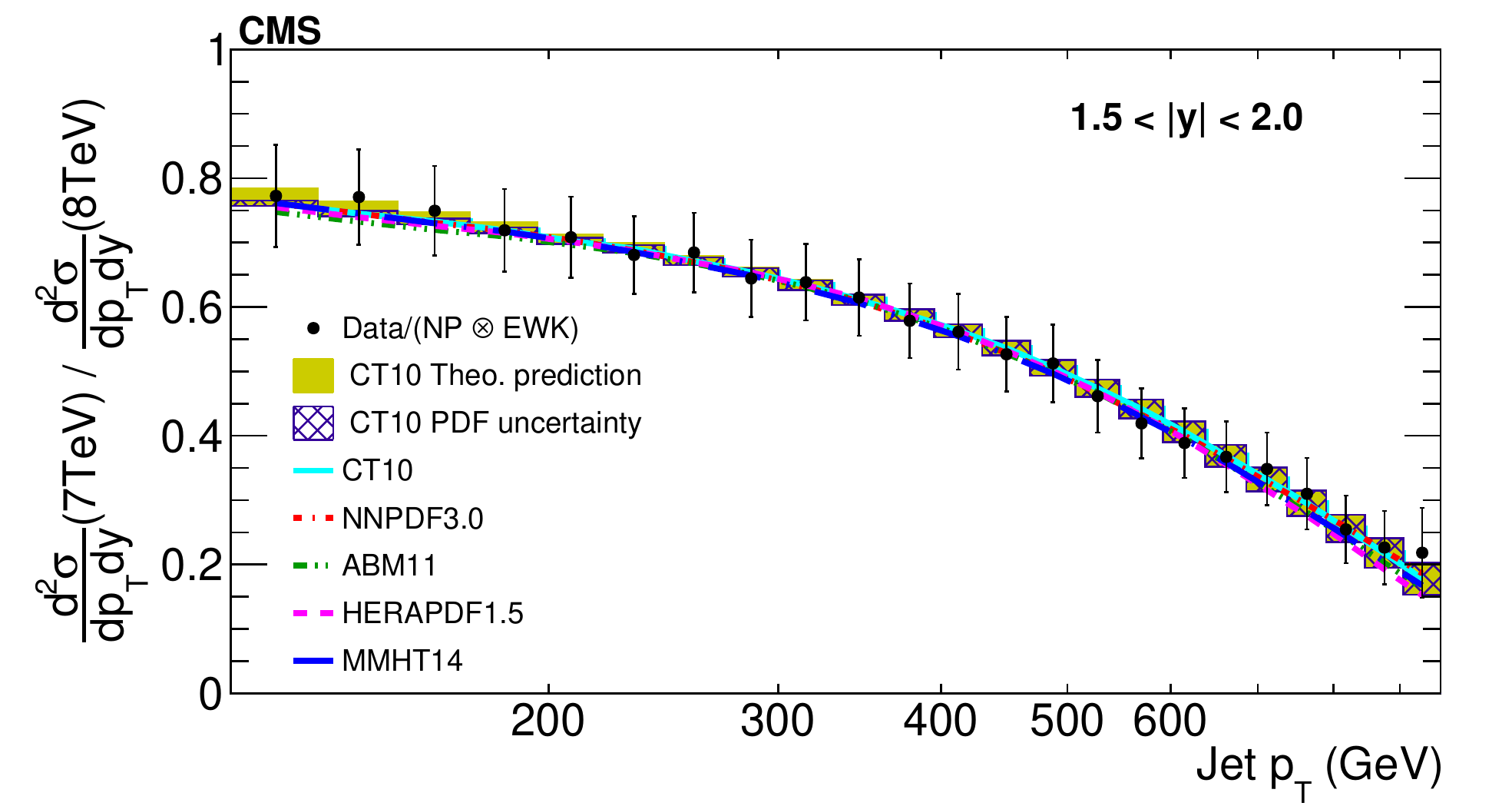}
\includegraphics[width=0.49\textwidth]{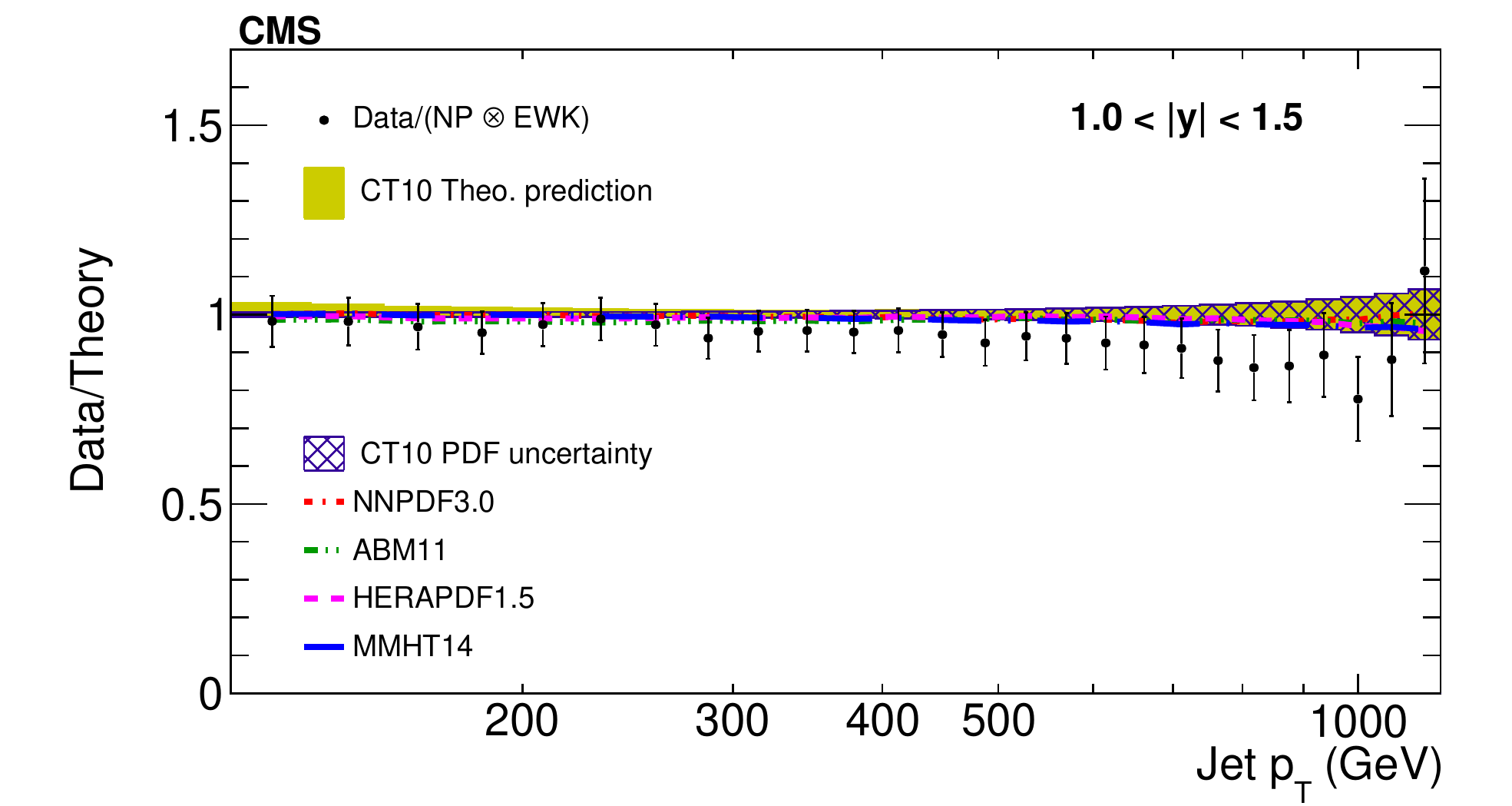}
\includegraphics[width=0.49\textwidth]{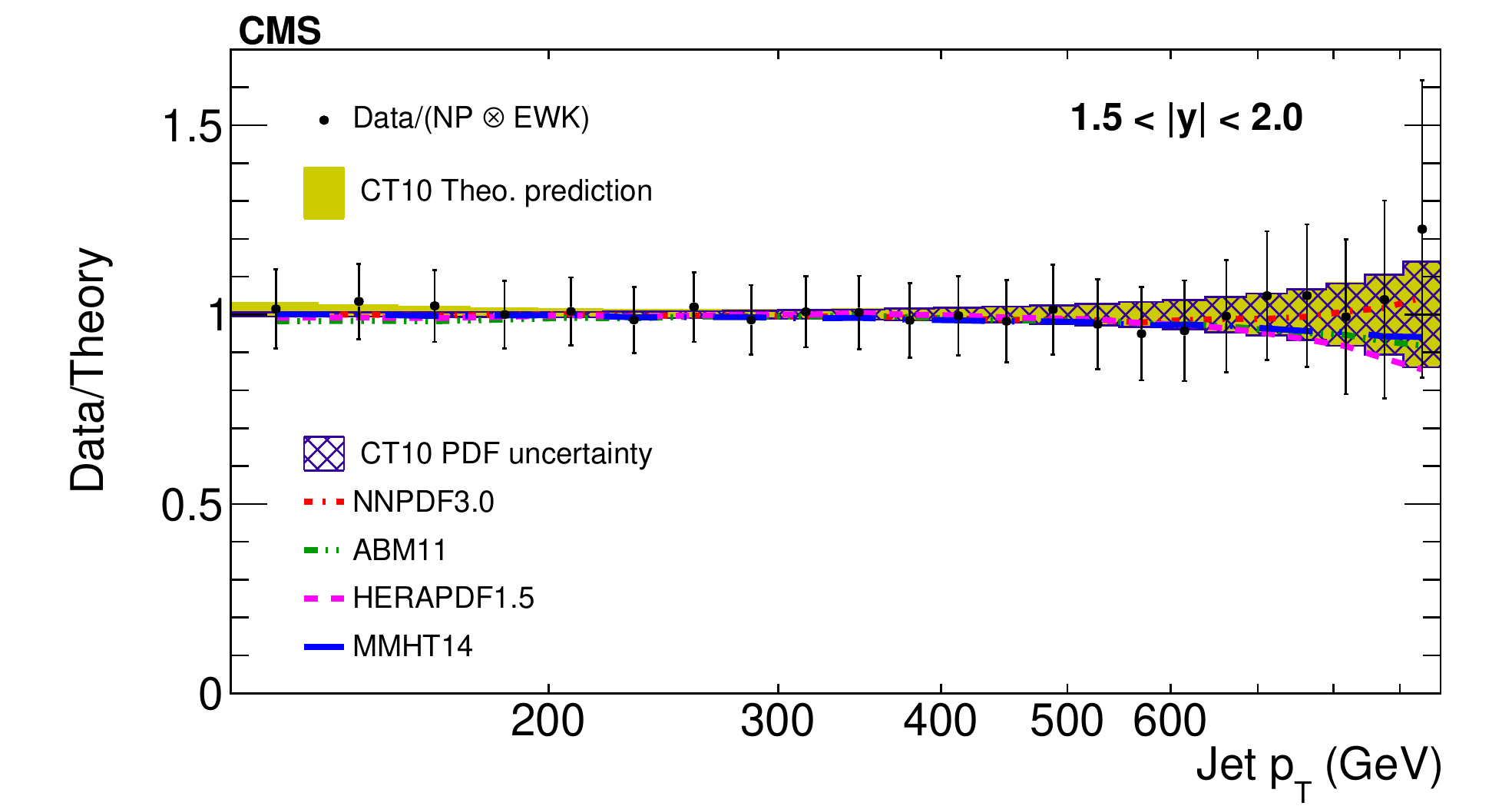}
\includegraphics[width=0.49\textwidth]{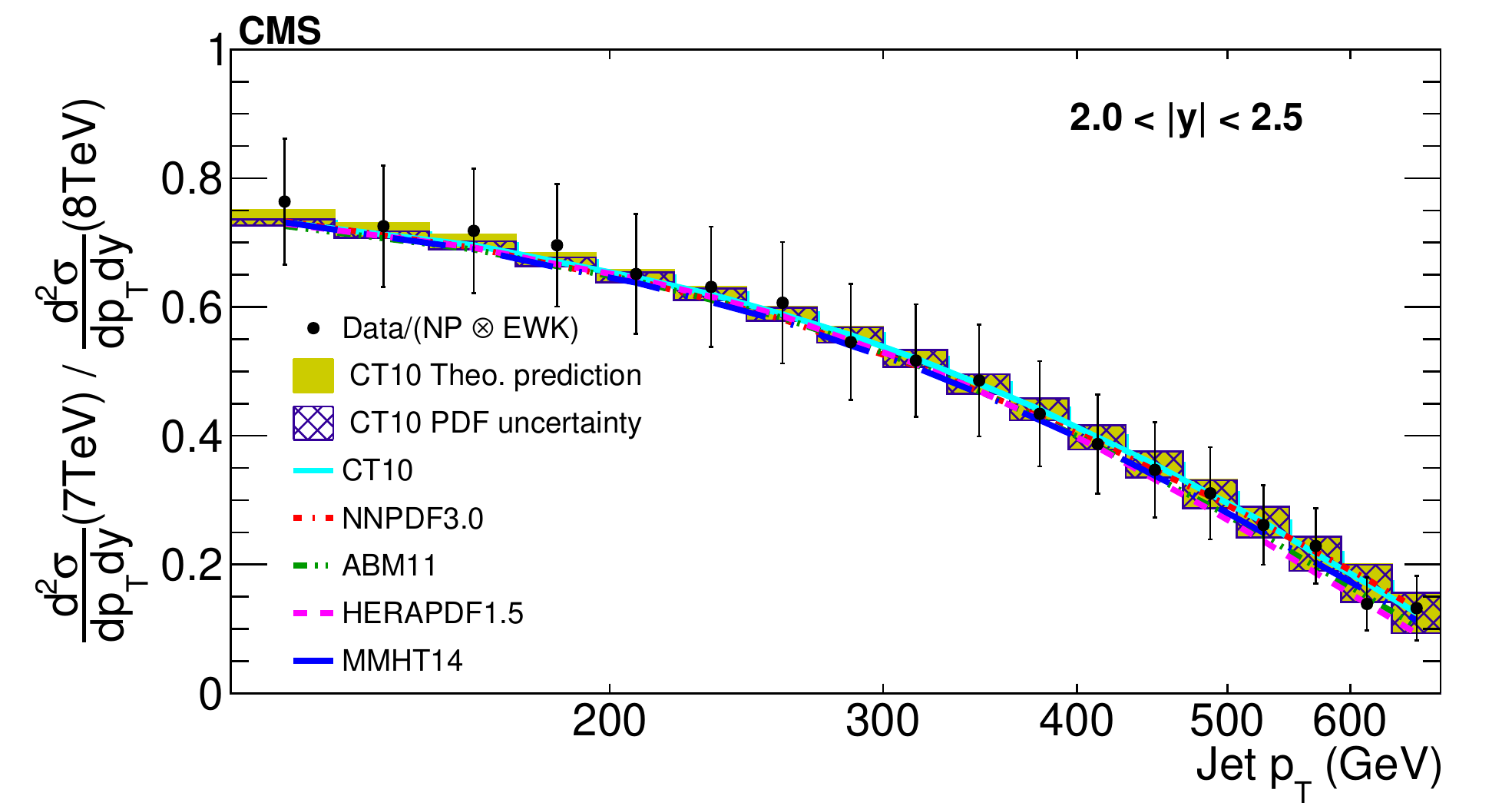}\\
\includegraphics[width=0.49\textwidth]{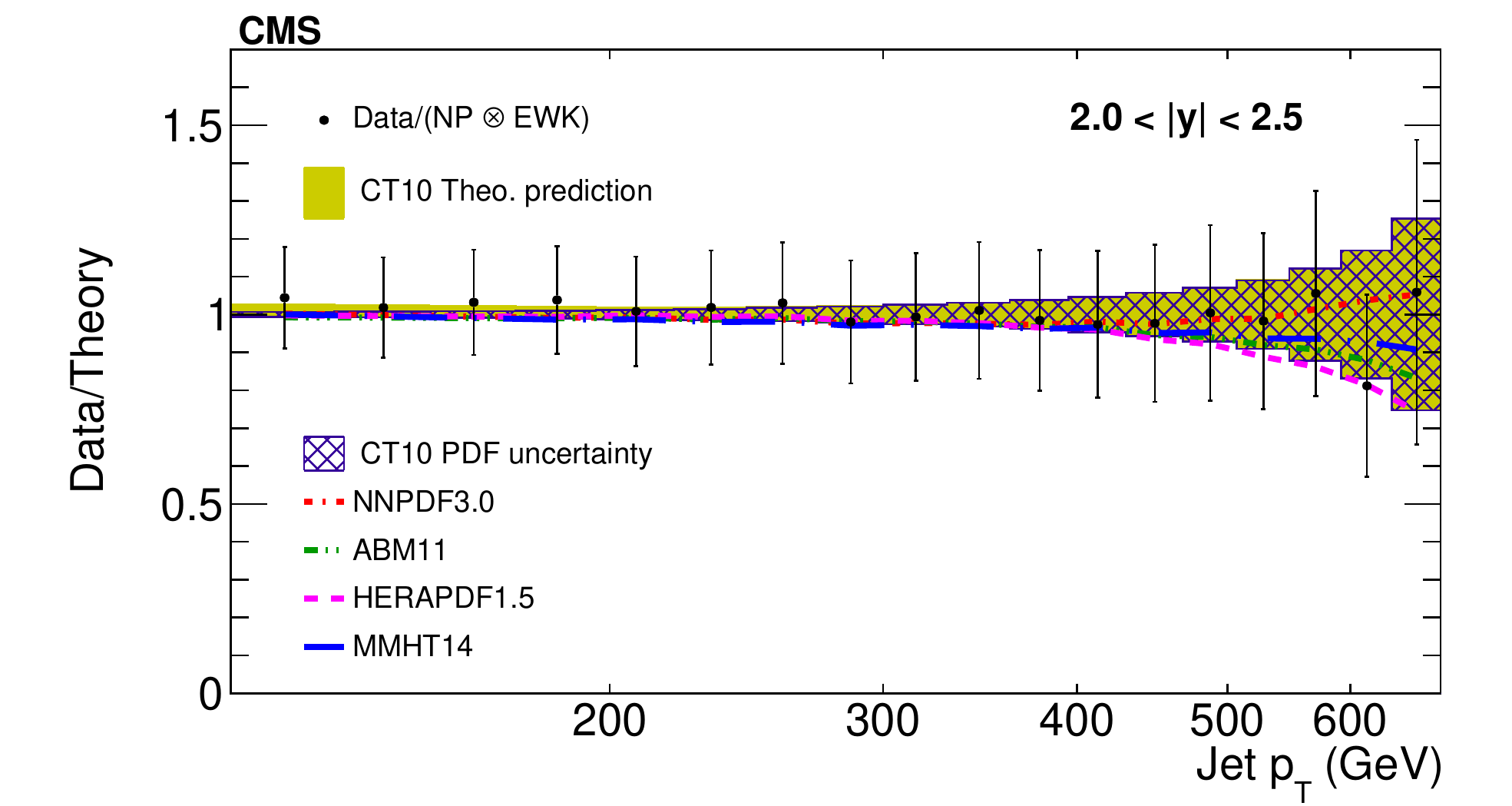}
\caption{The ratios of the inclusive jet production
cross sections at $\sqrt{s} =$ 7 and 8\TeV shown as a function of jet
\pt for the absolute rapidity $1.0 < \abs{y} < 1.5$ (top left), $1.5 < \abs{y}
< 2.0$ (top right) and $2.0 < \abs{y} < 2.5$ (bottom).}
\label{fig:8TeVOv7TeV3}
\end{figure*}

The ratios of the cross sections measured at $\sqrt{s}= 7$ and 8\TeV
are shown in Figs.~\ref{fig:8TeVOv7TeV1}--\ref{fig:8TeVOv7TeV3} for
the various rapidity bins and they are compared with theoretical
predictions obtained using different PDF sets. A general agreement
between data and theoretical predictions is observed. Some
discrepancies are visible at high \pt, in particular in the $1.0 < \abs{y}
< 1.5$ range.  In the cross section ratio the central values of the
predictions are not strongly influenced by the choice of the
PDFs. However, the uncertainty is mostly dominated by PDF
uncertainties, which are represented here for CT10. The experimental
uncertainty in the ratio is considerably larger than the theoretical
uncertainty. Consequently, no significant constraints on PDFs can be
expected from the inclusive jet cross section ratio of 7 to 8\TeV.

\begin{figure*}[ht!]
\centering
\includegraphics[width=0.49\textwidth]{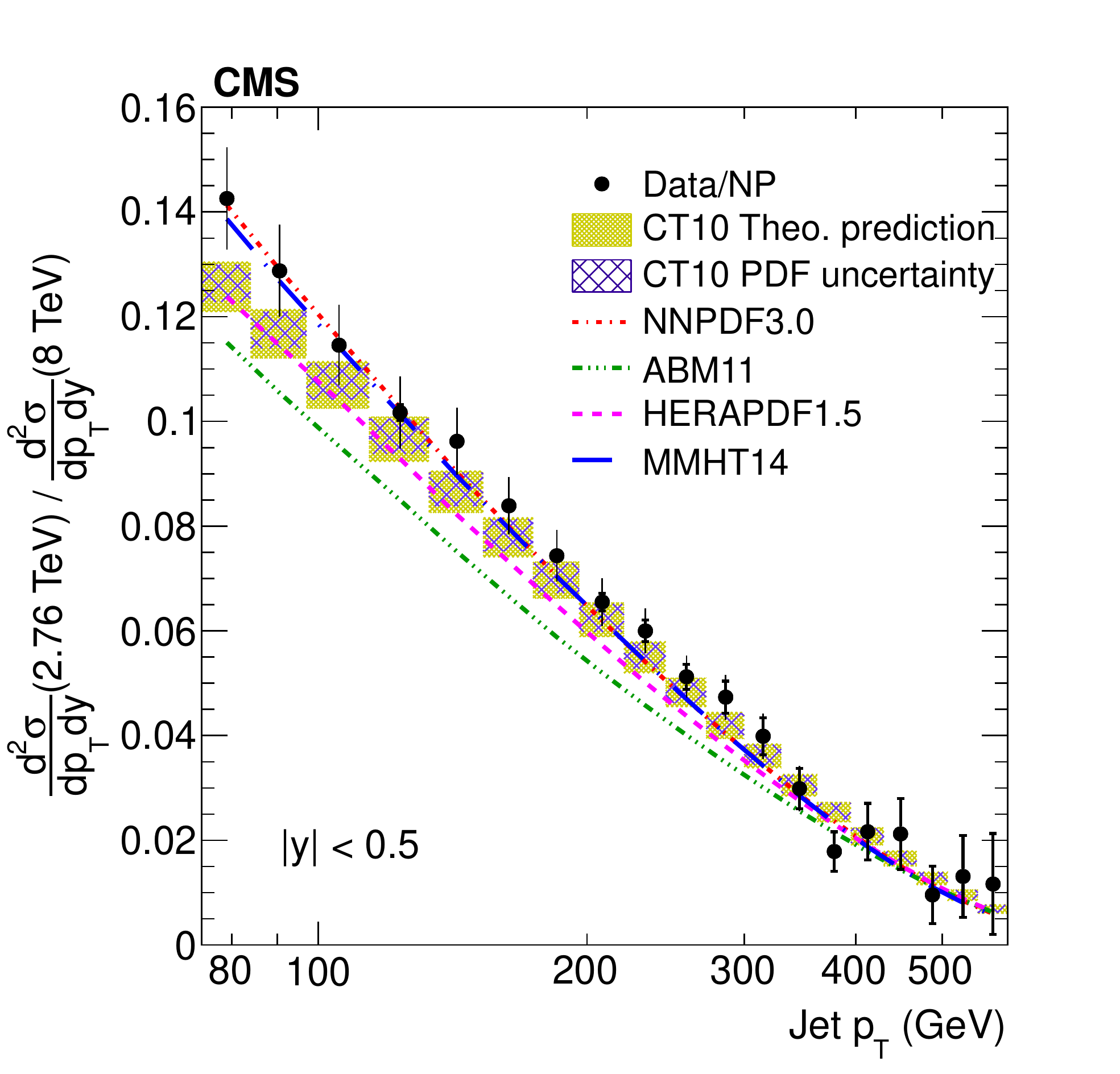}
\includegraphics[width=0.49\textwidth]{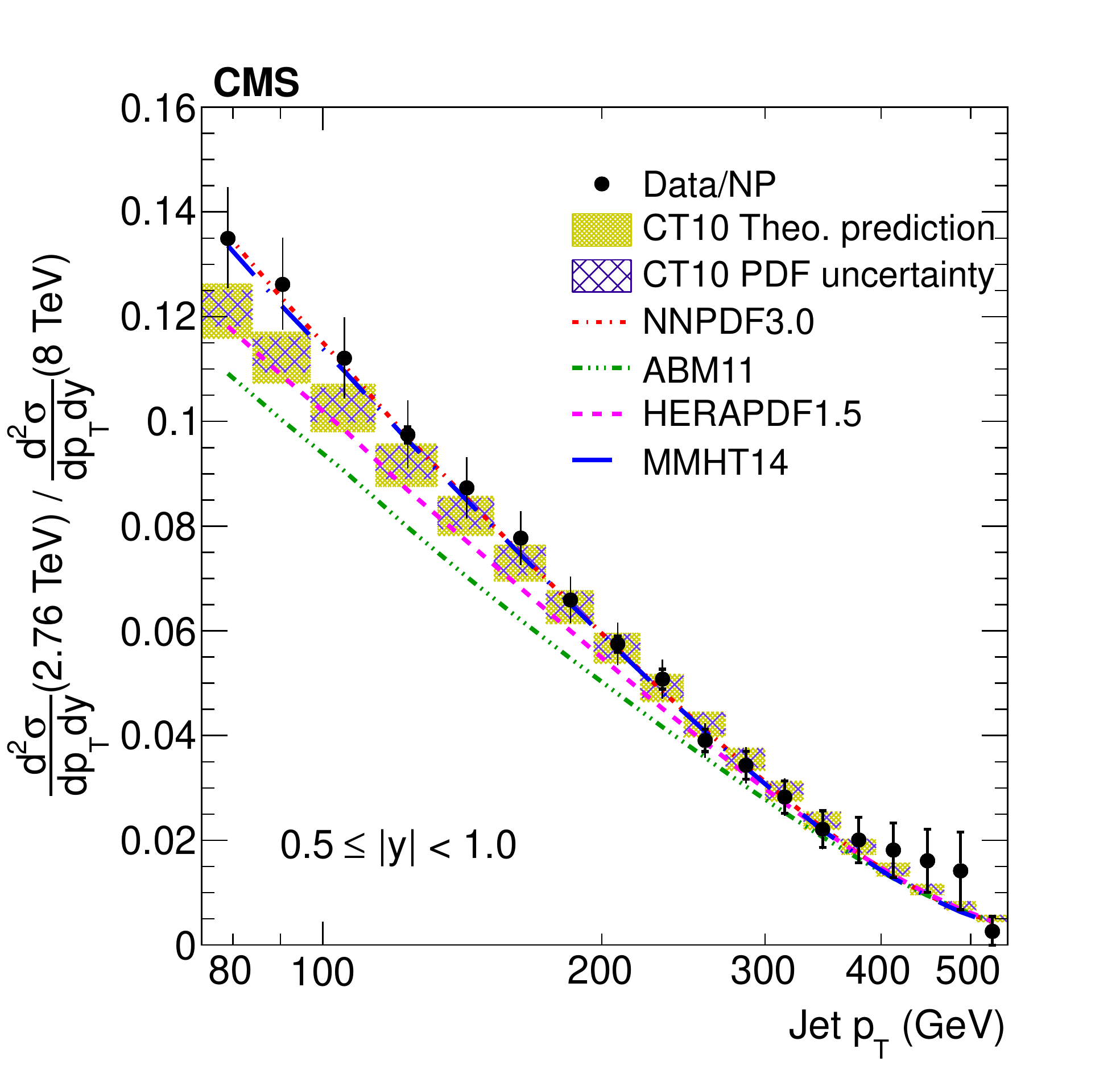} \\
\includegraphics[width=0.49\textwidth]{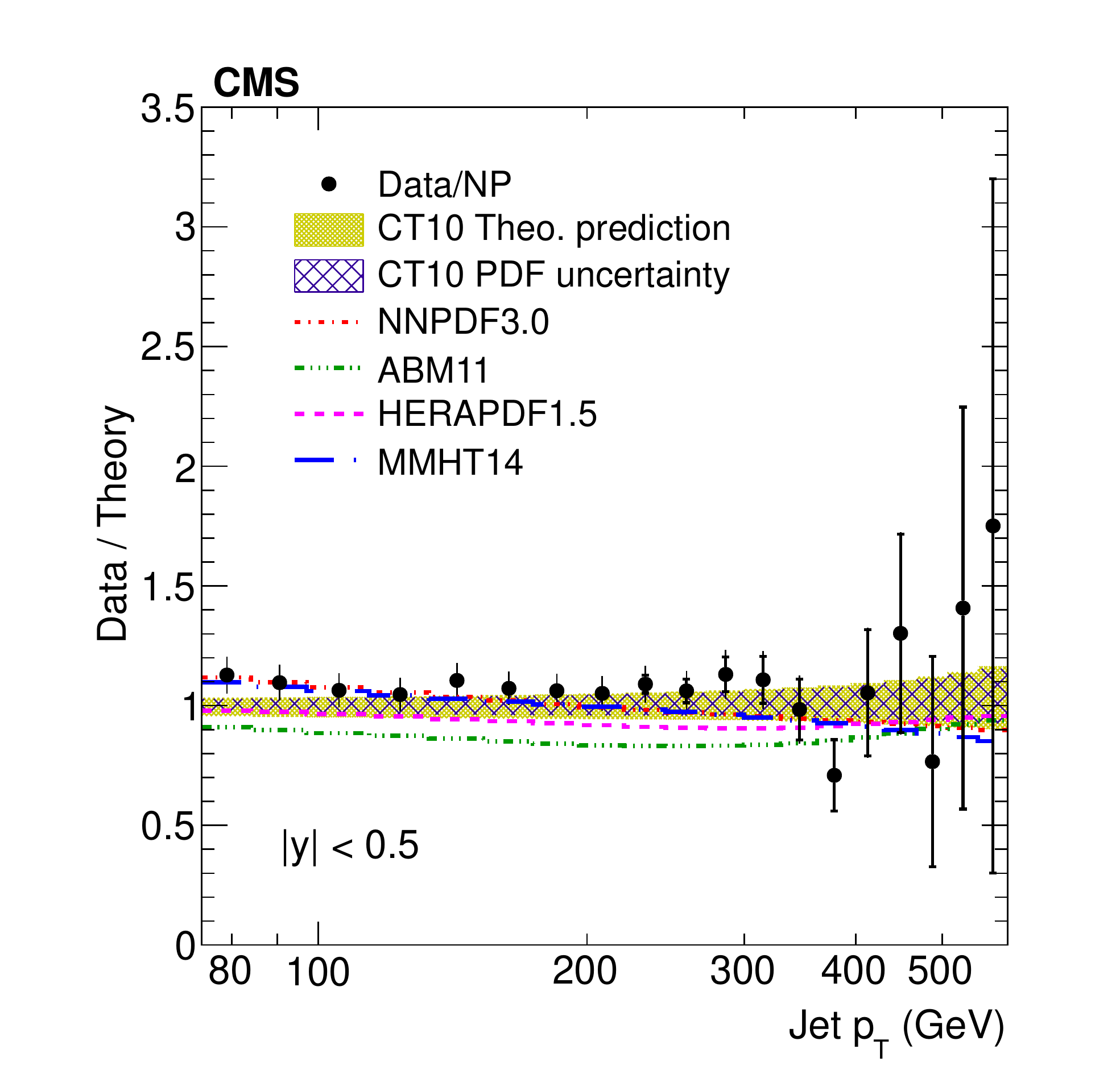}
\includegraphics[width=0.49\textwidth]{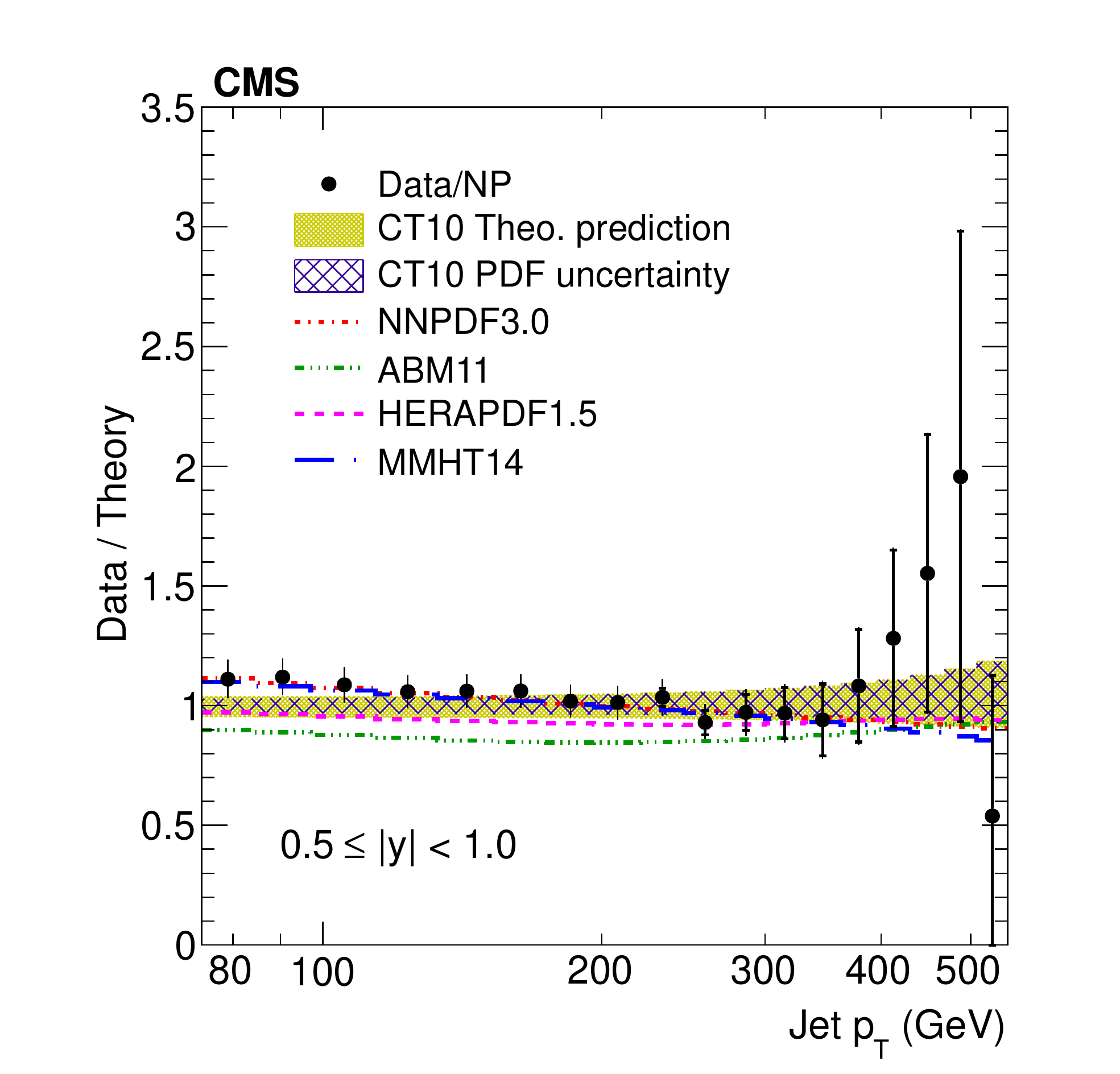} \\
\caption{The ratios (top panels) of the inclusive jet production
cross sections at $\sqrt{s}=2.76$ and 8\TeV are shown as
a function of jet \pt for the absolute rapidity range $\abs{y} < 0.5$
(left) and $0.5 < \abs{y} < 1.0$ (right). The data (closed symbols)
are shown with their statistical (inner error bar) and total
(outer error bar) uncertainties. For comparison, the NLO pQCD
prediction by using the CT10 PDF is shown with its total uncertainty (light
shaded band), while the contribution of the PDF
uncertainty is presented by the hatched band. Predictions that use
alternative PDF sets are shown by lines of different styles
without uncertainties. The data to theory ratios (bottom panels)
are shown using the same notations for the respective absolute rapidity
ranges.}
\label{fig:8TeVOv276TeV1}
\end{figure*}

\begin{figure*}[hbpt]
\centering
\includegraphics[width=0.45\textwidth]{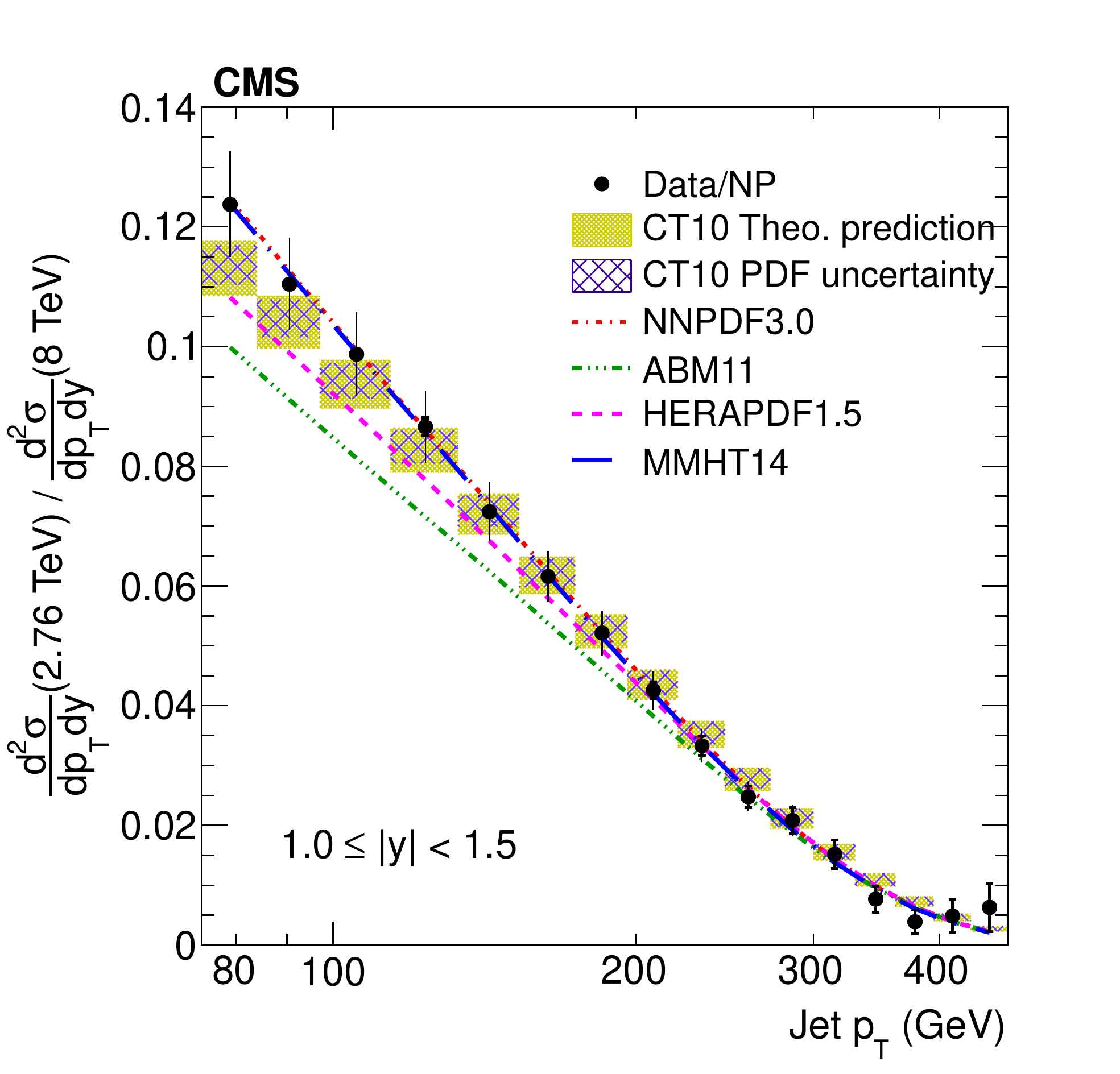}
\includegraphics[width=0.45\textwidth]{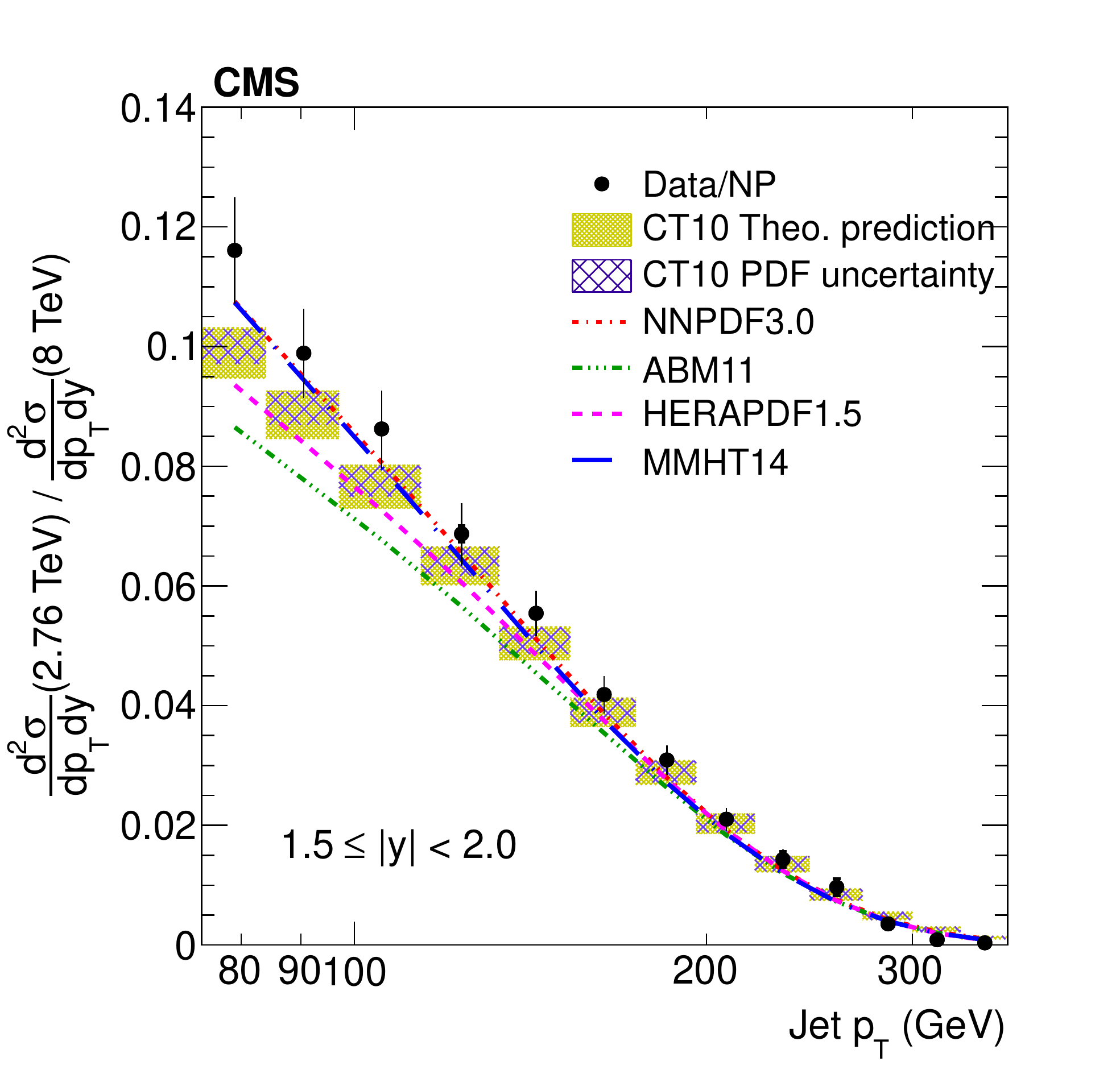} \\
\includegraphics[width=0.45\textwidth]{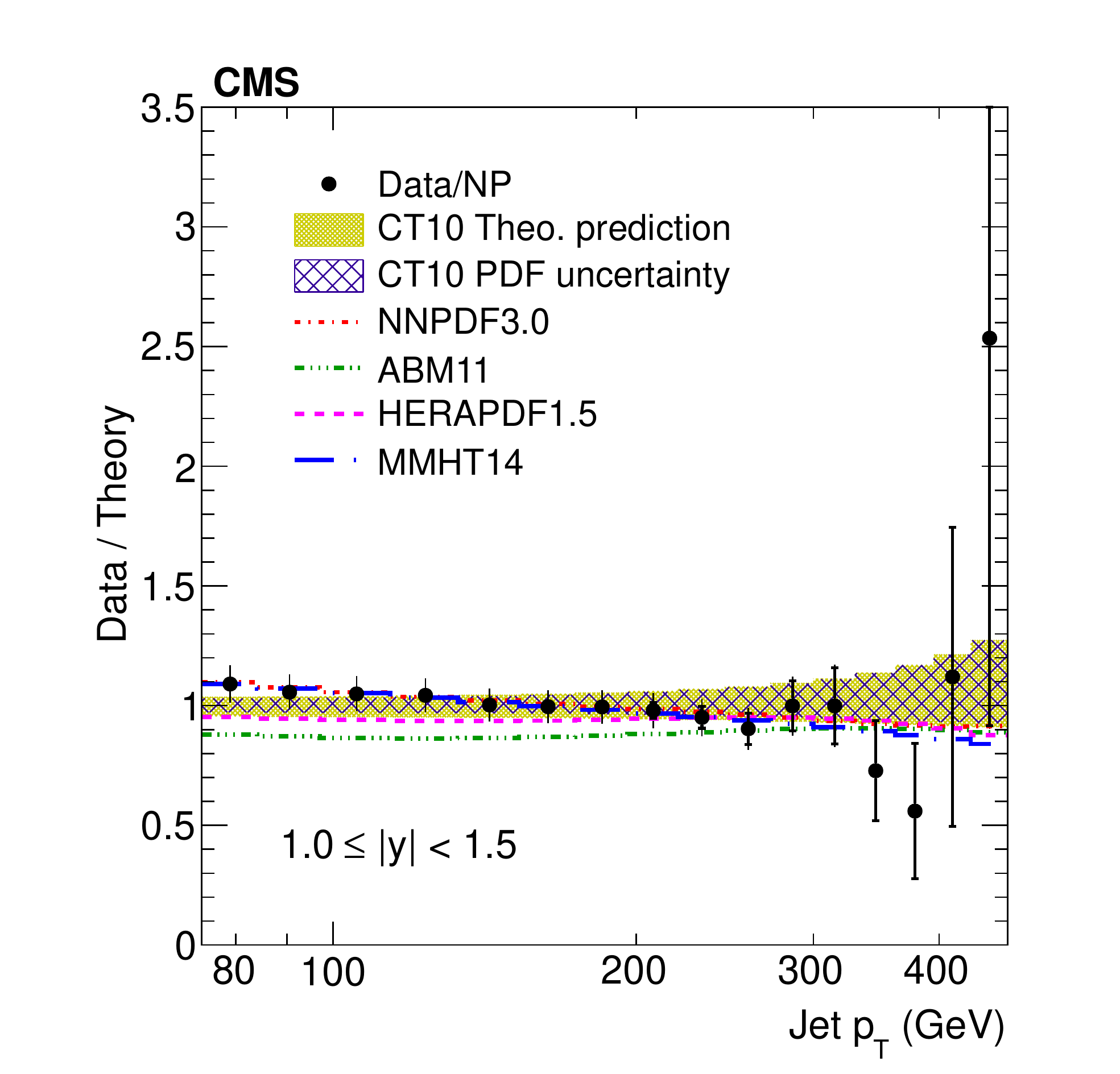}
\includegraphics[width=0.45\textwidth]{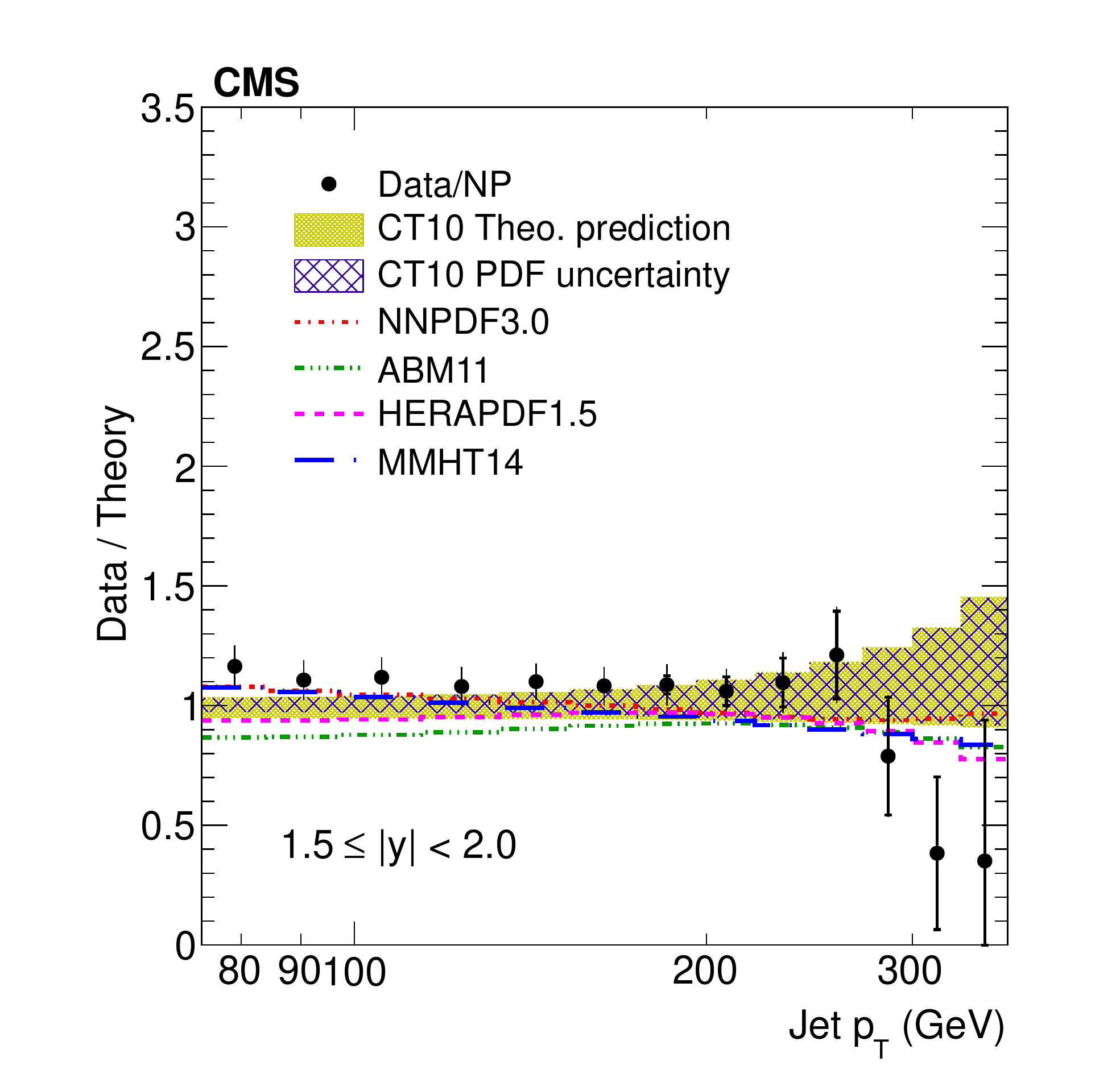} \\
\caption{The ratios of the inclusive jet production cross sections
at $\sqrt{s} =$ 2.76 and 8\TeV shown as a function of jet \pt for
the absolute rapidity ranges $1.0 < \abs{y} < 1.5$ and $1.5 < \abs{y} <
2.0$.}
\label{fig:8TeVOv276TeV2}
\end{figure*}

\begin{figure*}[hbpt]
\centering
\includegraphics[width=0.45\textwidth]{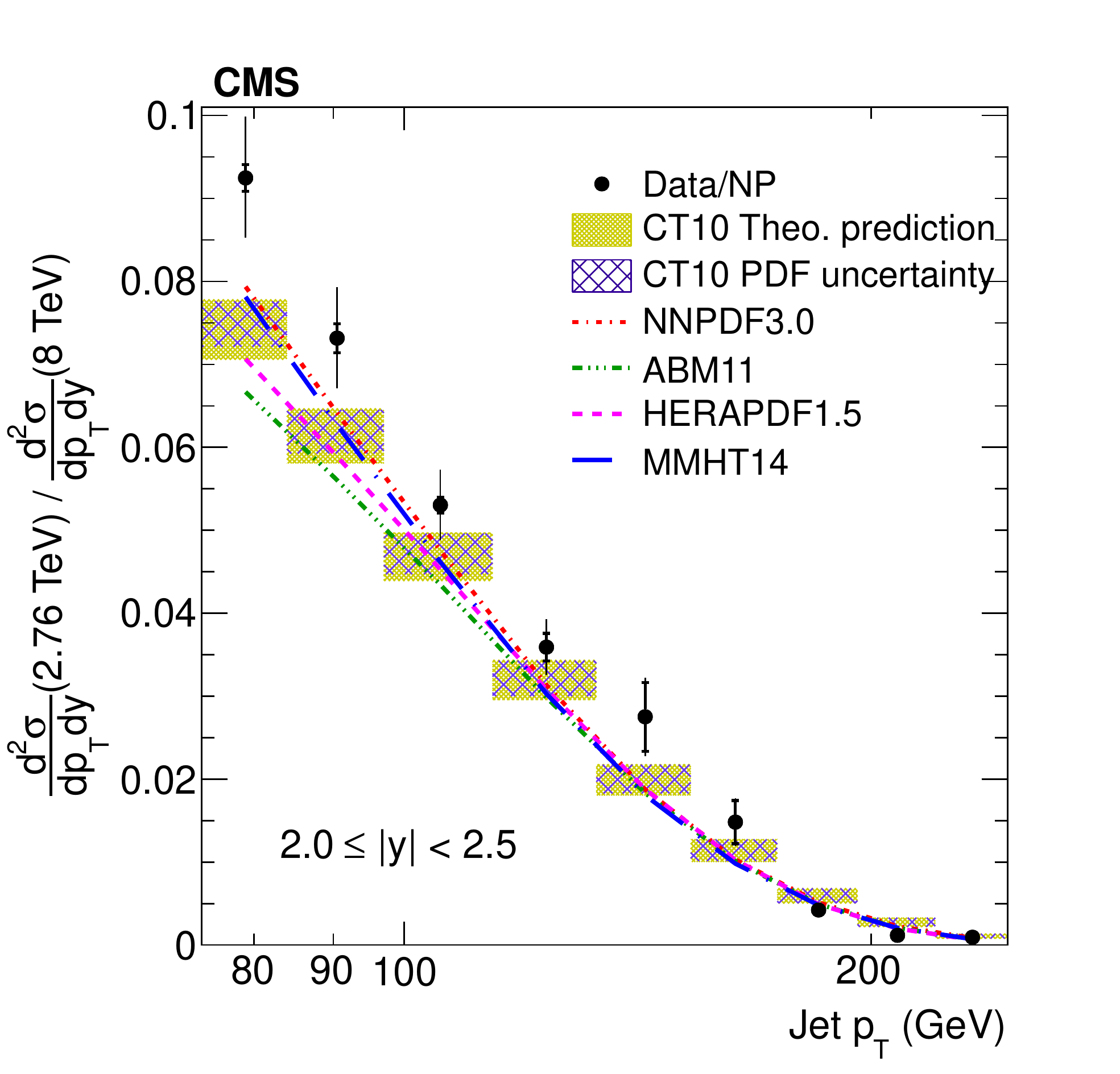}
\includegraphics[width=0.45\textwidth]{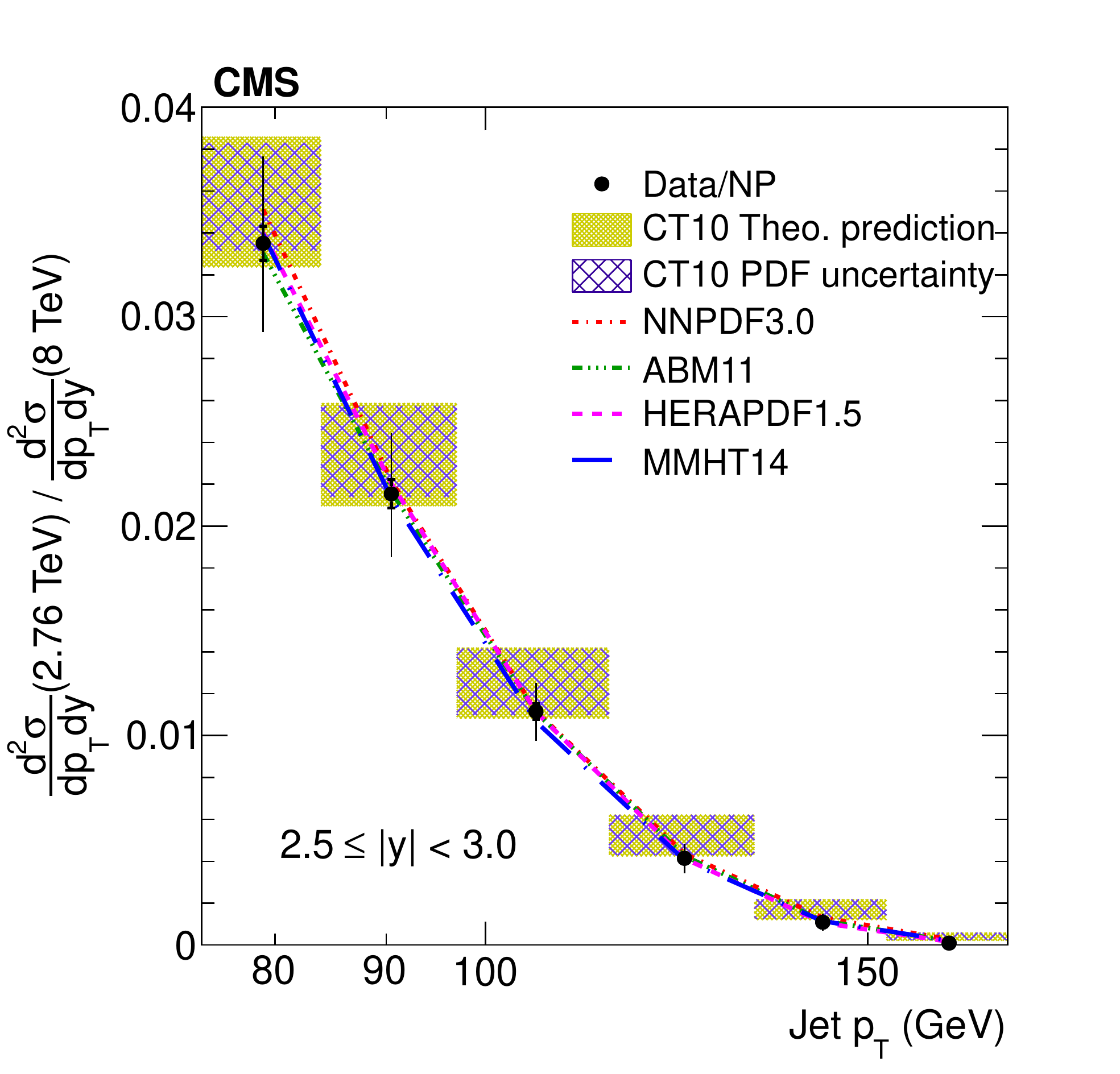} \\
\includegraphics[width=0.45\textwidth]{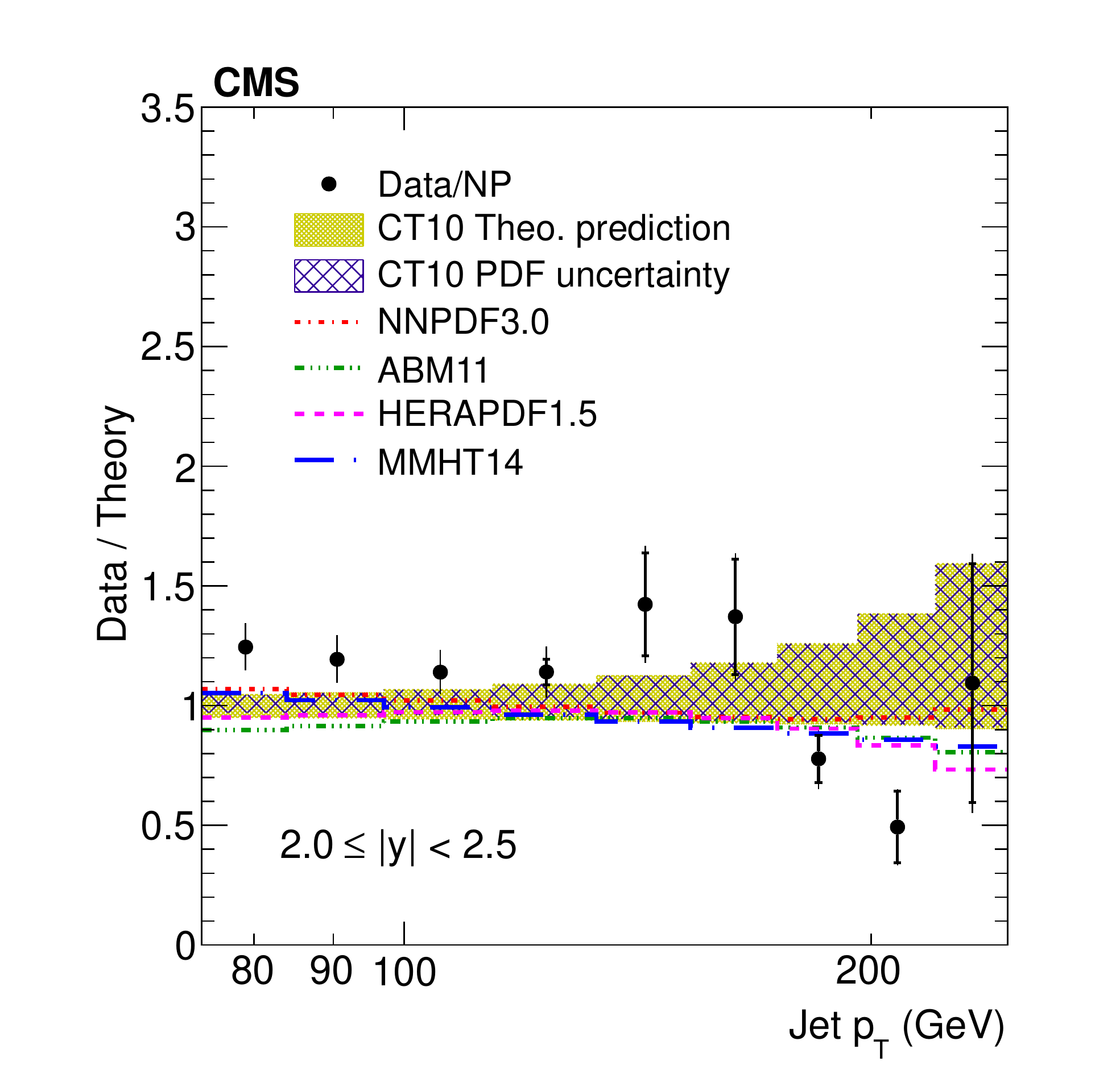}
\includegraphics[width=0.45\textwidth]{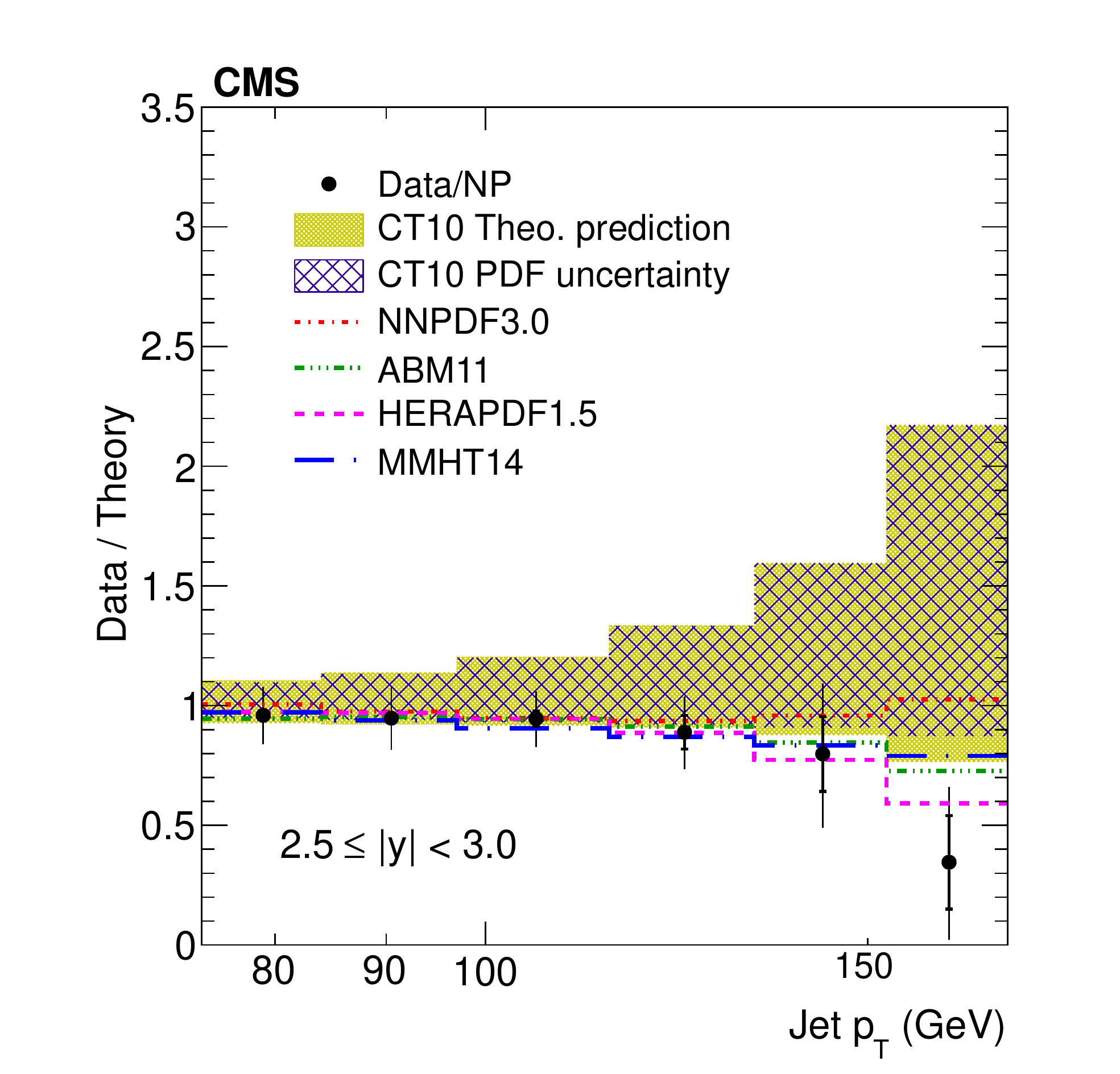} \\
\caption{The ratios of the inclusive jet production cross sections
at $\sqrt{s} =$ 2.76 and 8\TeV shown as a function of jet \pt for
the absolute rapidity ranges $2.0 < \abs{y} < 2.5$ and $2.5 < \abs{y} <
3.0$.}
\label{fig:8TeVOv276TeV3}
\end{figure*}

The ratios of the cross sections measured at 2.76\TeV to those
measured at 8\TeV are determined in a similar way.  Results are
presented in Figs.~\ref{fig:8TeVOv276TeV1}--\ref{fig:8TeVOv276TeV3},
and compared to theoretical predictions that use different PDF
sets. In general, the predictions describe the data well. The central
value of the theoretical prediction and its uncertainty are completely
dominated by the choice of and the uncertainty in the PDFs,
demonstrating the strong sensitivity of the 2.76 to 8\TeV cross
section ratio to the description of the proton structure.


\section{Determination of \texorpdfstring{\as}{alpha(S)}}
\label{sec:alphas}

Measurements of jet production at hadron colliders can be used to
determine the strong coupling constant \as, as has been previously
from the CMS 7\TeV inclusive jet
measurement~\cite{Khachatryan:2014waa}, and from Tevatron
measurements~\cite{D0Alpha1, D0Alpha2, ALPHASCDF}. The procedure to
extract \as in Ref.~\cite{Khachatryan:2014waa} is adopted here. Only
the high-\pt jet data are used, since the sensitivity of the \as
predictions increases with jet \pt. The determination of \as is
performed by minimizing the $\chi^2$ between the data and the theory
prediction.  The NLO theory prediction, corrected for nonperturbative
and electroweak effects, is used. At NLO, the dependence of the
differential inclusive jet production cross section $\rd\sigma/\rd\pt$
on \as is given by:
\begin{equation}
\frac{\rd\sigma}{\rd\pt} =
\as^2(\mu_{\mathrm{R}})\hat{X}^{(0)}(\mu_{\mathrm{F}},\pt)[1 +
\as(\mu_{\mathrm{R}})\mathrm{K1}(\mu_{\mathrm{R}},
\mu_{\mathrm{F}},\pt)] ,
\label{eqn:SigmaNLO}
\end{equation}
where \as is the strong coupling,
$\hat{X}^{(0)}(\mu_{\mathrm{F}},\pt)$ represents the LO contribution
to the cross section and
$\mathrm{K1}(\mu_{\mathrm{R}},\mu_{\mathrm{F}},\pt)$ is an NLO
correction term.  A comparison with the measured spectrum gives an
estimate of the input value of \as for which the cross section,
predicted from theory, has the best agreement with data.

The extraction of \as is performed by a least squares minimization of
the function
\begin{equation}
\chi^2(\alpsmz) = \Bigl(D - T(\alpsmz)\Bigr)^TC^{-1}\Bigl(D - T(\alpsmz)\Bigr),
\label{eq:chi2}
\end{equation}
where $D$ is the array of measured values of the double-differential
inclusive jet cross section for the different bins in \pt and $\abs{y}$,
$T(\alpsmz)$ is the corresponding set of theoretical cross sections
for a given value of $\alpsmz$, and $C$ is the covariance matrix
including all the experimental and theoretical uncertainties involved
in the measurement. The total covariance matrix $C$ is built from the
individual components as follows:
\begin{equation}
C= C^\text{stat} + C^\text{unfolding}+\sum C^\mathrm{JES} +
C^\text{uncor} + C^\text{lumi} + C^\mathrm{PDF}+ C^\mathrm{NP},
\label{eqn:CovMat}
\end{equation}
where:
\begin{itemize}
\item $C^\text{stat}$ is the statistical covariance matrix, taking into
account the correlation between different \pt bins of the same
rapidity range due to unfolding. Different rapidity ranges are
considered as uncorrelated among themselves;
\item $C^\text{unfolding}$ includes the uncertainty induced by
the JER parameterization in the unfolding procedure;
\item $C^\mathrm{JES}$ includes the uncertainty due to JES uncertainties,
obtained as the sum of 24 independent matrices, one for each source of
uncertainty;
\item $C^\text{uncor}$ includes all uncorrelated systematic
uncertainties such as trigger and jet identification inefficiencies,
and time dependence of the jet \pt resolution;
\item $C^\text{lumi}$ includes the 2.6\% luminosity uncertainty;
\item $C^\mathrm{PDF}$ is related to uncertainties in the PDF used in
the theoretical prediction;
\item $C^\mathrm{NP}$ includes the uncertainty due to nonperturbative corrections in
the theoretical prediction.
\end{itemize}
The unfolding, JES, lumi, PDF, and NP
systematic uncertainties are considered as 100\% correlated among all
\pt and $\abs{y}$ bins.

\begin{figure*}[hb!]
\centering
\includegraphics[width=0.4\textwidth]{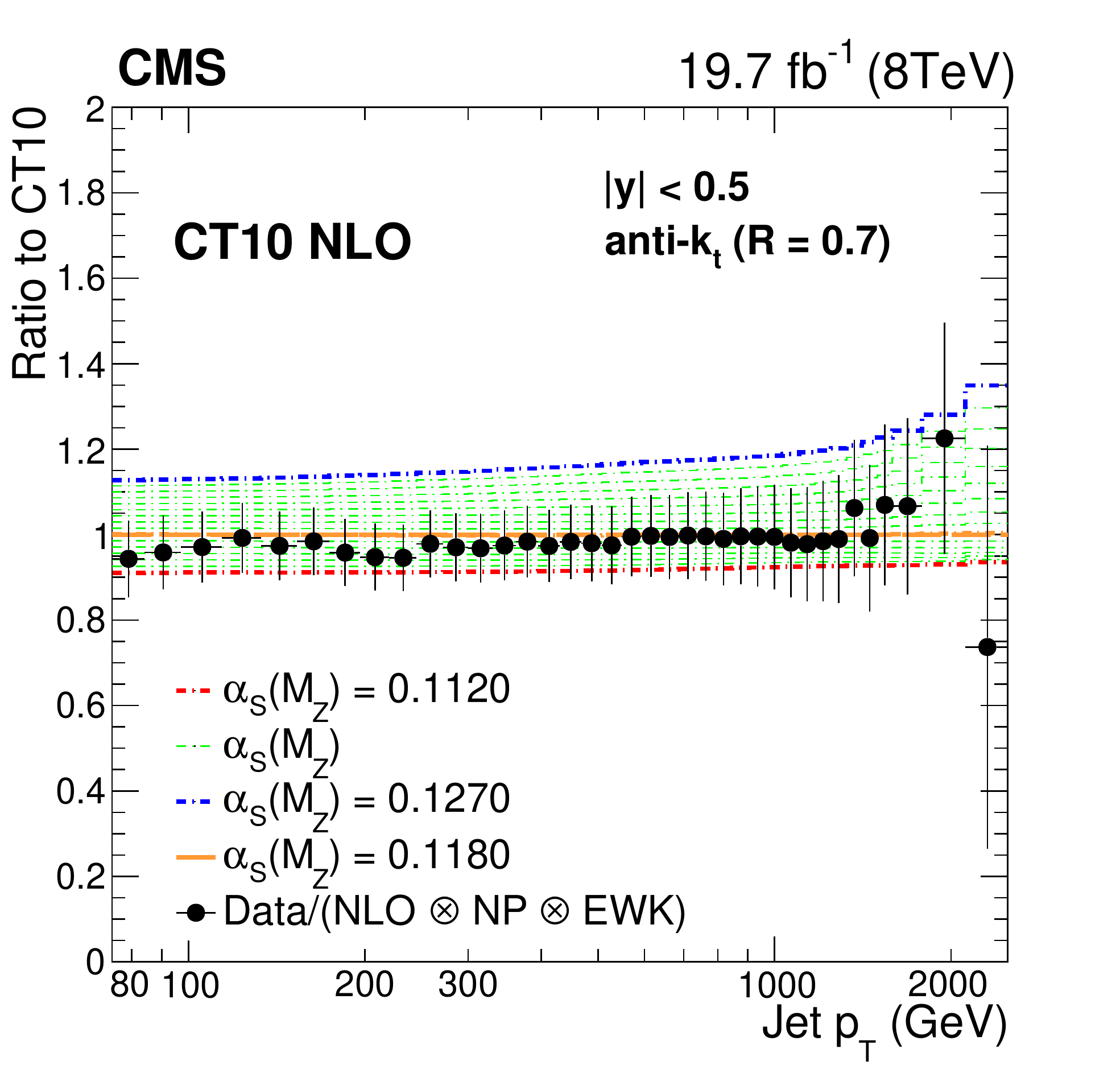}
\includegraphics[width=0.4\textwidth]{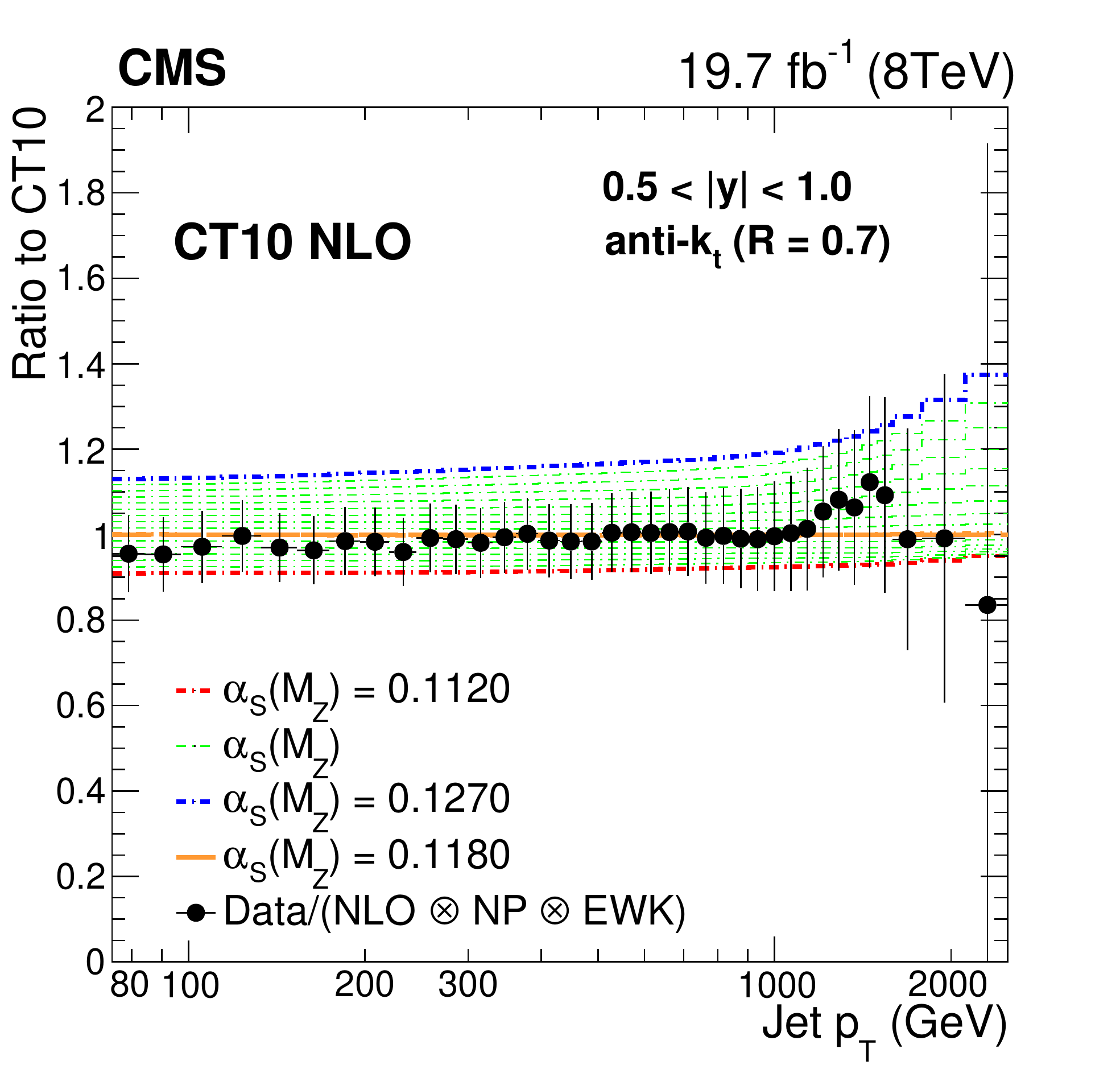}
\includegraphics[width=0.4\textwidth]{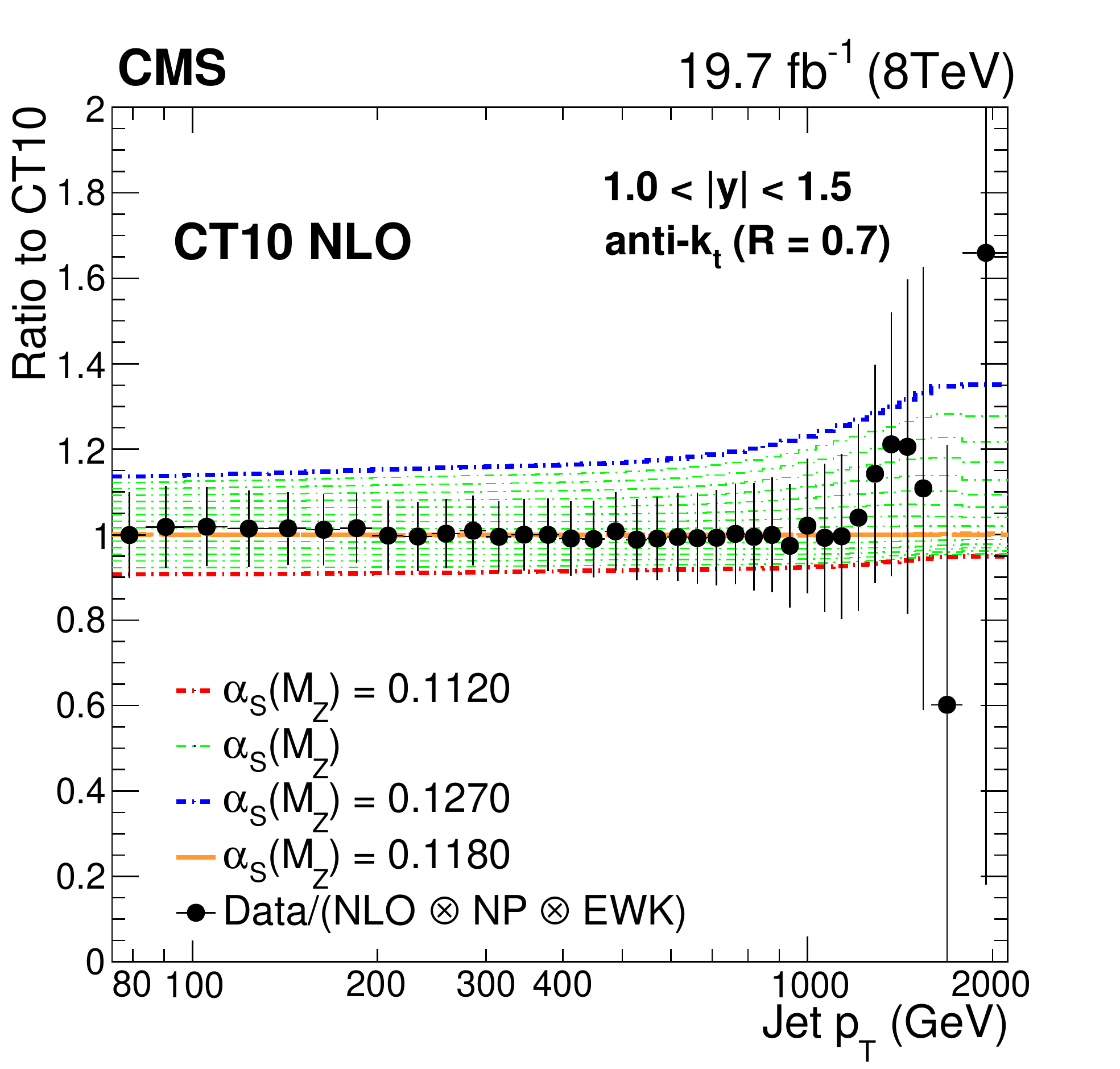}
\includegraphics[width=0.4\textwidth]{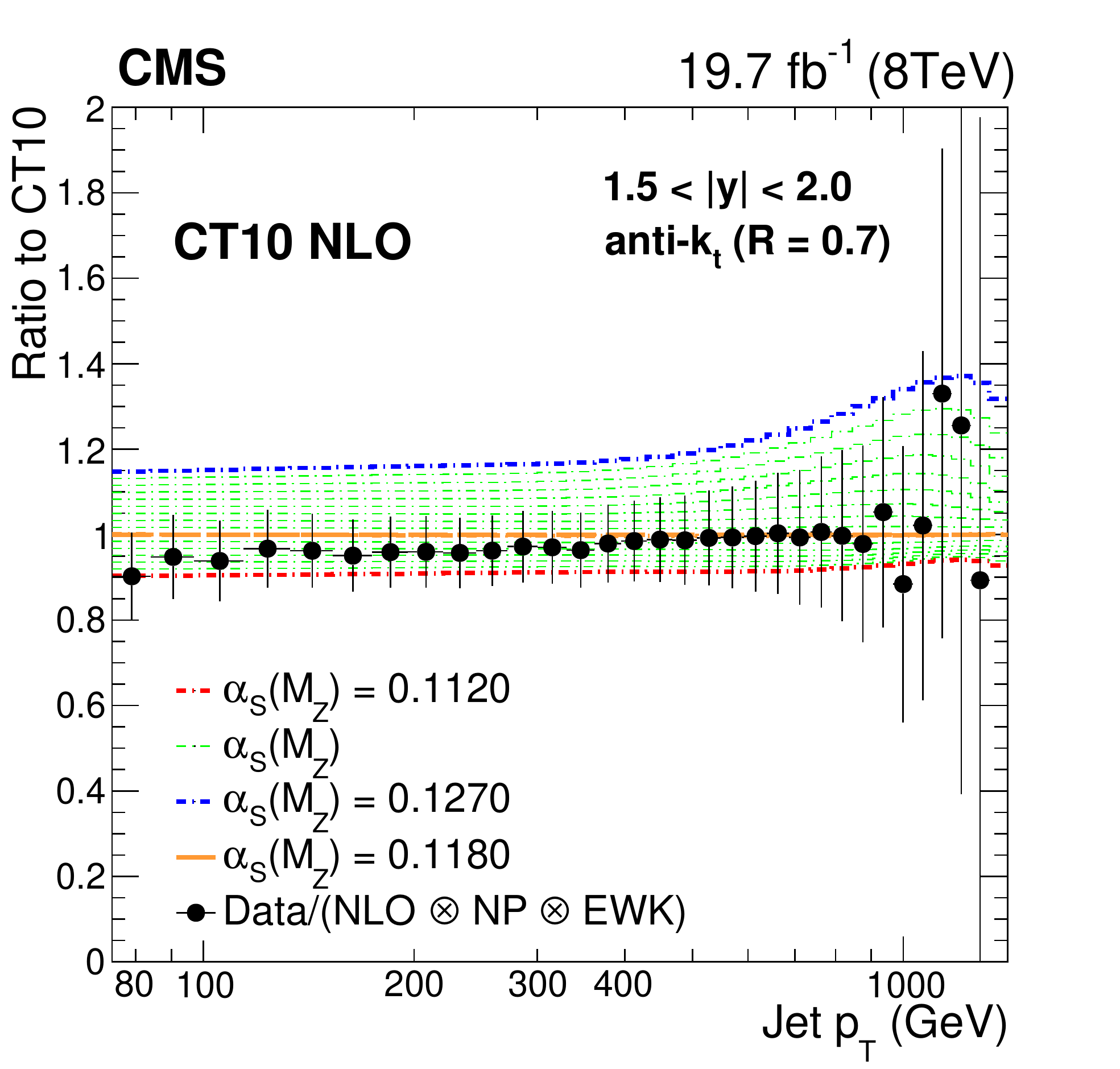}
\includegraphics[width=0.4\textwidth]{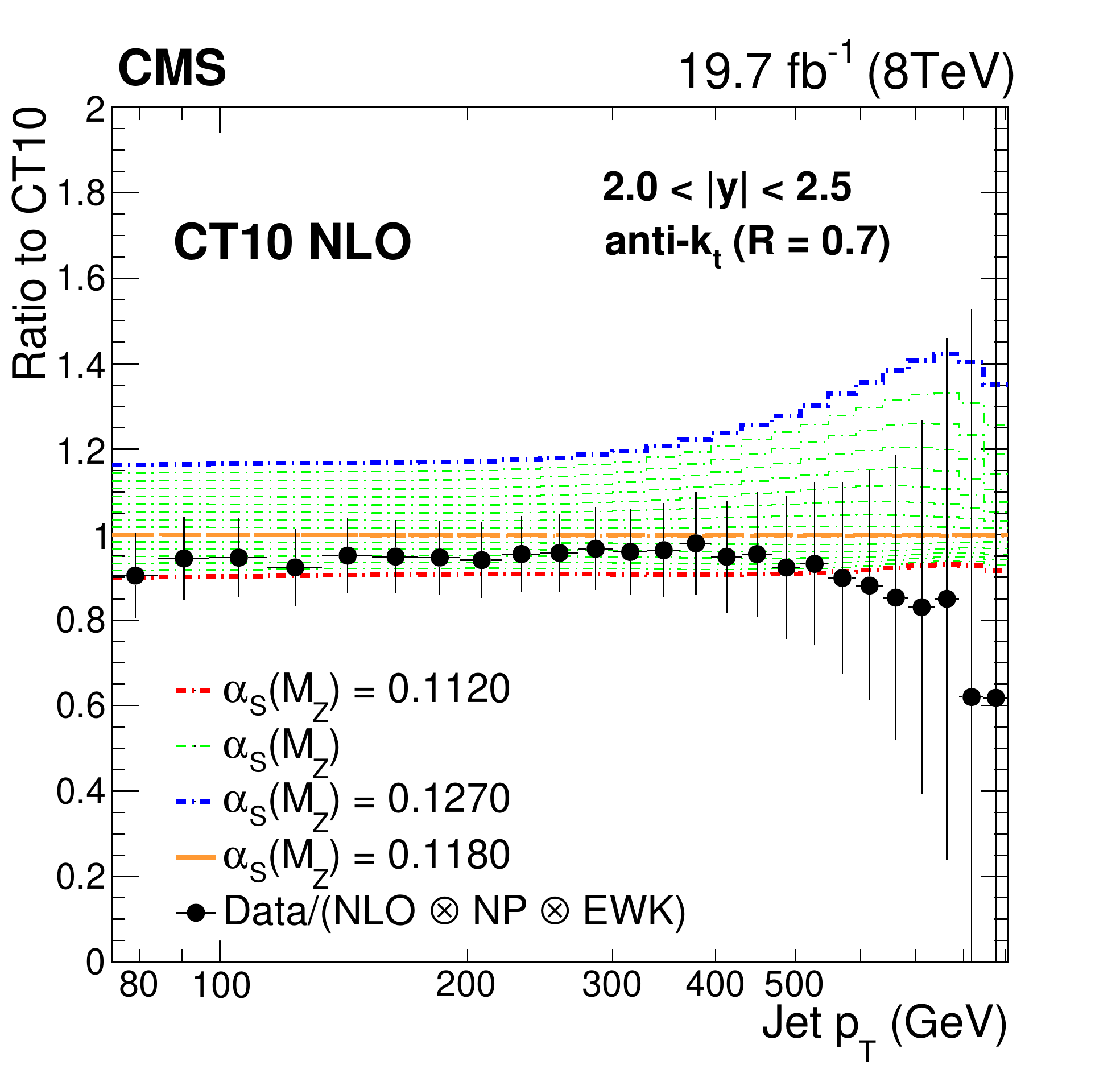}
\includegraphics[width=0.4\textwidth]{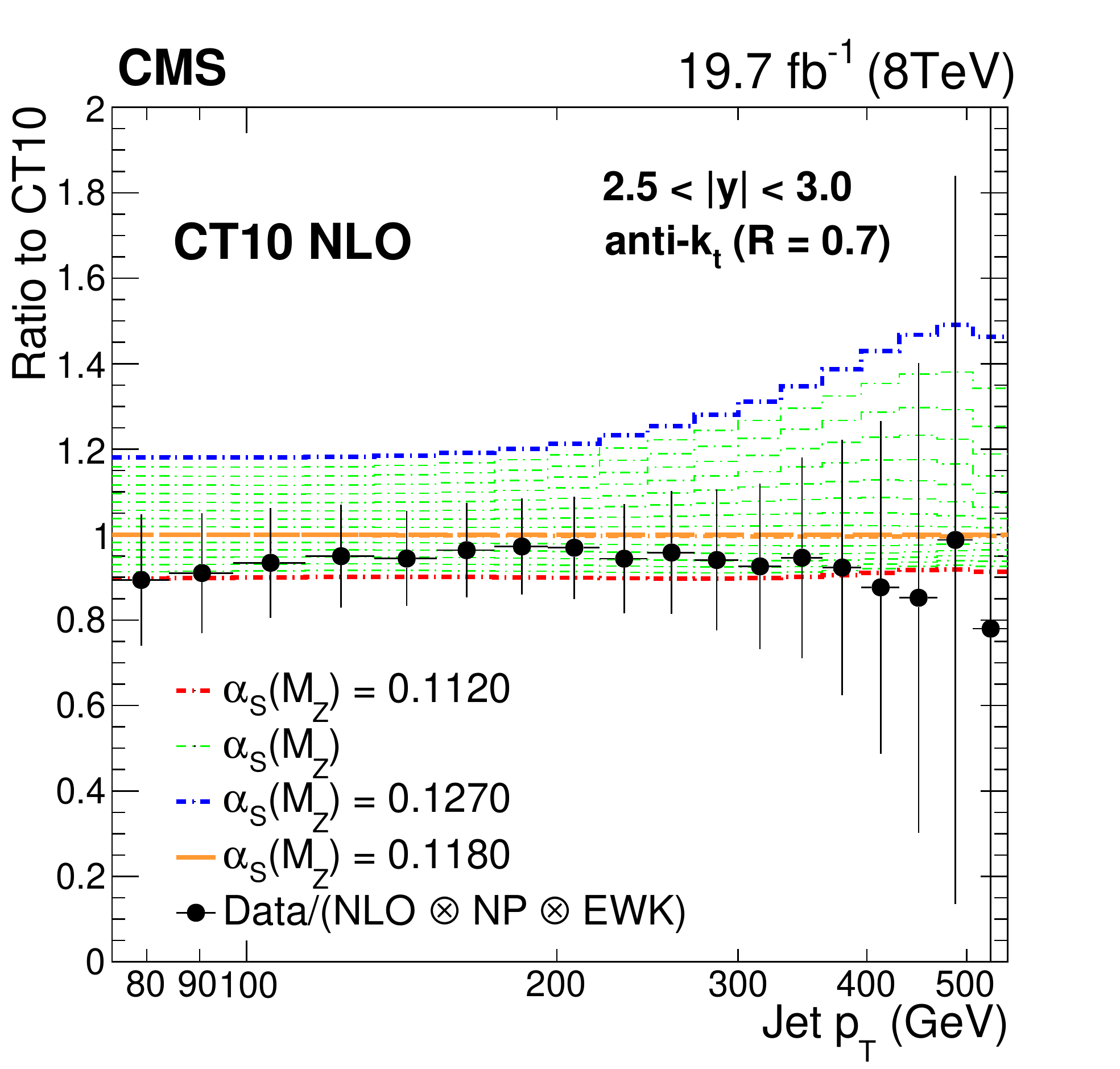}
\caption{Ratio of data over theory prediction (closed circles)
using the CT10 NLO PDF set, with the default \alpsmz value of
0.118. Dashed lines represent the ratios of the predictions
obtained with the CT10 PDF set evaluated with different \alpsmz
values, to the central one. The error bars correspond to the total
uncertainty of the data.}
\label{fig:alphaSsens}
\end{figure*}

The extraction of \as uses the CT10 NLO PDF set in the theoretical
calculation, since it provides the best agreement with measured cross
sections, as shown in Section~\ref{ref:compa}. This PDF set
provides variants corresponding to 16 different \alpsmz values in the
range 0.112--0.127 in steps of 0.001.  The sensitivity of the theory
prediction to the \as choice in the PDF is illustrated in
Fig.~\ref{fig:alphaSsens}.

\begin{figure*}[ht]
\centering
\includegraphics[width=0.6\textwidth]{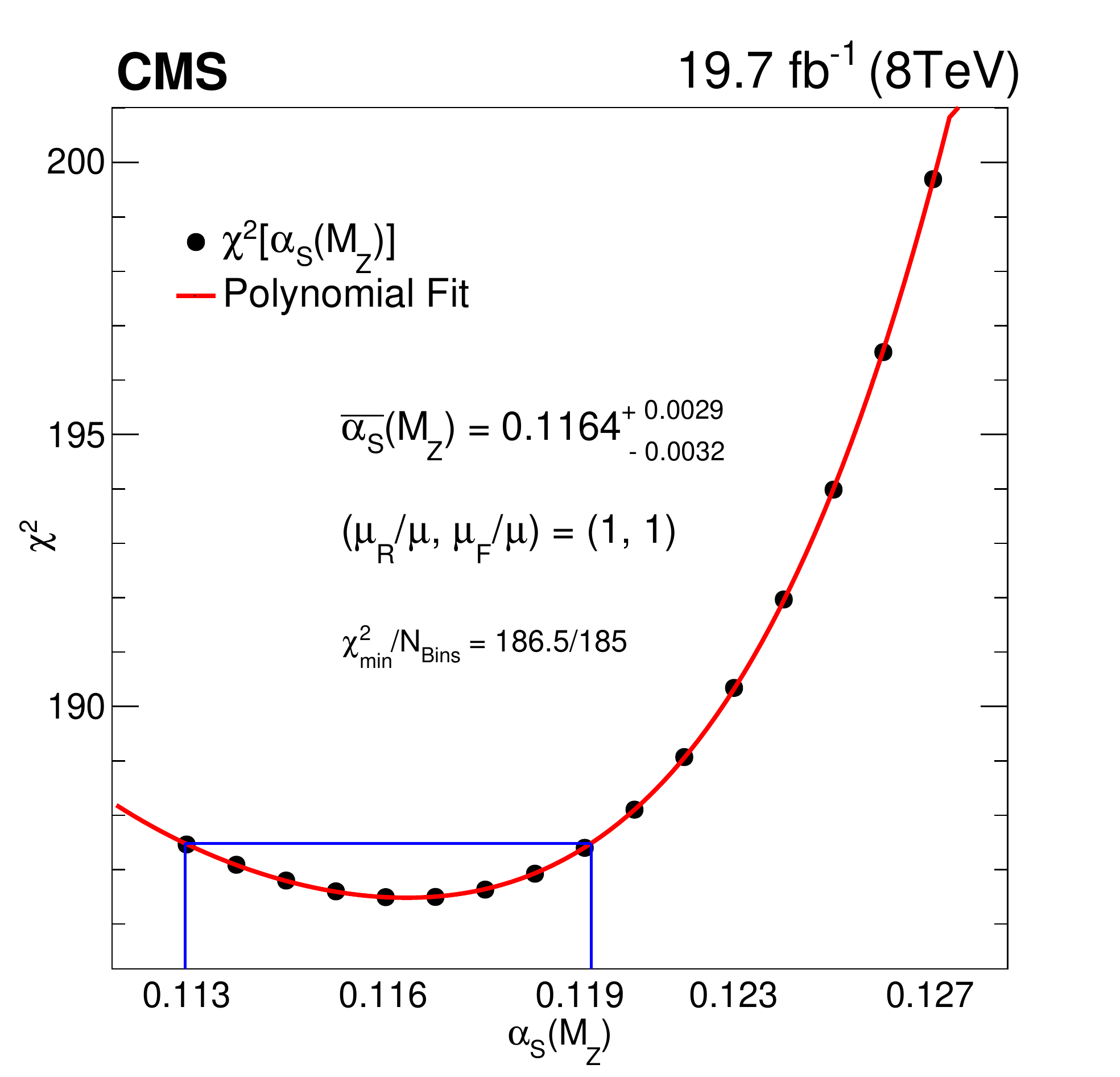}
\caption{The $\chi^2$ minimization with respect to \alpsmz by using
the CT10 NLO PDF set and data from all rapidity bins. The
uncertainty is obtained from the \alpsmz values for which $\chi^2$
is increased by one with respect to the minimum value, indicated by
the box. The curve corresponds to a fourth-degree polynomial fit
through the available $\chi^2$ points.}
\label{fig:ChiSquare}
\end{figure*}

The $\chi^2$ in Eq.~(\ref{eq:chi2}) is computed, combining all \pt
and $\abs{y}$ intervals, for each of the variants corresponding to a
different \as value, as shown in Fig.~\ref{fig:ChiSquare}. The
variation of $\chi^2$ with \as is fitted with a fourth-order
polynomial, and the minimum ($\chi^2_\text{min}$) corresponds to the
best \alpsmz value. Uncertainties are determined using the $\Delta
\chi^2 =1$ criterion.  The individual contribution from each
uncertainty source listed in Eq.~(\ref{eqn:CovMat}) is estimated as
the quadratic difference between the main result and the result of an
alternative fit, in which that particular source is left out of the
covariance matrix definition.

The uncertainties due to the choice of the renormalization and
factorization scales are evaluated by variations of the default
$\mu_{\mathrm{R}}$, $\mu_{\mathrm{F}}$ values, set to jet \pt, in the
following six combinations:
($\mu_{\mathrm{R}}/\pt$,$\mu_{\mathrm{F}}/\pt$) = (0.5,0.5), (0.5,1),
(1,0.5), (1,2), (2,1), and (2,2). The $\chi^2$ minimization with
respect to \alpsmz is repeated in each case, and the maximal upwards
and downwards deviations of \alpsmz from the central result are taken
as the corresponding uncertainties.

In Table~\ref{table:ct10AlphaNLO}, the fitted values of \as are
presented for each rapidity bin, separately, and for the whole
range. The contribution to the uncertainty due to each individual
source is also given, together with the best $\chi^2_\text{min}$
value for each separate fit. The largest source of uncertainty in the
determination of \as is due to the choice of renormalization and
factorization scales, pointing to the need for including higher order
corrections in the theoretical calculations.

\begin{table}[htb]
\renewcommand{\arraystretch}{1.5}
\centering
\topcaption{Results for \alpsmz extracted using the CT10 NLO PDF
set. The fitted value for each $\abs{y}$ bin; the corresponding
uncertainty components due to PDF, scale, and nonperturbative corrections; and
the total experimental uncertainty is shown. The last row of the
table shows the results of combined fitting of all the $\abs{y}$ bins
simultaneously.}
\resizebox{\textwidth}{!}{
\begin{tabular}{cccccccc}
\hline
$\abs{y}$ bin     & & Fitted \alpsmz &    PDF unc. &           scale unc. &                NP unc. &                exp unc. &         $\chi^{2}_{\text{min}}/N_{\text{Bins}}$ \\
\hline
0.0--0.5 & &    0.1155 &       $^{+0.0027}_{-0.0027}$ &  $^{+0.0070}_{-0.0026}$ &    $^{+0.0003}_{-0.0003}$ & $^{+0.0025}_{-0.0025}$    &    $48.6/37$ \\
0.5--1.0 & &    0.1156 &       $^{+0.0025}_{-0.0026}$ &  $^{+0.0069}_{-0.0026}$ &    $^{+0.0003}_{-0.0003}$ & $^{+0.0026}_{-0.0025}$    &    $28.4/37$ \\
1.0--1.5 & &    0.1177 &       $^{+0.0024}_{-0.0026}$ &  $^{+0.0062}_{-0.0027}$ &    $^{+0.0002}_{-0.0002}$ & $^{+0.0024}_{-0.0026}$    &    $19.3/36$ \\
1.5--2.0 & &    0.1163 &       $^{+0.0025}_{-0.0029}$ &  $^{+0.0040}_{-0.0019}$ &    $^{+0.0002}_{-0.0002}$ & $^{+0.0023}_{-0.0027}$    &    $65.6/32$ \\
2.0--2.5 & &    0.1164 &       $^{+0.0020}_{-0.0022}$ &  $^{+0.0046}_{-0.0024}$ &    $^{+0.0002}_{-0.0002}$ & $^{+0.0019}_{-0.0022}$    &    $38.3/25$ \\
2.5--3.0 & &    0.1158 &       $^{+0.0029}_{-0.0030}$ &  $^{+0.0049}_{-0.0025}$ &    $^{+0.0006}_{-0.0006}$ & $^{+0.0036}_{-0.0038}$    &    $14.3/18$ \\
\hline
Combined  & &    0.1164 &       $^{+0.0025}_{-0.0029}$ &  $^{+0.0053}_{-0.0028}$ &    $^{+0.0001}_{-0.0001}$           & $^{+0.0014}_{-0.0015}$    &    $186.5/185$ \\
\hline
\end{tabular}
}
\label{table:ct10AlphaNLO}
\end{table}

The best value obtained, by using the CT10 NLO PDF set, is
\begin{equation*}
\alpsmz(\mathrm{NLO}) =
0.1164^{+0.0025}_{-0.0029}(\mathrm{PDF})^{+0.0053}_{-0.0028}(\mathrm{scale})\pm
0.0001(\mathrm{NP})^{+0.0014}_{-0.0015}(\mathrm{exp}) = 0.1164^{+0.0060}_{-0.0043} \,.
\end{equation*}

Alternatively, the value of \alpsmz is also determined using the
NNPDF3.0 NLO PDF, resulting in $\alpsmz =
0.1172^{+0.0083}_{-0.0075}$. These values of \alpsmz are
compatible with the current world average $\alpsmz =
0.1181\pm0.0011$~\cite{Olive:2016xmw}.

The value of \as depends on the scale $Q$ at which it is
evaluated, decreasing as $Q$ increases. The measured \pt interval
74--2500\GeV is divided into nine different ranges as shown in the
first column in Table~\ref{table:ct10alphaRun}, and \alpsmz is
determined for each of them.

\begin{figure*}[hbpt]
\centering
\includegraphics[width=0.9\textwidth]{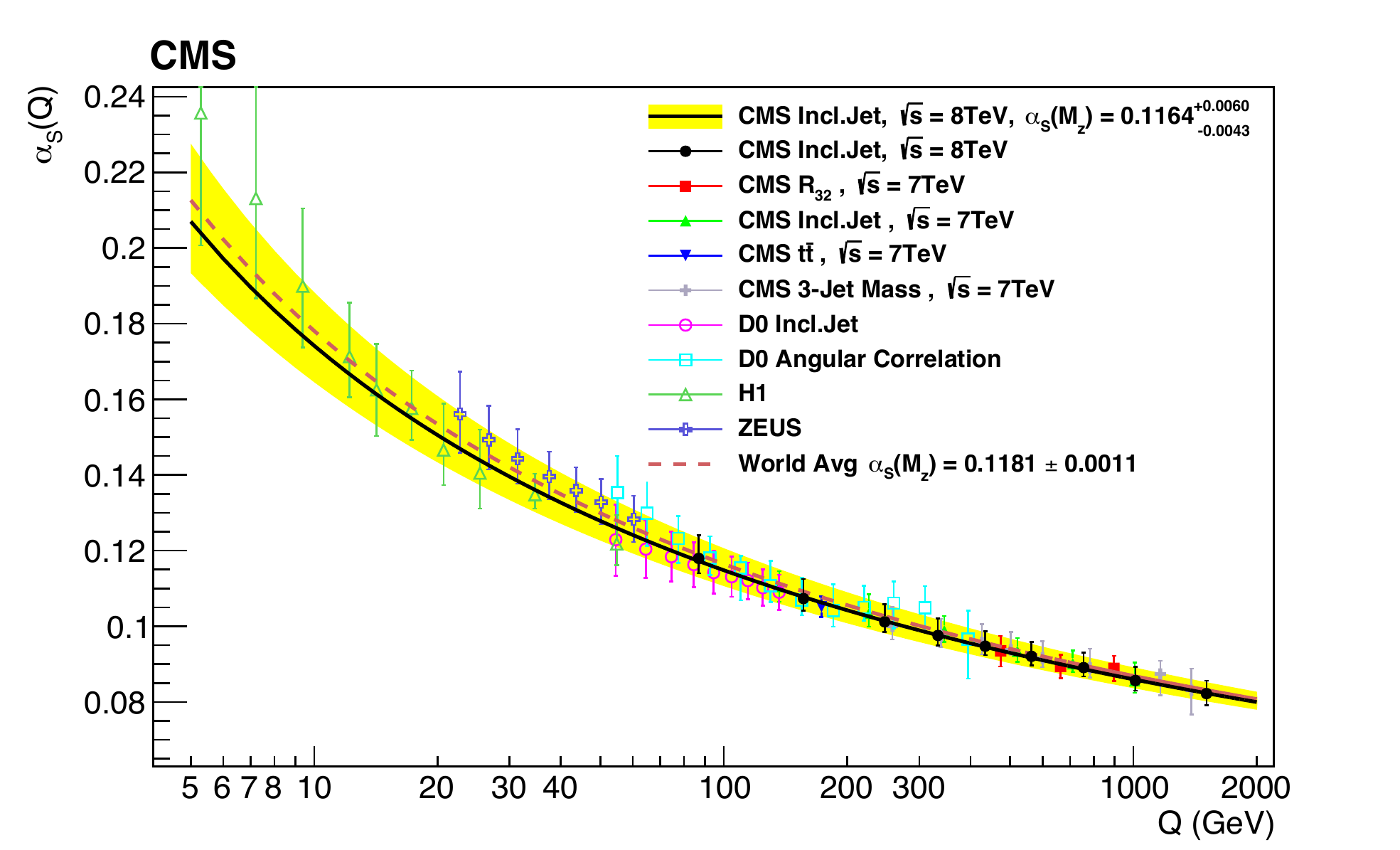}
\caption{The running \asq as a function of the scale $Q$ is shown, as
obtained by using the CT10 NLO PDF set. The solid line and the
uncertainty band are obtained by evolving the extracted \alpsmz
values by using the 2-loop 5-flavour renormalization group
equations. The dashed line represents the evolution of the world
average value. The black dots in the figure show the numbers
obtained from the $\sqrt{s}=8\TeV$ inclusive jet
measurement. Results from other CMS~\cite{R32, CMSTTBar,
CMS-PAS-SMP-12-027}, D0~\cite{D0Alpha1, D0Alpha2},
H1~\cite{Andreev:2016tgi}, and ZEUS~\cite{ZEUSAlpha}
measurements are superimposed.}
\label{fig:AlphasRun}
\end{figure*}

The $Q$ scale corresponding to each \pt range is evaluated as the
cross section weighted average \pt for that range. The extracted
\alpsmz values are evolved to the $Q$ scale corresponding to the
range, using the 2-loop 5-flavour renormalization group (RG)
evolution equation, resulting in the \asq values listed in
Table~\ref{table:ct10alphaRun}. The same RG equation is used to obtain
the corresponding uncertainties. The contributions to both the
experimental and theoretical uncertainties are shown in
Table~\ref{table:ct10alphaUnc}.  A comparison of these results with
those from the CMS~\cite{R32, CMSTTBar,
CMS-PAS-SMP-12-027}, D0~\cite{D0Alpha1, D0Alpha2},
H1~\cite{Andreev:2016tgi}, and ZEUS~\cite{ZEUSAlpha} experiments is
shown in Fig.~\ref{fig:AlphasRun}. The present measurement is in very
good agreement with results obtained by previous experiments. The
present analysis constrains the \asq running for $Q$ between 86\GeV
and 1.5\TeV, which is the highest scale at which \as has been measured,
to date.

\begin{table}[htb]
\renewcommand{\arraystretch}{1.5}
\centering
\topcaption{The extracted \alpsmz values, the corresponding \asq
values at the $Q$ scale for each \pt range, and
$\chi^{2}_{\text{min}}/{N}_\text{Bins}$ are
shown. Uncertainties are given for both \as values.}
\begin{tabular}{cccccc}
\hline
\pt range (\GeVns)  & $Q$ (\GeVns) & \alpsmz &  \asq  & $\chi^{2}_{\text{min}}/{N_\text{Bins}}$ \\
\hline
\x74--133                 & \x86.86   &       0.1171 $^{+0.0060}_{-0.0039}$    &     0.1180 $^{+0.0061}_{-0.0040}$ &     26.04/24           \\
133--220                & 156.52  &       0.1159 $^{+0.0061}_{-0.0037}$    &     0.1073 $^{+0.0052}_{-0.0032}$ &     19.47/24           \\
220--300                & 247.10  &       0.1161 $^{+0.0062}_{-0.0036}$    &     0.1012 $^{+0.0047}_{-0.0027}$ &     12.39/18           \\
300--395                & 333.27  &       0.1163 $^{+0.0064}_{-0.0039}$    &     0.0976 $^{+0.0045}_{-0.0027}$ &     19.48/18           \\
395--507                & 434.72  &       0.1167 $^{+0.0061}_{-0.0036}$    &     0.0947 $^{+0.0039}_{-0.0024}$ &     17.12/18           \\
507--686                & 563.77  &       0.1170 $^{+0.0064}_{-0.0039}$    &     0.0921 $^{+0.0038}_{-0.0024}$ &     23.25/21           \\
686--905                & 755.97  &       0.1171 $^{+0.0070}_{-0.0040}$    &     0.0891 $^{+0.0039}_{-0.0023}$ &     24.76/20           \\
\x905--1410               & 1011.02\x &       0.1160 $^{+0.0070}_{-0.0050}$    &     0.0857 $^{+0.0037}_{-0.0027}$ &     24.68/28           \\
1410--2500              & 1508.04\x &       0.1162 $^{+0.0070}_{-0.0062}$    &     0.0822 $^{+0.0034}_{-0.0031}$ &     18.79/14           \\
\hline
\end{tabular}
\label{table:ct10alphaRun}
\end{table}

\begin{table}[htb]
\renewcommand{\arraystretch}{1.5}
\centering
\topcaption{Composition of the uncertainty in \alpsmz fit results in
ranges of \pt. For each range, the corresponding statistical and
experimental systematic uncertainties and the components of the
theoretical uncertainty are shown. The numbers are obtained by
using the CT10 NLO PDF set.}
\begin{tabular}{ccccccc}
\hline
\pt range (\GeVns)  &    PDF unc.               &      scale unc.           &     NP unc.  &    stat unc.              &    syst unc.              &   exp unc.             \\
\hline
\x74--133                & $^{+0.0007}_{-0.0007}$   &  $^{+0.0054}_{-0.0028}$  & $^{+0.0004}_{-0.0004}$ & $^{+0.0016}_{-0.0015}$   &  $^{+0.0020}_{-0.0021}$   & $^{+0.0026}_{-0.0026}$ \\
133--220               & $^{+0.0009}_{-0.0009}$   &  $^{+0.0056}_{-0.0029}$  & $^{+0.0003}_{-0.0003}$ & $^{+0.0008}_{-0.0008}$   &  $^{+0.0019}_{-0.0019}$   & $^{+0.0021}_{-0.0021}$ \\
220--300               & $^{+0.0013}_{-0.0013}$   &  $^{+0.0058}_{-0.0028}$  & $^{+0.0003}_{-0.0003}$ & $^{+0.0003}_{-0.0003}$   &  $^{+0.0018}_{-0.0019}$   & $^{+0.0018}_{-0.0018}$ \\
300--395               & $^{+0.0016}_{-0.0017}$   &  $^{+0.0060}_{-0.0030}$  & $^{+0.0003}_{-0.0003}$ & $^{+0.0004}_{-0.0004}$   &  $^{+0.0016}_{-0.0016}$   & $^{+0.0017}_{-0.0017}$ \\
395--507               & $^{+0.0018}_{-0.0019}$   &  $^{+0.0056}_{-0.0027}$  & $^{+0.0002}_{-0.0003}$ & $^{+0.0007}_{-0.0008}$   &  $^{+0.0014}_{-0.0014}$   & $^{+0.0016}_{-0.0016}$ \\
507--686               & $^{+0.0021}_{-0.0022}$   &  $^{+0.0058}_{-0.0029}$  & $^{+0.0002}_{-0.0003}$ & $^{+0.0006}_{-0.0007}$   &  $^{+0.0014}_{-0.0013}$   & $^{+0.0015}_{-0.0015}$ \\
686--905               & $^{+0.0024}_{-0.0025}$   &  $^{+0.0062}_{-0.0031}$  & $^{+0.0002}_{-0.0002}$ & $^{+0.0014}_{-0.0016}$   &  $^{+0.0015}_{-0.0014}$   & $^{+0.0021}_{-0.0022}$ \\
\x905--1410              & $^{+0.0026}_{-0.0028}$   &  $^{+0.0058}_{-0.0027}$  & $^{+0.0001}_{-0.0002}$ & $^{+0.0021}_{-0.0026}$   &  $^{+0.0017}_{-0.0017}$   & $^{+0.0027}_{-0.0031}$ \\
1410--2500             & $^{+0.0029}_{-0.0032}$   &  $^{+0.0050}_{-0.0033}$  & $^{+0.0001}_{-0.0001}$ & $^{+0.0035}_{-0.0037}$   &  $^{+0.0019}_{-0.0020}$   & $^{+0.0040}_{-0.0042}$ \\
\hline
\end{tabular}
\label{table:ct10alphaUnc}
\end{table}

\section{The QCD analysis of the inclusive jet measurements}
\label{sec:QCD}

The CMS inclusive jet measurements at $\sqrt{s} = 7\TeV$ probe the
gluon and valence-quark distributions in the kinematic range $x
>0.01$~\cite{Khachatryan:2014waa}.  In this paper, we use the
inclusive jet cross section measurements at $\sqrt{s} = 8\TeV$ for
$\pt > 74\GeV$ in a QCD analysis at NLO together with the combined
measurements of neutral- and charged-current cross sections of deep
inelastic electron (positron)-proton scattering at
HERA~\cite{Abramowicz:2015mha}.  The correlations of the experimental
uncertainties for the jet measurements and DIS cross sections are
taken into account. The DIS measurements and the CMS jet cross section
data are treated as uncorrelated.  The theoretical predictions for the
cross sections of jet production are calculated at NLO by using the
\NLOJETPP program~\cite{Nagy:2001fj,Nagy:2003tz} as implemented into
the \fastNLO package~\cite{Britzger:2012bs}.  The open-source QCD fit
framework for PDF determination {\sc
heraf}itter~\cite{Alekhin:2014irh, herafitter}, version 1.1.1, is
used with the parton distributions evolved by using the DGLAP
equations~\cite{Gribov:1972ri,Altarelli:1977zs,Curci:1980uw,Furmanski:1980cm,Moch:2004pa,Vogt:2004mw}
at NLO, as implemented in the {\sc qcdnum}
program~\cite{Botje:2010ay}.

The Thorne--Roberts general mass variable flavour number scheme at
NLO~\cite{Thorne:2006qt,Martin:2009ad} is used for the treatment of
the heavy-quark contributions with the heavy-quark masses $m_{\cPqc}
= 1.47\GeV$ and $m_{\cPqb} = 4.5\GeV$. The renormalization and
factorization scales are set to $Q$, which denotes the four-momentum
transfer in case of the DIS data and the jet \pt in case of the CMS
jet cross sections.

The strong coupling constant is set to \alpsmz = 0.118, as in the
HERAPDF2.0 analysis~\cite{Abramowicz:2015mha} and following the global
PDF analyses, for example, in Ref.~\cite{NNPDF30}.  The $Q^2$ range of
HERA data is restricted to $Q^2 \geq Q^2_{\textrm{min}} = 7.5\GeV^2$.

The procedure for the determination of the PDFs follows the approach
used in the previous QCD analysis~\cite{Khachatryan:2014waa} with the
jet cross section measurements at $\sqrt{s} = 7\TeV$ replaced by those
at 8\TeV. At the initial scale of the QCD evolution $Q_0^2 = 1.9\GeV^2$, the parton distributions are represented by:

\begin{eqnarray}
x\cPg(x) &=& A_{\cPg} x^{B_{\cPg}}\,(1-x)^{C_{\cPg}} \, (1+E_{\cPg} x^2) - A'_{\cPg} ~ x^{B'_{\cPg}} ~ \,(1-x)^{C'_{\cPg}} ,
\label{eq:g}\\
x\cPqu_v(x) &=& A_{\cPqu_v}x^{B_{\cPqu_v}}\,(1-x)^{C_{\cPqu_v}}\,(1+D_{\cPqu_v}x+E_{\cPqu_v}x^2) ,
\label{eq:uv}\\
x\cPqd_v(x) &=& A_{\cPqd_v}x^{B_{\cPqd_v}}\,(1-x)^{C_{\cPqd_v}} \,(1+D_{\cPqd_v}x),
\label{eq:dv}\\
x\overline{U}(x)&=& A_{\overline{U}}x^{B_{\overline{U}}}\, (1-x)^{C_{\overline{U}}}\,(1+D_{\overline{U}}x),
\label{eq:Ubar}\\
x\overline{D}(x)&=& A_{\overline{D}}x^{B_{\overline{D}}}\, (1-x)^{C_{\overline{D}}}\,(1+D_{\overline{D}}x+E_{\overline{D}}x^2).
\label{eq:Dbar}
\end{eqnarray}

The normalization parameters $A_{\cPqu_{\textrm{v}}}$,
$A_{\cPqd_\textrm{v}}$, $A_\cPg$ are determined by the QCD sum rules;
the $B$ parameter is responsible for small-$x$ behavior of the PDFs;
and the parameter $C$ describes the shape of the distribution as
$x\to1$.  A flexible form for the gluon distribution is adopted here,
where the (fixed) choice of $C'_\cPg=25$ is motivated by the approach
of the MSTW group~\cite{Thorne:2006qt,Martin:2009ad}.  Additional
constraints $B_{\overline{\textrm{U}}} = B_{\overline{\textrm{D}}}$
and $A_{\overline{\textrm{U}}} = A_{\overline{\textrm{D}}}(1 -
f_\cPqs)$ are imposed with $f_\cPqs$ being the strangeness fraction,
$f_\cPqs = \cPaqs/( \cPaqd + \cPaqs)$, fixed to
$f_\cPqs=0.31\pm0.08$, as in Ref.~\cite{Martin:2009ad}, consistent
with the determination of the strangeness fraction made by using the CMS
measurements of $\PW$+charm
production~\cite{Chatrchyan:2013mza}. Additional $D$ and $E$
parameters allow probing the sensitivity of results on the specific
selected functional form. The
parameters in Eqs.(~\ref{eq:g})--(\ref{eq:Dbar}) are selected by
first fitting with all $D$ and $E$ parameters set to zero. The other
parameters are then included in the fit one at a time. The improvement
in $\chi^2$ of the fits is monitored
and the procedure is stopped when no further improvement is
observed. This leads to an 18-parameter fit.

The PDF uncertainties are estimated in a way similar to the earlier
CMS analyses~\cite{Chatrchyan:2013mza, Khachatryan:2014waa} according
to the general approach of {\sc HERAPDF1.0}~\cite{Aaron:2009aa} in
which experimental, model, and parameterization uncertainties are
taken into account. The experimental uncertainties originate from the
measurements included in the analysis and are determined by using the
Hessian~\cite{Pumplin:2001ct} method, applying a tolerance criterion
of $\Delta\chi^2 =1$. Alternatively, the Monte Carlo
method~\cite{Giele:1998gw, Giele:2001mr} to determine the PDF
uncertainties is used.

Model uncertainties arise from variations in the values assumed for
the charm and bottom quark masses $m_\cPqc$ and $m_\cPqb$, with $1.41\leq m_\cPqc\leq 1.53\GeV$  and $4.25\leq m_\cPqb\leq 4.75\GeV$, following
Ref.~\cite{Abramowicz:2015mha}, and the value of $Q^2_{\text{min}}$
imposed on the HERA data, which is varied within the interval $5.0 \leq
Q^2_{\text{min}}\leq 10.0\GeV^2$. The strangeness fraction $f_\cPqs$
is varied by its uncertainty.

The parameterization uncertainty is estimated by extending the
functional form of all PDFs with additional
parameters. The uncertainty is constructed as an envelope built from
the maximal differences between the PDFs resulting from all the
parameterization variations and the central fit at each $x$ value.

The total PDF uncertainty is obtained by adding experimental, model,
and parameterization uncertainties in quadrature.  In the following,
the quoted uncertainties correspond to 68\% confidence level. The
global and partial $\chi^2$ values for each data set are listed in
Table~\ref{chi2_pdffit_table}, where the $\chi^2$ values illustrate a
general agreement among all the data sets. The somewhat high
$\chi^2/N_{\mathrm{dof}}$ values for the combined DIS data are very
similar to those observed in Ref.~\cite{Abramowicz:2015mha}, where
they are investigated in detail.

\begin{table}[!h]
\centering
\renewcommand{\arraystretch}{1.2}
\normalsize
\topcaption{Partial $\chi^2/N_{\text{dp}}$ per number of data points
$N_{\textrm{dp}}$ and the global $\chi^2$ per degree of freedom,
$N_{\text{dof}}$, as obtained in the QCD analysis of HERA DIS data
and the CMS measurements of inclusive jet production at $\sqrt{s}=8$
TeV.}
\begin{tabular}[h]{l l | c}
\hline
Data sets &   & Partial $\chi^2/N_{\textrm{dp}}$ \\
\hline
HERA I+II neutral current& $\Pep \Pp$,  $E_\Pp=920\GeV$&   $376/332$  \\
HERA I+II neutral current& $\Pep \Pp$,  $E_\Pp=820\GeV$&    $61/63$ \\
HERA I+II neutral current& $\Pep \Pp$,  $E_\Pp=575\GeV$&   $197/234$ \\
HERA I+II neutral current& $\Pep \Pp$,  $E_\Pp=460\GeV$&   $204/187$ \\
HERA I+II neutral current& $\Pem \Pp$                &   $219/159$ \\
HERA I+II charged current& $\Pep \Pp$                &    $41/39 $ \\
HERA I+II charged current& $\Pem \Pp$                &    $50/42 $ \\
\hline
CMS inclusive jets 8\TeV    &  $\phantom{0.}0<y<0.5$            & $53/36$ \\
&  $0.5<y<1.0$          & $34/36$ \\
&  $1.0<y<1.5$          & $35/35$ \\
&  $1.5<y<2.0$          & $52/29$ \\
&  $2.0<y<2.5$          & $49/24$ \\
&  $2.5<y<3.0$          & $4.9/18$ \\
\hline
Correlated  $\chi^2$        &                       &  $94$ \\
Global $\chi^2/N_{\text{dof}}$ &                       & $1471/1216$\\
\hline
\end{tabular}
\linespread{1.0}
\label{chi2_pdffit_table}
\end{table}

\begin{figure}[h]
\center
\includegraphics[width=0.3\textwidth]{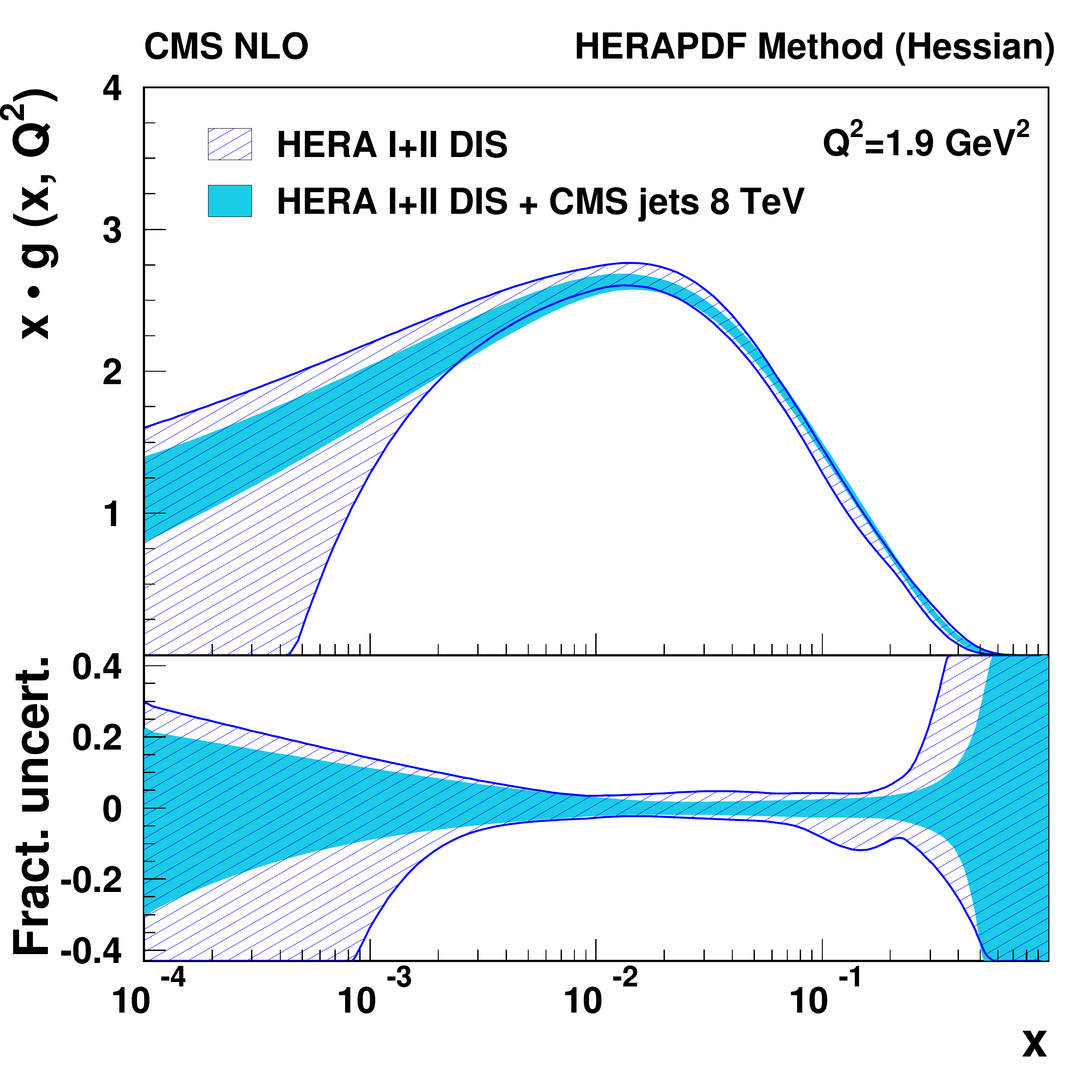}
\includegraphics[width=0.3\textwidth]{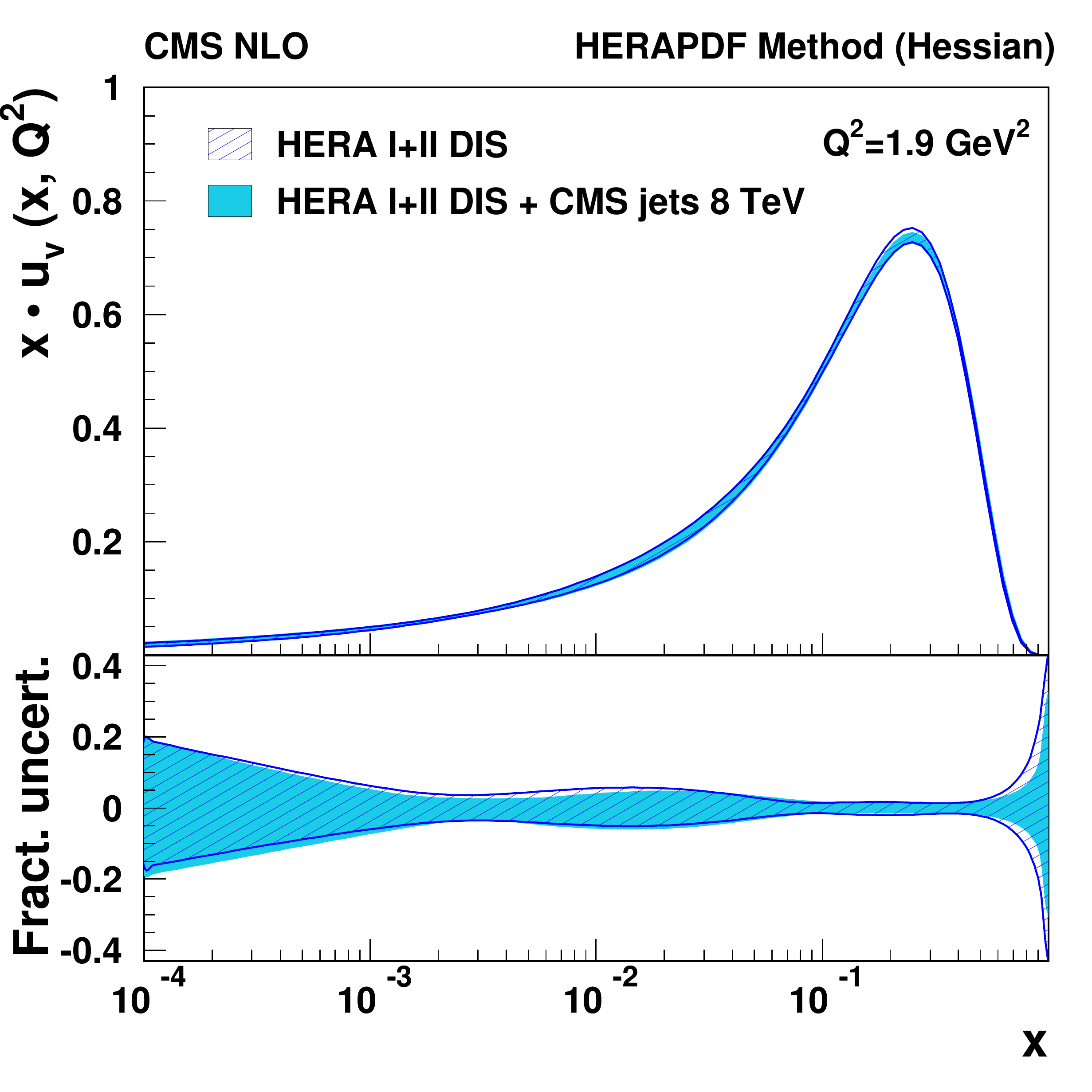}
\includegraphics[width=0.3\textwidth]{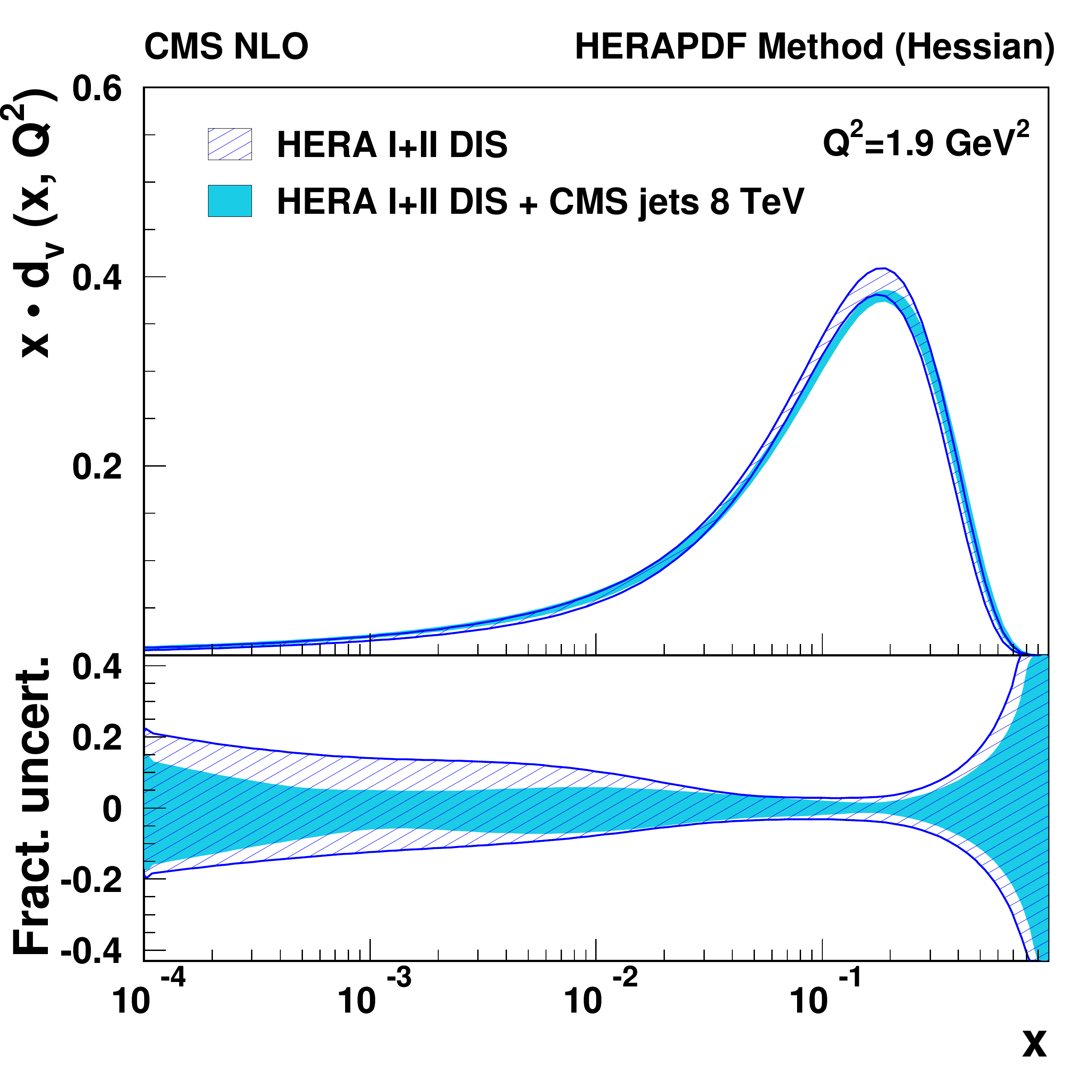}
\setlength{\unitlength}{1cm}
\caption{Gluon (left), $\cPqu$-valence quark (middle), and
$\cPqd$-valence quark (right) distributions as functions of $x$ at
the starting scale $Q^2=1.9\GeV^2$. The results of the fit to the
HERA data and inclusive jet measurements at 8\TeV (shaded band), and
to HERA data only (hatched band) are compared with their total
uncertainties as determined by using the HERAPDF method. In the
bottom panels the fractional uncertainties are shown.}
\label{heraplusjets_HERAPDF_1.9}
\end{figure}

\begin{figure}[h]
\center
\includegraphics[width=0.3\textwidth]{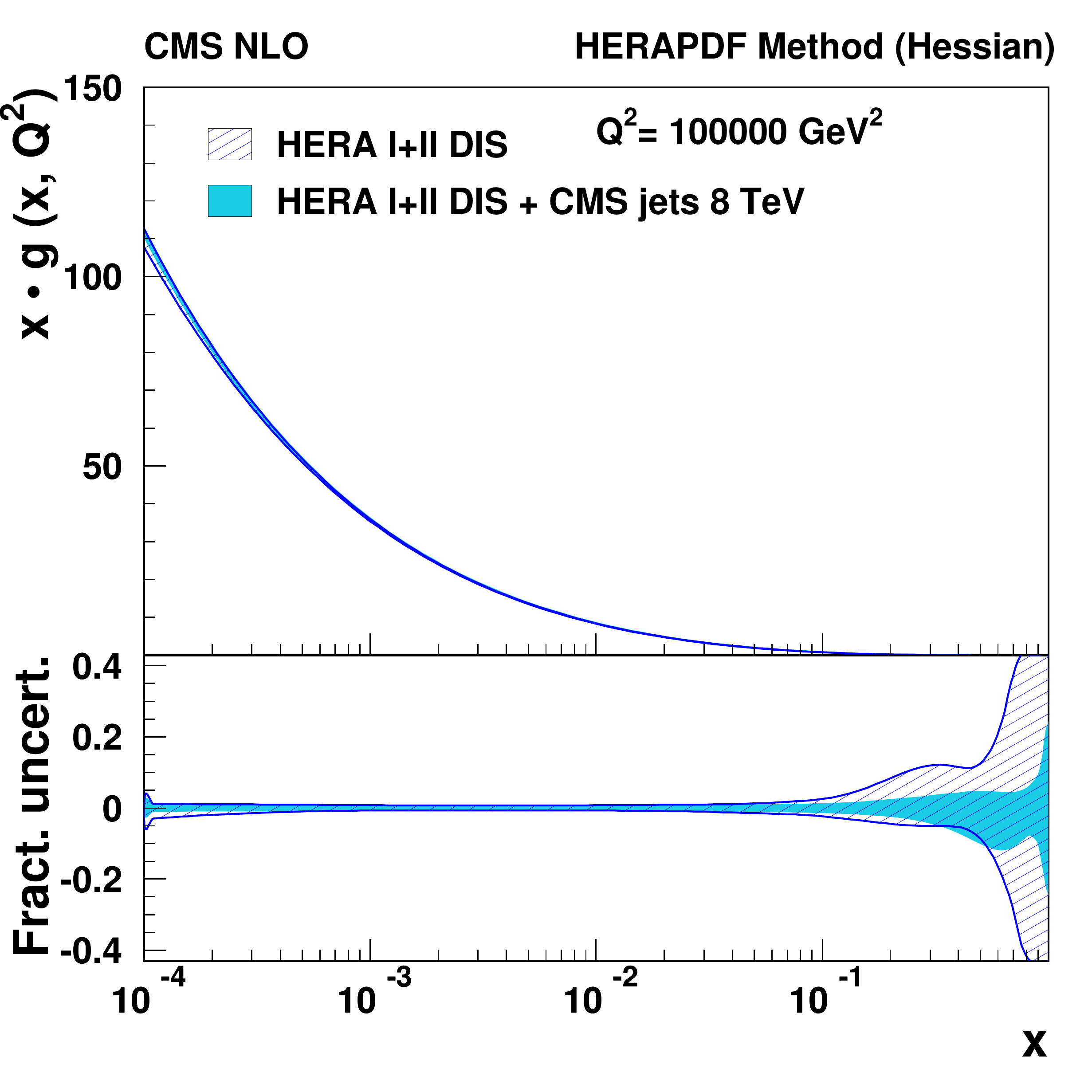}
\includegraphics[width=0.3\textwidth]{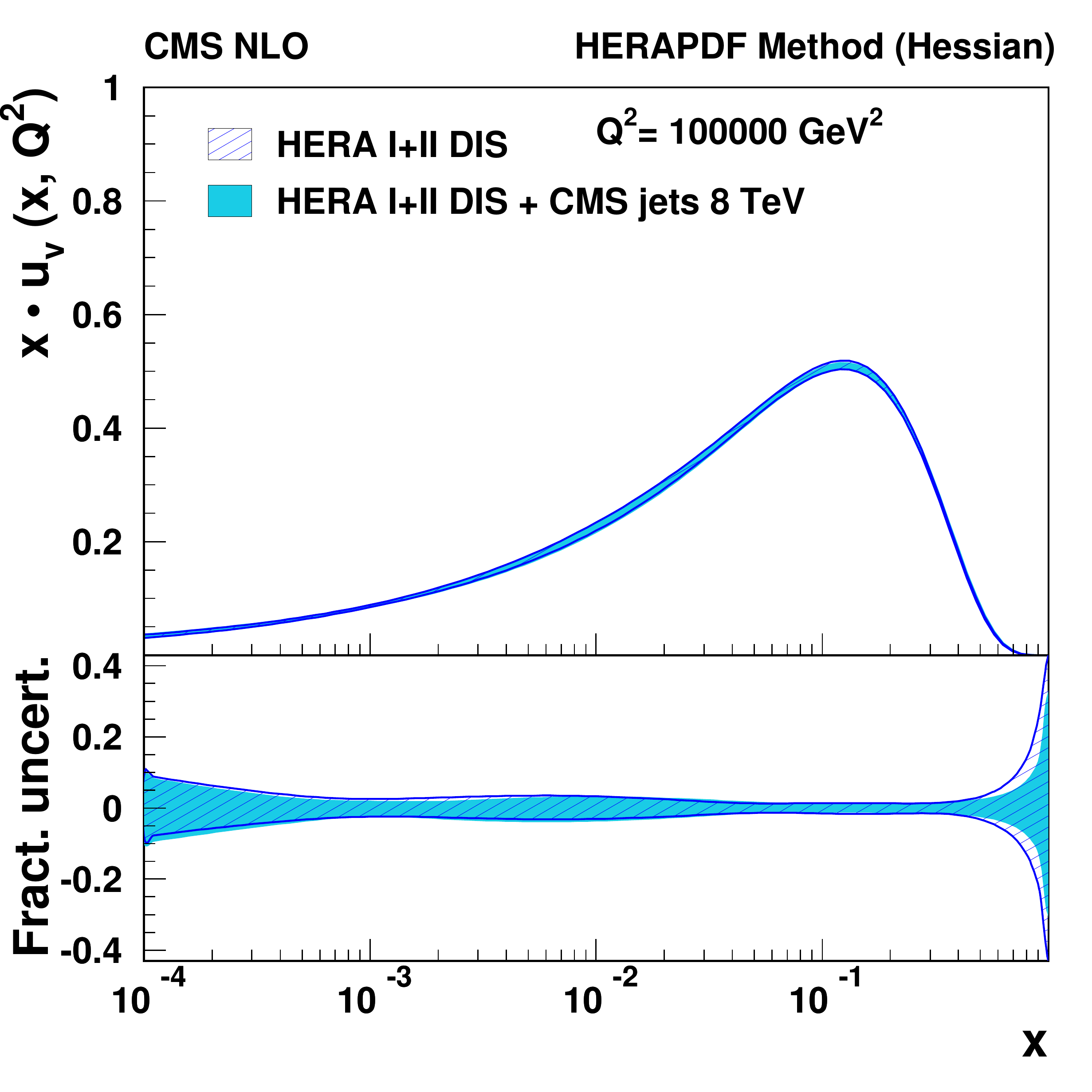}
\includegraphics[width=0.3\textwidth]{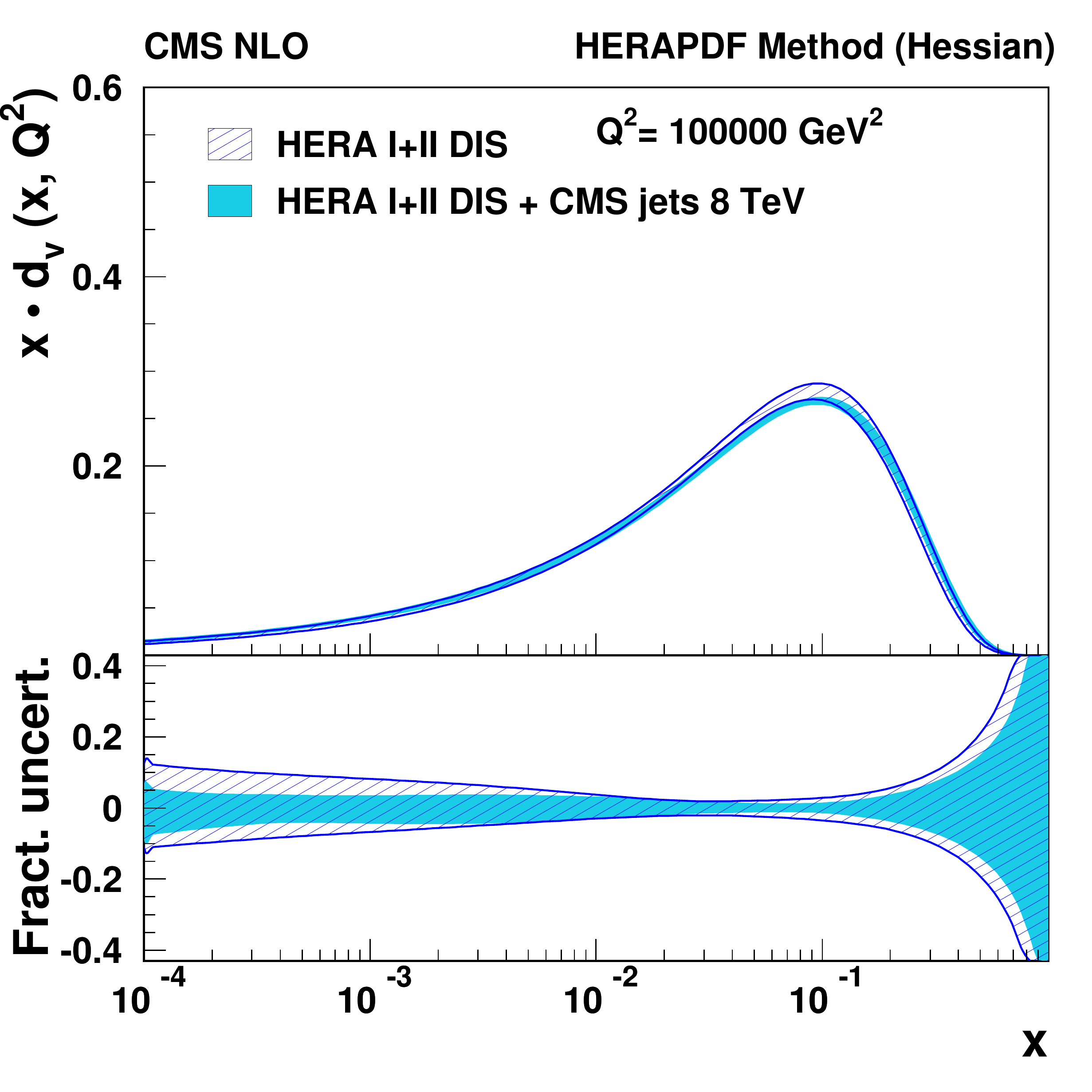}
\setlength{\unitlength}{1cm}
\caption{Same as Fig.~\ref{heraplusjets_HERAPDF_1.9}, but for the
scale of $Q^2=10^5\GeV^2$.}
\label{heraplusjets_HERAPDF_100}
\end{figure}

\begin{figure}[h]
\center
\includegraphics[width=0.4\textwidth]{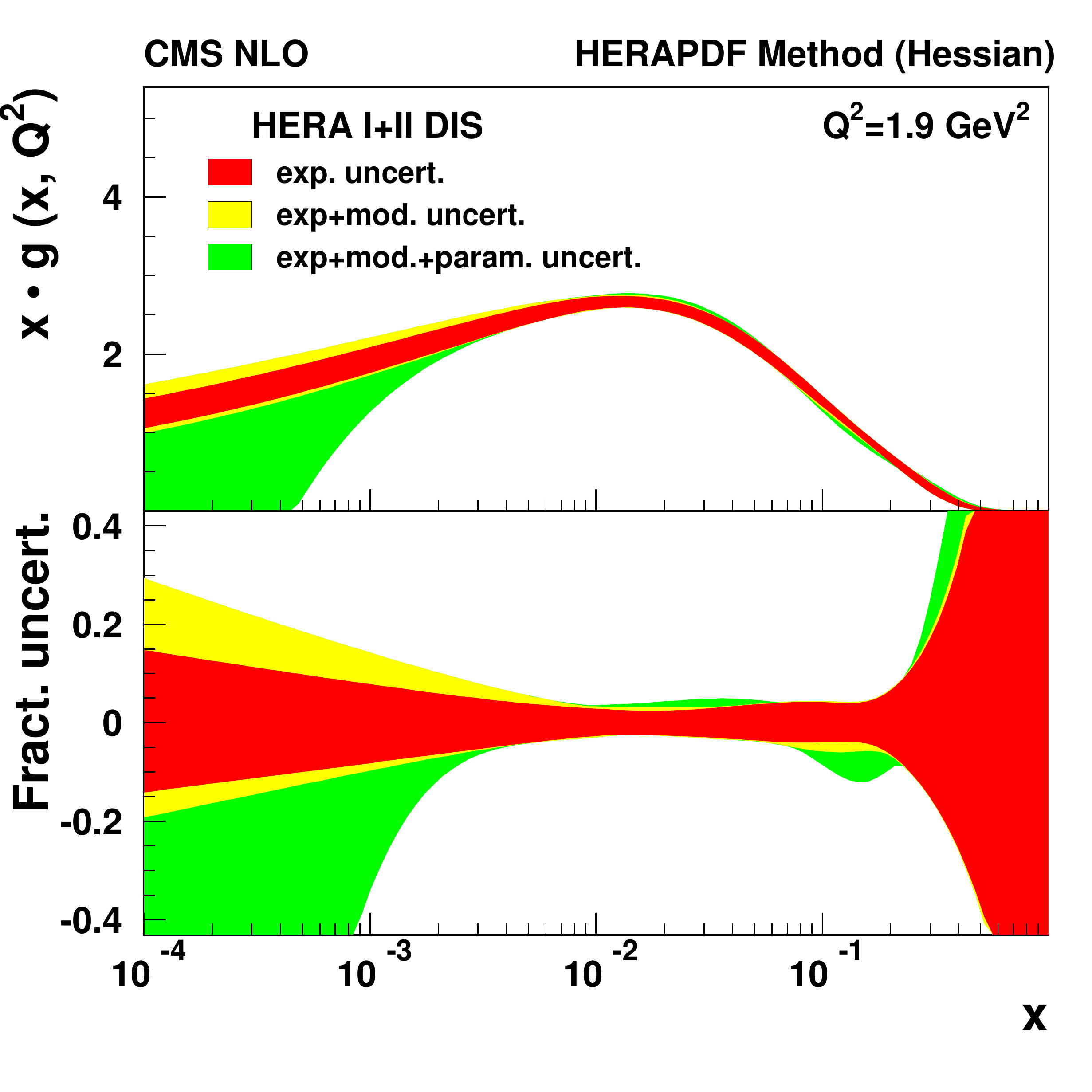}
\includegraphics[width=0.4\textwidth]{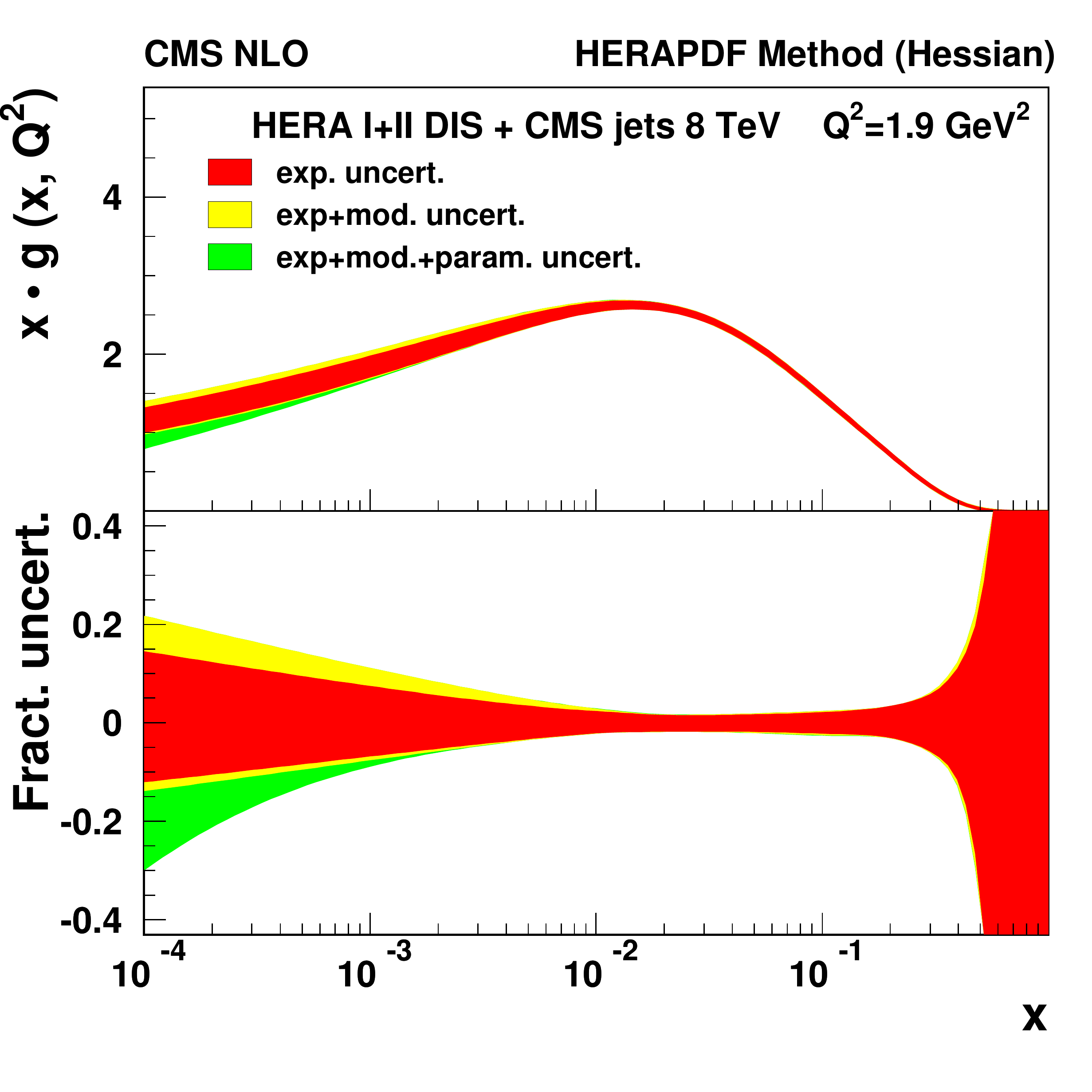}
\setlength{\unitlength}{1cm}
\caption{Gluon PDF distribution as a function of $x$ at the starting
scale $Q^2=1.9\GeV^2$ as derived from HERA inclusive DIS (left)
and in combination with CMS inclusive jet data (right). Different
contributions to the PDF uncertainty are represented by bands of
different shades. In the bottom panels the fractional uncertainties
are shown.}
\label{heraplusjets_unc_HERAPDF}
\end{figure}

Together with HERA DIS cross section data, the inclusive jet
measurements provide important constraints on the gluon and
valence-quark distributions in the kinematic range studied. These
constraints are illustrated in Figs.~\ref{heraplusjets_HERAPDF_1.9}
and \ref{heraplusjets_HERAPDF_100}, where the distributions of the
gluon and valence quarks are shown at the scales of $Q^2=1.9$ and
$10^{5}\GeV^2$, respectively. The results obtained using the Monte
Carlo method to determine the PDF uncertainties are consistent with
those obtained with the Hessian method. The uncertainties for the
gluon distribution, as estimated by using the HERA\-PDF method for
HERA-only and HERA+CMS jet analyses, are shown in
Fig.~\ref{heraplusjets_unc_HERAPDF}. The parameterization uncertainty
is significantly reduced once the CMS jet measurements are included.


The same QCD analysis has been performed using both the low- and
high-\pt measurements of the jet cross sections at 8\TeV and including
the systematic correlations of the two CMS data sets.
The PDFs obtained with the addition of the low-\pt jet cross sections
are consistent with those from the high-\pt jet cross sections alone;
the low-\pt jet cross sections do not, however, improve the PDF
uncertainties significantly.

\begin{figure}[h]
\center
\includegraphics[width=0.4\textwidth]{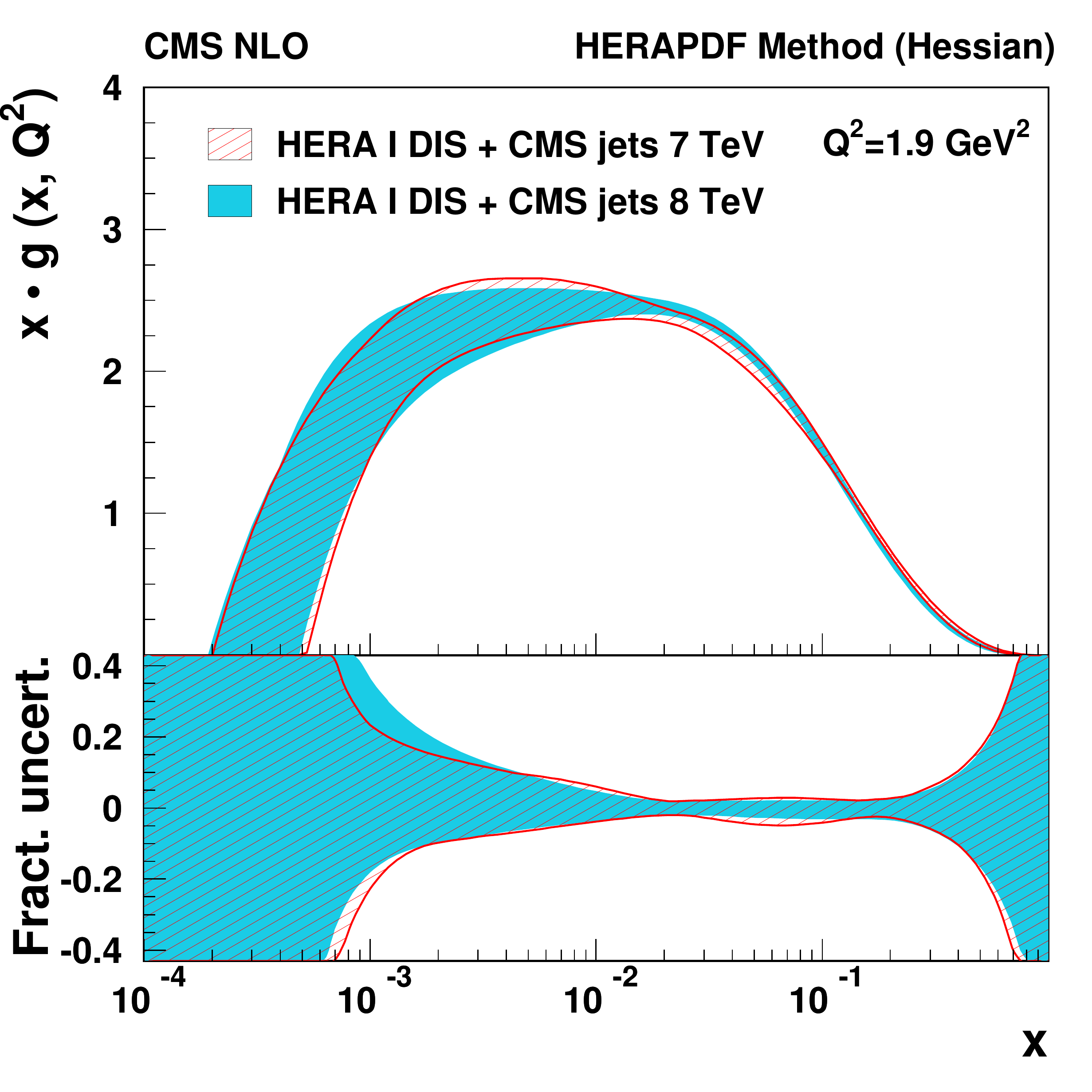}
\includegraphics[width=0.4\textwidth]{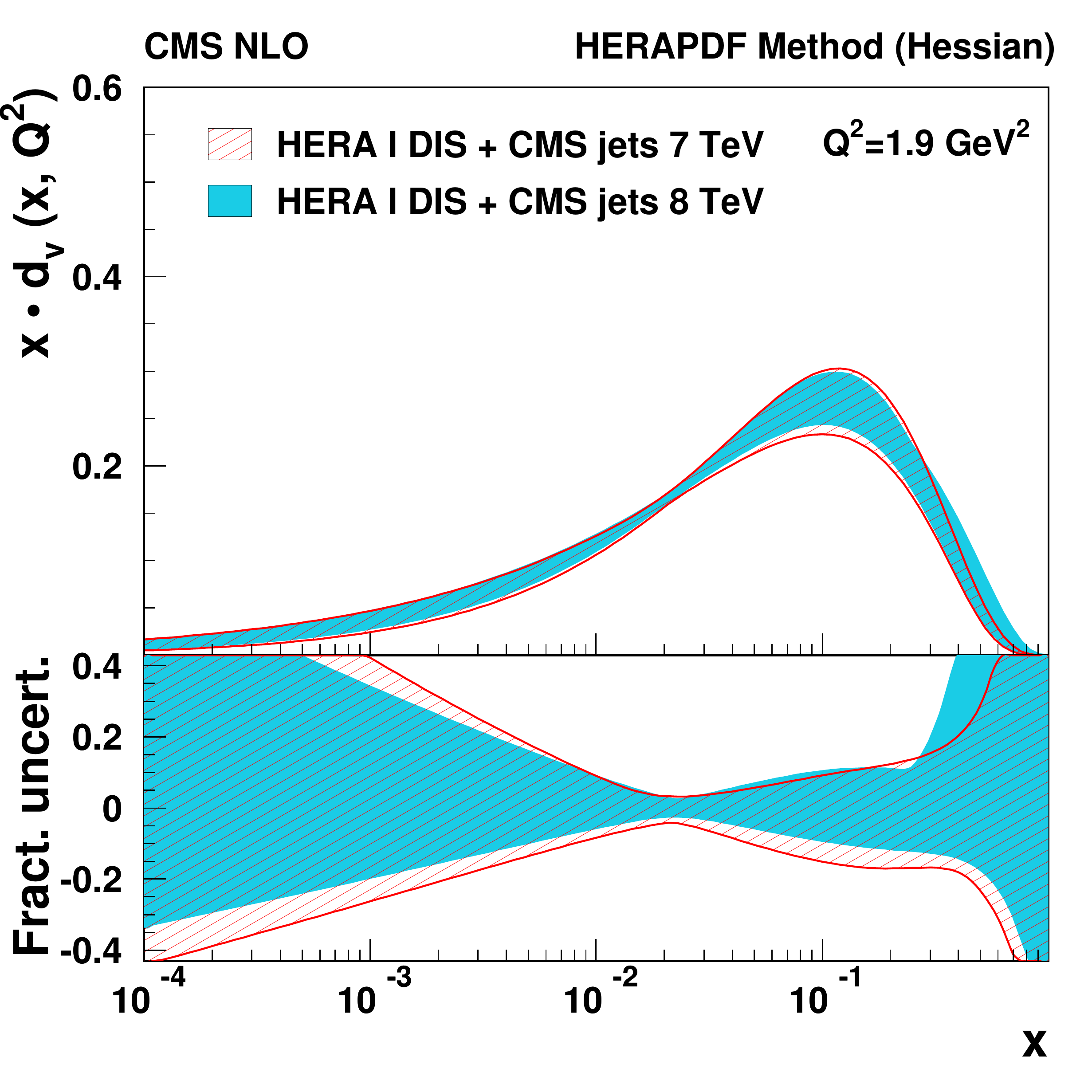}
\setlength{\unitlength}{1cm}
\caption{Gluon (left) and $\cPqd$-valence quark (right) distributions
as functions of $x$ at the starting scale $Q^2=1.9\GeV^2$. The
results of the 13-parameter fit~\cite{Khachatryan:2014waa} to the
subset~\cite{Aaron:2009aa} of the combined HERA data and inclusive
jet measurements at 7\TeV (hatched band), and, alternatively, 8\TeV
(shaded band) are compared with their total uncertainties, as
determined by using the HERAPDF method. In the bottom panels the
fractional uncertainties are shown.}
\label{hera_78}
\end{figure}

The gluon PDFs obtained from the 8\TeV jet cross sections are compared
to those from the 7\TeV cross sections~\cite{Khachatryan:2014waa} in
Fig.~\ref{hera_78}.  The results are very similar.

The extraction of the PDFs from the jet cross sections depends on the
value of \as.  Consequently, the PDF fits
are repeated taking \as to be a free parameter. In this way, the PDFs
and the strong coupling constant are determined simultaneously,
diminishing the correlation between the gluon PDF and \as. The
experimental, model, and parameterization uncertainties of \alpsmz are
obtained in a manner similar to the procedure for determining
uncertainties of the PDFs. The uncertainty due to missing higher-order
corrections in the theoretical predictions for jet production cross
sections is estimated by varying the renormalization and factorization
scales. The scales are varied independently by a factor of two with
respect to the default choice of $\mu_{\mathrm{R}}$ and $\mu_{\mathrm{F}}$
equal to the \pt of the jet and the combined fit of PDFs and \alpsmz
is repeated for each variation of the scale choice in the following
six combinations: ($\mu_{\mathrm{R}}/\pt$, $\mu_{\mathrm{F}}/\pt$) =
(0.5,0.5), (0.5,1), (1,0.5), (1,2), (2,1), and (2,2).
The scale for the HERA DIS data is not changed. The maximal
observed upward and downward changes of \alpsmz with respect to the
default are then taken as the scale uncertainty. The strong coupling
constant is $\alpsmz = 0.1185 ^{+0.0019}_{-0.0021}\,
(\text{exp})^{+0.0002}_{-0.0015} \,
(\text{model})^{+0.0000}_{-0.0004}(\text{param})^{+0.0022}_{-0.0018}\,
(\text{scale})$.  Within the uncertainties, this value is consistent
with the one determined in Section~\ref{sec:alphas} and is an
important cross-check of the \alpsmz obtained by using the fixed
PDF. The scale uncertainties in \alpsmz obtained simultaneously with
the PDFs are smaller due to consistent treatment of the scales in the
PDFs and the theory prediction for the jet cross sections in the
simultaneous fit. The evaluation of scale
uncertainties is an open issue that is ignored in all global PDF fits
to date. There is no recommended procedure for the determination of
the scale uncertainties in combined fits of PDFs and \alpsmz.

\section{Summary}

A measurement of the double-differential inclusive jet cross section
has been presented that uses data from proton-proton collisions at
$\sqrt{s}=8\TeV$ collected with the CMS detector and corresponding to
an integrated luminosity of $19.7\fbinv$. The result is presented as a
function of jet transverse momentum \pt and absolute rapidity $\abs{y}$
and covers a large range in jet \pt from $74\GeV$ up to $2.5\TeV$, in
six rapidity bins up to $\abs{y}=3.0$. The region of low jet \pt, in
particular the range from 21 to 74\GeV, has also been studied up to
$\abs{y}=4.7$, using a dedicated low-pileup 5.6\pbinv data
sample. The ratios to the cross sections measured at 2.76 and 7\TeV
have been also determined.

Detailed studies of experimental and theoretical sources of
uncertainty have been carried out. The dominant sources of
experimental systematic uncertainty are due to the jet energy scale,
unfolding, and the integrated luminosity measurement. These lead to
uncertainties of 5--45\% in the differential cross section
measurement. The theoretical predictions are most affected by PDF
uncertainties, and their range is strongly dependent on the \pt and
rapidity interval; at low \pt they are about 7\%, but their size
increases up to 40\% in the most central intervals and exceeds 200\%
in the outermost regions. Many uncertainties cancel in the ratio with
the corresponding results at 2.76 and 7\TeV, leading to uncertainties
ranging from 5\% to 25\%, both for the measurement and for the
theoretical predictions. Perturbative QCD, supplemented by a small
nonperturbative and electroweak corrections, describes the data over a
wide range of jet \pt and $y$.

The strong coupling constant is extracted from the high-\pt jet cross
section measurements using the probed \pt range and six different
rapidity bins. The best fitted value is $\alpsmz =
0.1164^{+0.0060}_{-0.0043}$ using the CT10 NLO PDF set.  The running
of the strong coupling constant as a function of the energy scale $Q$,
\asq, measured for nine different values of energy scale between
86\GeV and 1.5\TeV, is in good agreement with previous experiments and
extends the measurement to the highest values of the energy scale.

This measurement of the double-differential jet cross section probes
hadronic parton-parton interaction over a wide range of $x$ and
$Q$. The QCD analysis of these data together with HERA DIS
measurements illustrates the potential of the high-\pt jet cross
sections to provide important constraints on the gluon PDF in a new
kinematic regime.

\begin{acknowledgments}

\hyphenation{Bundes-ministerium Forschungs-gemeinschaft Forschungs-zentren} We congratulate our colleagues in the CERN accelerator departments for the excellent performance of the LHC and thank the technical and administrative staffs at CERN and at other CMS institutes for their contributions to the success of the CMS effort. In addition, we gratefully acknowledge the computing centres and personnel of the Worldwide LHC Computing Grid for delivering so effectively the computing infrastructure essential to our analyses. Finally, we acknowledge the enduring support for the construction and operation of the LHC and the CMS detector provided by the following funding agencies: the Austrian Federal Ministry of Science, Research and Economy and the Austrian Science Fund; the Belgian Fonds de la Recherche Scientifique, and Fonds voor Wetenschappelijk Onderzoek; the Brazilian Funding Agencies (CNPq, CAPES, FAPERJ, and FAPESP); the Bulgarian Ministry of Education and Science; CERN; the Chinese Academy of Sciences, Ministry of Science and Technology, and National Natural Science Foundation of China; the Colombian Funding Agency (COLCIENCIAS); the Croatian Ministry of Science, Education and Sport, and the Croatian Science Foundation; the Research Promotion Foundation, Cyprus; the Secretariat for Higher Education, Science, Technology and Innovation, Ecuador; the Ministry of Education and Research, Estonian Research Council via IUT23-4 and IUT23-6 and European Regional Development Fund, Estonia; the Academy of Finland, Finnish Ministry of Education and Culture, and Helsinki Institute of Physics; the Institut National de Physique Nucl\'eaire et de Physique des Particules~/~CNRS, and Commissariat \`a l'\'Energie Atomique et aux \'Energies Alternatives~/~CEA, France; the Bundesministerium f\"ur Bildung und Forschung, Deutsche Forschungsgemeinschaft, and Helmholtz-Gemeinschaft Deutscher Forschungszentren, Germany; the General Secretariat for Research and Technology, Greece; the National Scientific Research Foundation, and National Innovation Office, Hungary; the Department of Atomic Energy and the Department of Science and Technology, India; the Institute for Studies in Theoretical Physics and Mathematics, Iran; the Science Foundation, Ireland; the Istituto Nazionale di Fisica Nucleare, Italy; the Ministry of Science, ICT and Future Planning, and National Research Foundation (NRF), Republic of Korea; the Lithuanian Academy of Sciences; the Ministry of Education, and University of Malaya (Malaysia); the Mexican Funding Agencies (BUAP, CINVESTAV, CONACYT, LNS, SEP, and UASLP-FAI); the Ministry of Business, Innovation and Employment, New Zealand; the Pakistan Atomic Energy Commission; the Ministry of Science and Higher Education and the National Science Centre, Poland; the Funda\c{c}\~ao para a Ci\^encia e a Tecnologia, Portugal; JINR, Dubna; the Ministry of Education and Science of the Russian Federation, the Federal Agency of Atomic Energy of the Russian Federation, Russian Academy of Sciences, and the Russian Foundation for Basic Research; the Ministry of Education, Science and Technological Development of Serbia; the Secretar\'{\i}a de Estado de Investigaci\'on, Desarrollo e Innovaci\'on and Programa Consolider-Ingenio 2010, Spain; the Swiss Funding Agencies (ETH Board, ETH Zurich, PSI, SNF, UniZH, Canton Zurich, and SER); the Ministry of Science and Technology, Taipei; the Thailand Center of Excellence in Physics, the Institute for the Promotion of Teaching Science and Technology of Thailand, Special Task Force for Activating Research and the National Science and Technology Development Agency of Thailand; the Scientific and Technical Research Council of Turkey, and Turkish Atomic Energy Authority; the National Academy of Sciences of Ukraine, and State Fund for Fundamental Researches, Ukraine; the Science and Technology Facilities Council, UK; the US Department of Energy, and the US National Science Foundation.

Individuals have received support from the Marie-Curie programme and the European Research Council and EPLANET (European Union); the Leventis Foundation; the A. P. Sloan Foundation; the Alexander von Humboldt Foundation; the Belgian Federal Science Policy Office; the Fonds pour la Formation \`a la Recherche dans l'Industrie et dans l'Agriculture (FRIA-Belgium); the Agentschap voor Innovatie door Wetenschap en Technologie (IWT-Belgium); the Ministry of Education, Youth and Sports (MEYS) of the Czech Republic; the Council of Science and Industrial Research, India; the HOMING PLUS programme of the Foundation for Polish Science, cofinanced from European Union, Regional Development Fund, the Mobility Plus programme of the Ministry of Science and Higher Education, the National Science Center (Poland), contracts Harmonia 2014/14/M/ST2/00428, Opus 2013/11/B/ST2/04202, 2014/13/B/ST2/02543 and 2014/15/B/ST2/03998, Sonata-bis 2012/07/E/ST2/01406; the Tha\-lis and Aristeia programmes cofinanced by EU-ESF and the Greek NSRF; the National Priorities Research Program by Qatar National Research Fund; the Programa Clar\'in-COFUND del Principado de Asturias; the Rachadapisek Sompot Fund for Postdoctoral Fellowship, Chulalongkorn University and the Chulalongkorn Academic into Its 2nd Century Project Advancement Project (Thailand); and the Welch Foundation, contract C-1845.

\end{acknowledgments}

\bibliography{auto_generated}
\cleardoublepage \appendix\section{The CMS Collaboration \label{app:collab}}\begin{sloppypar}\hyphenpenalty=5000\widowpenalty=500\clubpenalty=5000\textbf{Yerevan Physics Institute,  Yerevan,  Armenia}\\*[0pt]
V.~Khachatryan, A.M.~Sirunyan, A.~Tumasyan
\vskip\cmsinstskip
\textbf{Institut f\"{u}r Hochenergiephysik der OeAW,  Wien,  Austria}\\*[0pt]
W.~Adam, E.~Asilar, T.~Bergauer, J.~Brandstetter, E.~Brondolin, M.~Dragicevic, J.~Er\"{o}, M.~Flechl, M.~Friedl, R.~Fr\"{u}hwirth\cmsAuthorMark{1}, V.M.~Ghete, C.~Hartl, N.~H\"{o}rmann, J.~Hrubec, M.~Jeitler\cmsAuthorMark{1}, A.~K\"{o}nig, I.~Kr\"{a}tschmer, D.~Liko, T.~Matsushita, I.~Mikulec, D.~Rabady, N.~Rad, B.~Rahbaran, H.~Rohringer, J.~Schieck\cmsAuthorMark{1}, J.~Strauss, W.~Treberer-Treberspurg, W.~Waltenberger, C.-E.~Wulz\cmsAuthorMark{1}
% ...
\vskip\cmsinstskip
% ...
1:~~Also at Vienna University of Technology, Vienna, Austria\\
% ...

\end{sloppypar}
\end{document}